\documentclass[iop]{emulateapj}

\usepackage{graphicx}
\usepackage{amsmath}

\newcommand{\ms}{\ifmmode{\rm m\thinspace s^{-1}}\else m\thinspace s$^{-1}$\fi}
\newcommand{\kms}{\ifmmode{\rm km\thinspace s^{-1}}\else km\thinspace s$^{-1}$\fi}
\newcommand{\kepler}{{\it Kepler}}
\newcommand{\Kepler}{{\it KEPLER}}
\newcommand\blender{{\tt BLENDER}}
\newcommand\cofiam{{\tt CoFiAM}}

\shorttitle{Validation of small HZ planets}
\shortauthors{Torres et al.}

\begin{document}

\submitted{To appear in The Astrophysical Journal}

\title{Validation of twelve small \Kepler\ transiting planets in the
  habitable zone}

\author{
Guillermo Torres\altaffilmark{1,2},
David M.\ Kipping\altaffilmark{1},
Francois Fressin\altaffilmark{1},
Douglas A.\ Caldwell\altaffilmark{3},
Joseph D.\ Twicken\altaffilmark{3},
Sarah Ballard\altaffilmark{4,5},
Natalie M.\ Batalha\altaffilmark{6},
Stephen T.\ Bryson\altaffilmark{6},
David R.\ Ciardi\altaffilmark{7},
Christopher E.\ Henze\altaffilmark{6},
Steve B.\ Howell\altaffilmark{6},
Howard T.\ Isaacson\altaffilmark{8},
Jon M.\ Jenkins\altaffilmark{6},
Philip S.\ Muirhead\altaffilmark{9,10},
Elisabeth R.\ Newton\altaffilmark{1},
Erik A.\ Petigura\altaffilmark{8},
Thomas Barclay\altaffilmark{6},
William J.\ Borucki\altaffilmark{6},
Justin R.\ Crepp\altaffilmark{11},
Mark E.\ Everett\altaffilmark{12},
Elliott P.\ Horch\altaffilmark{13},
Andrew W.\ Howard\altaffilmark{14},
Rea Kolbl\altaffilmark{8},
Geoffrey W.\ Marcy\altaffilmark{8},
Sean McCauliff\altaffilmark{15}, and
Elisa V.\ Quintana\altaffilmark{6}
}

\altaffiltext{1}{Harvard-Smithsonian Center for Astrophysics, 60
  Garden Street, Cambridge, MA 02138, USA}

\altaffiltext{2}{E-mail: gtorres@cfa.harvard.edu}

\altaffiltext{3}{SETI Institute/NASA Ames Research Center, Moffett
  Field, CA 94035, USA}

\altaffiltext{4}{University of Washington, Seattle, WA 98195, USA}

\altaffiltext{5}{NASA Carl Sagan Fellow}

\altaffiltext{6}{NASA Ames Research Center, Moffett Field, CA 94035,
  USA}

\altaffiltext{7}{NASA Exoplanet Science Institute/Caltech, Pasadena,
  CA 91125, USA}

\altaffiltext{8}{Astronomy Department, University of California,
  Berkeley, CA 94720, USA}

\altaffiltext{9}{Department of Astronomy, Boston University, Boston, MA
  02215, USA}

\altaffiltext{10}{Hubble Postdoctoral Fellow}

\altaffiltext{11}{University of Notre Dame, Notre Dame, IN 46556,
  USA}

\altaffiltext{12}{National Optical Astronomy Observatory, Tucson, AZ
  85719, USA}

\altaffiltext{13}{Department of Physics, Southern Connecticut State
  University, New Haven, CT 06515, USA}

\altaffiltext{14}{Institute for Astronomy, University of Hawaii at
Manoa, Honolulu, HI 96822, USA}

\altaffiltext{15}{Orbital Sciences Corporation/NASA Ames Research
Center, Moffett Field, CA 94035, USA}

\begin{abstract}

We present an investigation of twelve candidate transiting planets
from \kepler\ with orbital periods ranging from 34 to 207 days,
selected from initial indications that they are small and potentially
in the habitable zone (HZ) of their parent stars. Few of these objects are known. The expected Doppler
signals are too small to confirm them by demonstrating that their
masses are in the planetary regime. Here we verify their planetary
nature by validating them statistically using the \blender\ technique,
which simulates large numbers of false positives and compares the
resulting light curves with the \kepler\ photometry. This analysis was
supplemented with new follow-up observations (high-resolution optical
and near-infrared spectroscopy, adaptive optics imaging, and speckle
interferometry), as well as an analysis of the flux centroids. For
eleven of them (KOI-0571.05, 1422.04, 1422.05, 2529.02, 3255.01,
3284.01, 4005.01, 4087.01, 4622.01, 4742.01, and 4745.01) we show that
the likelihood they are true planets is far greater than that of a
false positive, to a confidence level of 99.73\% (3$\sigma$) or
higher. For KOI-4427.01 the confidence level is about 99.2\%
(2.6$\sigma$).  With our accurate characterization of the GKM host
stars, the derived planetary radii range from 1.1 to
2.7\,$R_{\earth}$.  All twelve objects are confirmed to be in the HZ,
and nine are small enough to be rocky. Excluding three of them that
have been previously validated by others, our study doubles the number
of known rocky planets in the HZ.  KOI-3284.01 (Kepler-438\,b) and KOI-4742.01 (Kepler-442\,b) are the
planets most similar to the Earth discovered to date when considering
their size and incident flux jointly.

\end{abstract}


\section{Introduction}
\label{sec:introduction}

Over the duration of its four-year mission the \kepler\ spacecraft has enabled the
identification of several thousand candidate transiting planets
\citep[\kepler\ Objects of Interest, or KOIs;][]{Borucki:11,
  Batalha:13, Burke:14}, and many more continue to be found from
reanalysis of the original data with increasingly sophisticated
methods. Only a tiny fraction of these candidates have been
``confirmed'' in the traditional sense of having had their masses
measured, either spectroscopically or by modeling their transit timing
variations (TTVs). Hundreds of others, mostly in multiple systems
(``multis''), have recently been shown statistically to have a very
high chance of being true planets \citep{Lissauer:14, Rowe:14}, even
if their masses are not currently known.

Public curiosity and scientific interest have motivated efforts in the
last few years to find and confirm rocky planets similar in size to
the Earth that are orbiting in the so-called habitable zone (HZ) of
their parent stars, usually taken in this context to be the region in
which water on the surface can be in a liquid state.  Recent estimates
by \cite{Petigura:13} suggest the rate of occurrence of Earth-size
planets in the HZ of Sun-like stars (which they defined broadly as
corresponding to an incident flux between $\slantfrac{1}{4}$ and 4
times that of the Earth) may be as high as $22 \pm 8$\% \citep[for
  other estimates see][]{Zsom:13, Pintr:14, Foreman:14}, but very few of these have actually been confirmed.  These two
conditions for habitability --- rocky nature and suitable location ---
are sometimes more challenging to establish unambiguously than it may
seem, as they require an accurate knowledge of the host star
properties, something that is not always trivial to achieve for the
faint \kepler\ targets.  Definitions of the HZ are actively debated
and have changed over time \citep[e.g.,][]{Kasting:93, Selsis:07,
  Kopparapu:13, Seager:13, Leconte:13, Zsom:13, Kopparapu:14}, as has
our understanding of the radius at which planets transition from being
rocky to being dominated by thick hydrogen/helium envelopes, which
presumably makes them unsuitable for life as we know it. The interface
is currently thought to be between 1.5 and 2\,$R_{\earth}$
\citep[e.g.,][]{Lopez:13, Weiss:14, Marcy:14, Rogers:14, Wolfgang:14},
but many uncertainties remain and it is possible that some planets
that are slightly larger may still be rocky.

These difficulties notwithstanding, a handful of transiting planets
that appear to meet these conditions have already emerged from the
\kepler\ sample. Among those considered at the time of publication to
be smaller than 2.5\,$R_{\earth}$ and in or near the HZ, the first to
be announced was Kepler-22\,b \citep{Borucki:12}, with a radius of
$R_p = 2.38 \pm 0.13$\,$R_{\earth}$. Others followed including
Kepler-61\,b \citep[$2.15 \pm 0.13$\,$R_{\earth}$;][]{Ballard:13},
Kepler-62\,e and f \citep[$1.61 \pm 0.05$\,$R_{\earth}$ and $1.41 \pm
  0.07$\,$R_{\earth}$;][]{Borucki:13}, Kepler-69\,c
\citep[$1.71_{-0.23}^{+0.34}$\,$R_{\earth}$;][]{Barclay:13}, and
recently Kepler-186\,f \citep[$1.11 \pm
  0.14$\,$R_{\earth}$;][]{Quintana:14}. All of these were
``validated'' statistically using techniques such as
\blender\ \citep{Torres:04, Torres:11, Fressin:12} by examining the
likelihood of different false positive scenarios to measure the
confidence in the planetary interpretation. Some of them are in
multi-planet systems; as mentioned above, \cite{Lissauer:14} have
developed a statistical argument by which they demonstrate that most
candidates in multis are very likely to be true planets. Based on this
reasoning and a careful examination of follow-up observations,
\cite{Lissauer:14} reported an additional small planet (Kepler-296\,f,
$2.31 \pm 0.12$\,$R_{\earth}$) that appears to be in the HZ, according
to \cite{Rowe:14}. Taking advantage of the same statistical framework
for validation, the latter authors announced one more small HZ planet
in a multi-planet system: Kepler-174\,d ($2.19 \pm
0.13$\,$R_{\earth}$).  Other small HZ candidates have been proposed
\citep[see, e.g.,][]{Muirhead:12, Dressing:13, Gaidos:13, Mann:13b,
  Petigura:13, Star:14} but have yet to be investigated in detail and
confirmed. A few additional examples also believed to be in this
special group were later shown to fail one or both requirements (size
and location), either because of incorrect stellar parameters
(originally based on photometry and later improved with spectroscopy)
or because of the discovery of other stars in the photometric aperture
that dilute the transits and change the inferred planetary radius
\citep[e.g.,][]{Star:14}.

In this paper we investigate a sample of twelve \kepler\ candidates
identified on the basis of preliminary analyses indicating that they
are small and likely to be in or near the HZ of their parent stars.
Our goal is to validate them as bona-fide planets, and to confirm
their key properties pertaining to habitability.  We describe our
efforts over the past year to obtain the follow-up observations
necessary to robustly characterize the host stars and to validate the
signals. In the interim, three of them that are in systems with five
candidates each (KOI-0571.05 = Kepler-186\,f, KOI-1422.04 =
Kepler-296\,f, and KOI-1422.05 = Kepler-296\,e) were the subject of
recent studies by others. KOI-1422.04 and KOI-1422.05 were announced
as planets by \cite{Lissauer:14} and \cite{Rowe:14} using their
statistical framework for validation of multis ``in bulk'', though
only the first was mentioned as being in the HZ.  KOI-0571.05 was
validated independently by \cite{Quintana:14}. These authors relied in
part on the fact that false positives are much less common in multis,
and on the validation of the other four planets in the same system by
the previous authors, in order to reach a sufficiently high level of
confidence for the fifth candidate.  Most of the other targets in our
sample are considerably more difficult to validate because they tend
to have long orbital periods and do not belong to multiple systems.
Nevertheless, as we show below, the application of the
\blender\ technique used in many of the previous discoveries is able
to achieve validations here as well. Our work has now essentially doubled the number of known small HZ planets.

Our paper is organized as follows: Sect.~\ref{sec:sample} and
Sect.~\ref{sec:photometry} describe our sample of candidates and the
\kepler\ photometry we use. The follow-up observations including
high-resolution imaging, optical spectroscopy, and near-infrared
spectroscopy are presented in Sect.~\ref{sec:followup}, where we also
report the complicating discovery of nearby stellar companions to four
of the targets. These close companions not only dilute the transits
but also bring ambiguity as to the precise location of the planets in
these systems. A description of the analysis of the flux centroids
that addresses some of those companions is given here as well.
Sect.~\ref{sec:stellarproperties} follows with a determination of the
stellar properties. The formal validation of our candidates is the
subject of Sect.~\ref{sec:validation}, after which we proceed with the
transit light curve fits to derive the planetary parameters
(Sect.~\ref{sec:fits}). In Sect.~\ref{sec:ap} we apply the powerful
technique of asterodensity profiling to investigate whether the
planets that are in multiple systems orbit the same star, to extract
eccentricity information from the light curves of our targets, and for
the four host stars with close stellar companions, to address the
possibility that the planets orbit the companions rather than the
intended targets. The topic of habitability is examined in
Sect.~\ref{sec:habitability}. We conclude with a discussion of our
findings and final remarks.

\section{Candidate identification}
\label{sec:sample}

Candidate transiting planets of special interest for this study were
initially identified as `threshold crossing events' \citep[TCEs;
  objects with a $7.1 \sigma$ or greater multiple-event transit
  detection statistic; see][]{Jenkins:02, Jenkins:10} from a transit
search conducted using nearly three years of \kepler\ data gathered
from quarters 1 to 12, with version 8.3 of the Science Operations
Center (SOC) processing pipeline \citep{Tenenbaum:13}. The analysis of
three years of data gave us the first chance to detect three transits
from planets in habitable-zone orbits around stars like the Sun, as
well as to significantly increase the signal-to-noise ratio (SNR) for
transit signals from habitable-zone orbits around cooler stars.  In
order to investigate the smaller planets potentially in the HZ, we
selected for further scrutiny the TCEs that met the criteria $R_p <
2\,R_{\earth}$ and $T_{\rm eq} <303$\,K, as well as a transit SNR
$>7$. Here $T_{\rm eq}$ represents the planetary equilibrium
temperature, calculated assuming an Earth-like albedo, and the stellar
properties relied on information from the \kepler\ Input Catalog
\citep[KIC;][]{Brown:11}.  We used data products from versions 8.3 and
9.0 of the SOC Data Validation (DV) pipeline module for the initial
triage and vetting \citep[for an overview of this process,
  see][]{Wu:10}\footnote{All DV results are available from the NASA
  Exoplanet Archive: {\tt
    http://exoplanetarchive.ipac.caltech.edu}\,.}.  We compared the
results of our vetting with the scores from early runs of an automated
vetting code being developed at the time \citep{McCauliff:14}. From
the 18,407 TCEs that passed through DV, we identified 385 TCEs around
338 unique stars that met our selection criteria.  From this list of
385 TCEs we identified ten that passed flux and centroid vetting
criteria and that had low false-alarm probabilities from the
auto-vetter.  During this vetting process, the SOC completed a transit
search of \kepler\ data from quarters 1--16 using version 9.1 of the
processing pipeline. With the addition of three quarters of data and
the improvements in the version 9.1 DV products, we promoted an
additional TCE (KOI-4742.01), which had marginal signal-to-noise ratio
in the 12-quarter run, and identified KOI-1422.05, the fifth KOI and
second habitable-zone candidate around this star.  The complete list
of targets selected for this study is given in
Table~\ref{tab:targets}.

\begin{deluxetable*}{llrcccc}
\tablewidth{0pc}
\tablecaption{Sample of KOIs in this study.\label{tab:targets}}
\tablehead{
\colhead{} &
\colhead{} &
\colhead{} &
\colhead{$K\!p$} &
\colhead{$b$} &
\colhead{Period} &
\colhead{Depth}
\\
\colhead{Candidate} &
\colhead{Name} &
\colhead{KID} &
\colhead{(mag)} &
\colhead{(deg)} &
\colhead{(days)} &
\colhead{(ppm)}
}
\startdata
  KOI-0571.05 & Kepler-186\,f &  8120608 &  14.625 & \phn8.2 &    129.94 & \phn540 \\
  KOI-1422.04 & Kepler-296\,f & 11497958 &  15.921 &    18.0 & \phn63.34 &    1070 \\
  KOI-1422.05 & Kepler-296\,e & 11497958 &  15.921 &    18.0 & \phn34.14 & \phn850 \\
  KOI-2529.02 & Kepler-436\,b &  8463346 &  15.856 & \phn6.5 & \phn64.00 &    1150 \\
  KOI-3255.01 & Kepler-437\,b &  8183288 &  14.352 & \phn9.0 & \phn66.65 & \phn520 \\
  KOI-3284.01 & Kepler-438\,b &  6497146 &  14.467 &    18.6 & \phn35.23 & \phn400 \\
  KOI-4005.01 & Kepler-439\,b &  8142787 &  14.560 &    19.9 &    178.14 & \phn700 \\
  KOI-4087.01 & Kepler-440\,b &  6106282 &  15.134 &    15.8 &    101.11 &    1010 \\
  KOI-4427.01 &               &  4172805 &  15.645 & \phn8.2 &    147.66 &    1220 \\
  KOI-4622.01 & Kepler-441\,b & 11284772 &  15.142 &    19.1 &    207.25 & \phn960 \\
  KOI-4742.01 & Kepler-442\,b &  4138008 &  14.976 &    15.0 &    112.31 & \phn560 \\
  KOI-4745.01 & Kepler-443\,b & 11757451 &  15.891 &    17.0 &    177.67 & \phn930 
\enddata

\tablecomments{Columns after the first indicate the \kepler\ planet
  designation, \kepler\ identification number, brightness in the
  \kepler\ passband, Galactic latitude, orbital period, and transit
  depth in parts per million relative to the out-of-transit stellar
  flux. For consistency in this paper we will refer to all objects by
  their original KOI names throughout, even though the validations
  described later earn them the official \kepler\ planet designations
  listed in the second column.}

\end{deluxetable*}

We note that there is significant contamination of the TCE list for
periods near the orbital period of the \kepler\ spacecraft (372 days)
caused by instrumental image artifacts on specific detector channels,
as has been noted by \citet{Tenenbaum:13}. The contamination is due to
moir{$\acute{\rm e}$} pattern noise injected during readout of some of
the detector chains \citep{Caldwell:10, Kolodziejczak:10}. The
moir{$\acute{\rm e}$} noise generates a host of single-event
transit-like signals for targets on the noisy channel.  When folded at
a period near one \kepler\ year these signals can give multiple-event
statistics above the 7.1$\sigma$ detection threshold, as the same
target star falls on the same noisy channel. We have not included any
KOIs from the moir{$\acute{\rm e}$} pattern noise channels in our
analysis here. Two of the KOIs (KOI-2529.02 and KOI-4005.01) do show
some of their transits on detector channel 56, which has increased
noise relative to the other channels on which these targets
fall. However, we included these KOIs in our sample because both show
transits on other channels and both passed all of our vetting checks.

\section{\Kepler\ photometry}
\label{sec:photometry}

For the analysis in this paper we have made use of the publicly
available \kepler\ data for each KOI from the
\href{http://archive.stsci.edu/}{Mikulski Archive for Space
  Telescopes} (MAST).\footnote{{\tt
    https://archive.stsci.edu/index.html}\,.} The observations span
\kepler\ quarters 1--17, corresponding to a period of four years
beginning in 2009 May. The downloaded data were made available as part
of Data Release 23 and were processed using SOC Pipeline version
9.1. While long-cadence (29.4\,min) data were available for all KOIs,
short-cadence (58.9\,s) data were only available for KOI-0571 and
KOI-1422. In the following we describe our further processing of the
simple aperture photometry (SAP) time series used for the detailed
light-curve analysis of all KOIs. Later in Sect.~\ref{sec:rotation}
and Sect.~\ref{sec:overview} we describe the slightly different
processing of the photometry used to extract stellar rotation
information and for the \blender\ analysis.

Prior to making use of the SAP measurements we removed instrumental
and stellar photometric variability in the \kepler\ data that can
disturb the transit light curve profile. This process was conducted in
two steps: pre-detrending cleaning, and long-term detrending. The
pre-detrending cleaning was carried out independently for each transit
event, restricted to plus or minus half an orbital period surrounding
the mid-transit time. We visually inspected each epoch and removed any
charge-trapping ramps, flare-like events, and instrumental
discontinuities in the data. We made no attempt to correct these
artifacts and simply excluded them from the photometry manually. We
then removed all of the transit signals of the other known candidates
in each system (for stars having multiple candidates) within $\pm0.6$
transit durations centered on the mid-times, using the reported
ephemerides. Finally, we cleaned the data of 3$\sigma$ outliers from a
moving median smoothing curve with a 20-point window.

Next we removed long-term trends, which can be due to instrumental
effects such as focus drift, or stellar effects such as rotational
modulations. For this task we used the Cosine Filtering with
Autocorrelation Minimization (\cofiam) algorithm, which was
specifically developed to protect the shape of a transit light
curve. We briefly highlight the main features of \cofiam\ and the
specific inputs used for this analysis, and we direct the reader to
\cite{Kippingetal:13} for a more detailed description of the technique.
It is essentially a Fourier-based method that removes periodicities
occurring at timescales greater than a pre-designated ``protected
timescale''. In this work, we selected three times the known transit
duration to be the protected timescale, which ensures that the transit
profile is not distorted in frequency space. \cofiam\ does not
directly attempt to remove high frequency noise, since the Fourier
transform of a trapezoidal-like light curve contains significant high
frequency power \citep{Waldmann:12}. It is able to explore many
different harmonics by trying longer protected timescales than the
nominal choice (we capped the maximum number of harmonics at 30) and
evaluate the autocorrelation at a pre-selected timescale (we used
30\,minutes) locally surrounding each transit. From these different
harmonics, we selected the harmonic order that minimizes this local
autocorrelation, as quantified using the Durbin-Watson statistic. This
``Autocorrelation Minimization'' component of \cofiam\ provides
optimized data for subsequent analysis. For each KOI we defined the
local transit data as being within six transit durations either side
of the mid-transit time, in order to provide an ample out-of-transit
baseline. These local data were divided through by the final
\cofiam\ function and then stitched together to form our final light
curve for analysis.

\section{Follow-up observations and centroid analysis}
\label{sec:followup}

\subsection{High-resolution imaging}
\label{sec:imaging}

The presence of other stars blended with our targets is a potential
concern, as these companions could be orbited by another object
causing eclipses that may be the source of the transit signals we
observe. Even if they are not eclipsing, the light of the companion
stars will attenuate the signal and lead to a bias in the planetary
radius determination if the effect is not accounted for.

Images from the $J$-band UK Infrared Telescope survey
\citep[UKIRT;][]{Lawrence:07} have indeed revealed companions to three
of our targets (KOI-2529, 3284, and 4427) that are close enough to
fall within the photometric aperture of \kepler\ (i.e., within a few
arc seconds), although as we describe later they are ruled out as the
source of the transits by our centroid analysis in
Sect.~\ref{sec:centroids}.  We list these companions in
Table~\ref{tab:ao}, based on information taken from the
\kepler\ Community Follow-up Observing Program (CFOP)
Web site\footnote{\tt
  https://cfop.ipac.caltech.edu/home/\,.}. Additional companions to
three other KOIs are seen in the UKIRT images at larger separations
(and are also excluded by the centroid analysis), but these stars are
all very faint ($\Delta m \gtrsim 5$), and have a negligible impact on
the inferred planetary sizes.

\begin{deluxetable}{cccll}
\tablewidth{0pc}
\tablecaption{Close companions to target stars.\label{tab:ao}}
\tablehead{
\colhead{} &
\colhead{Ang.\ sep.} &
\colhead{P.A.} &
\colhead{Mag.\ diff.\tablenotemark{a}} &
\colhead{}
\\
\colhead{Star} &
\colhead{(\arcsec)} &
\colhead{(deg)} &
\colhead{(mag)} &
\colhead{Source}
}
\startdata
KOI-1422 & 0.220 & 216.3 & $\Delta R = 1.72$\tablenotemark{b} & Speckle  \\
KOI-2529 & 0.402 & 350.4 & $\Delta K^{\prime} = 4.69$ & Keck AO \\
\nodata  & 5.05\phn  & \phn42.6 & $\Delta J = 3.75$ & UKIRT \\
KOI-3255 & \phm{\tablenotemark{c}}0.180\tablenotemark{c} & 336.5 & $\Delta K^{\prime} = 0.05$ & Keck AO \\
KOI-3284 & \phm{\tablenotemark{c}}0.438\tablenotemark{c} & 193.2  & $\Delta K^{\prime} = 2.03$ & Keck AO \\
\nodata  & \phm{\tablenotemark{d}}4.10\tablenotemark{d}\phn & 359.7 & $\Delta J = 2.90$ & UKIRT \\
KOI-4427 & 4.76\phn  & 274.8 & $\Delta J = 2.87$ & UKIRT 
\enddata

\tablenotetext{a}{For targets with both a close and a wide companion
  the magnitude differences for the wide companions are relative to
  the total brightness of the inner pair.}

\tablenotetext{b}{Also observed at 880\,nm (approximately Sloan $z$
  band), giving $\Delta z = 1.62$. The separation and P.A.\ in the
  table are the average of the two speckle bands. Similar results were
  reported by \cite{Star:14} from \emph{HST} observations.}

\tablenotetext{c}{This companion was also detected in our speckle
  imaging observations.}

\tablenotetext{d}{This companion was also detected in the $UBV$ survey
  of \cite{Everett:12}, with magnitude differences $\Delta B = 1.80$
  and $\Delta V = 2.01$. The separation and P.A.\ in the table are the
  average of the two surveys.}
\end{deluxetable}

The UKIRT images have a typical seeing-limited resolution of about
0\farcs8 or 0\farcs9. To explore the inner regions around our targets
beyond the reach of UKIRT we observed them with near-infrared adaptive
optics (AO) in the $J$ (1.248\,$\mu$m) and $K^{\prime}$
(2.124\,$\mu$m) filters using the NIRC2 imager \citep{Wizinowich:04,
  Johansson:08} on the Keck\,II, 10\,m telescope. KOI-0571 and
KOI-3255 were observed in August 2012, and the rest in August 2013 as
part of a general infrared AO survey of KOIs \citep[e.g.,][]{Adams:12,
  Rowe:14, Marcy:14}. KOI-1422 was not observed with AO for this
project as it has been the target of a separate effort using
\emph{HST} \citep{Gilliland:14, Star:14}; those authors found it to
have a close companion (see below). For KOI-0571, 3255, 2529, 3284,
and 4742 the targets themselves were used as natural guide stars; for
KOI-4005, 4087, 4427, 4622, and 4745 we used the laser guide-star AO
system. In all cases the observations were obtained in a 3-point
dither pattern to avoid the lower left quadrant of the NIRC2 array,
which displays elevated noise.  Five images were collected per dither
pattern position, each shifted 0\farcs5 from the previous dither
position to enable the use of the source frames for the creation of
the sky image.

\begin{figure}
\centering
\begin{tabular}{cc}

\includegraphics[width=4.2cm]{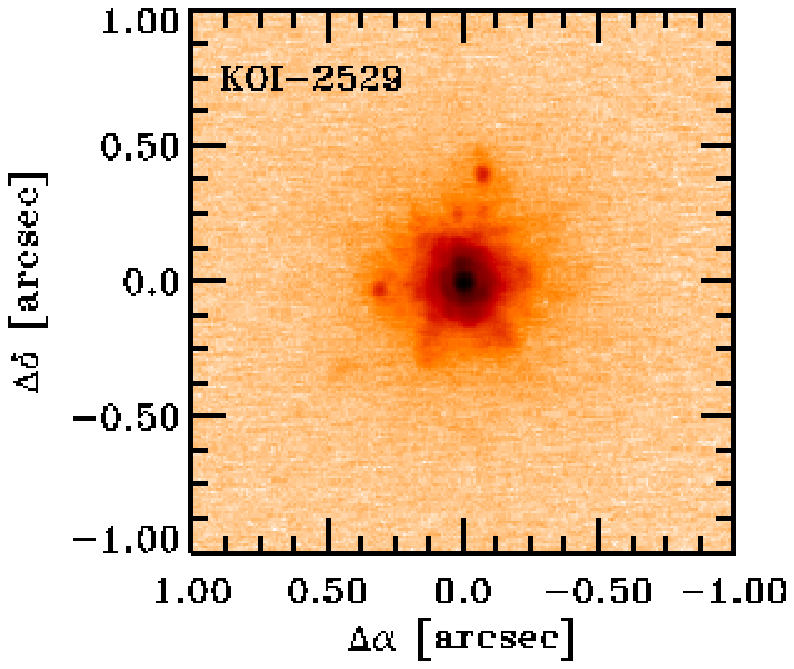} &
\includegraphics[width=4.2cm]{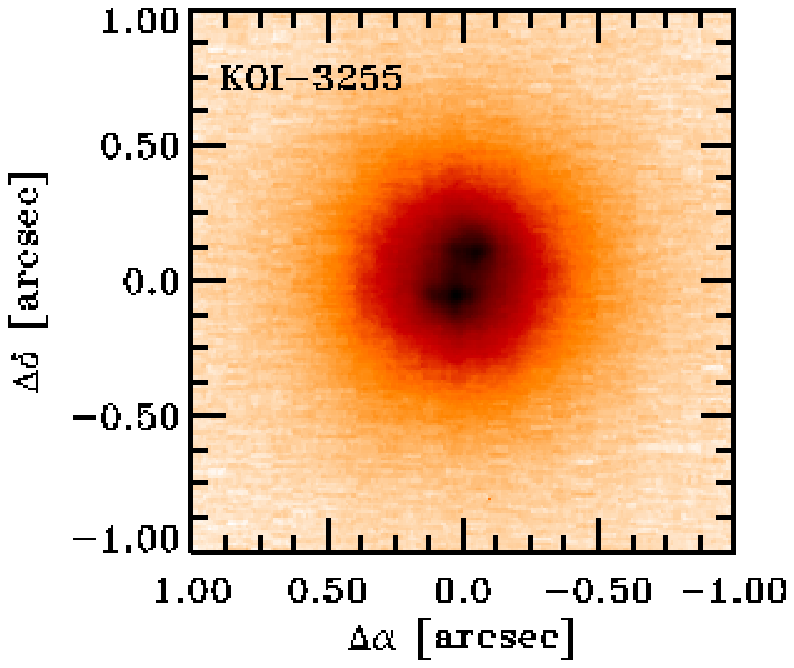}\\[1ex]
\multicolumn{2}{c}{\includegraphics[width=4.2cm]{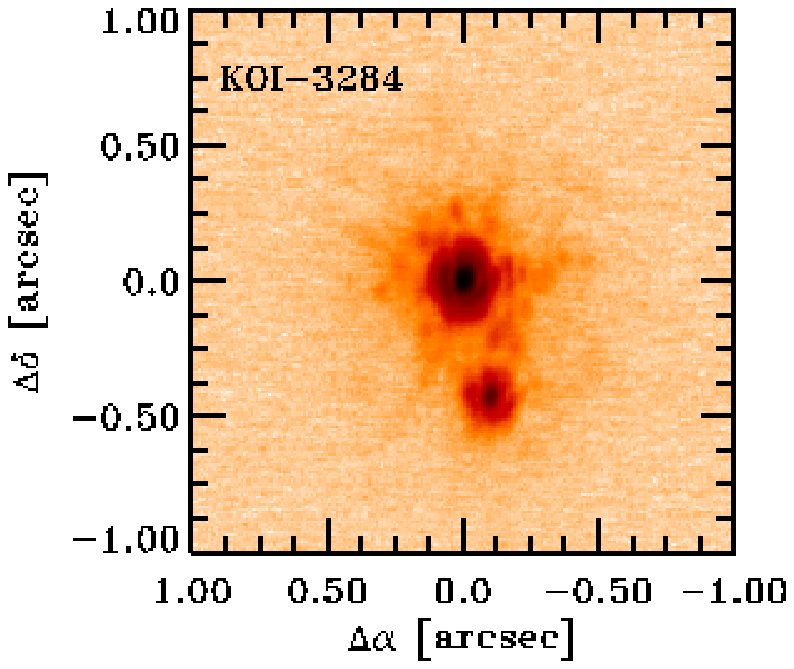}}\\
\end{tabular}

\figcaption[]{Images in the $K^{\prime}$ band from our Keck AO imaging
  observations showing close companions found around KOI-2529,
  KOI-3255, and KOI-3284 (see also Table~\ref{tab:ao}).\label{fig:ao}}

\end{figure}

In general the NIRC2 imaging detects all of the sources found in the
UKIRT $J$-band imaging within 5\arcsec\ of each target. The NIRC2
array has $1024 \times 1024$ pixels with a scale of about
10\,mas\,pix$^{-1}$, and a field of view of $10\farcs1 \times
10\farcs1$.  Each frame was dark-subtracted and flat-fielded, and the
sky frames were constructed for each target from the target frames
themselves by median filtering and co-adding the 15 or 25 dithered
frames. Individual exposure times varied depending on the brightness
of the target, but were long enough to obtain at least 5000 counts per
frame (NIRC2 has a gain of 4 electrons per DN); frame times were
typically 1--30 seconds.  Data reduction was performed with a custom
set of IDL routines. Close companions were detected around KOI-2529,
3255, and 3284 (see Figure~\ref{fig:ao}), two of which also have wider
companions seen in the UKIRT images. We report the relative positions
and brightness of these close companions also in Table~\ref{tab:ao}.

\setlength{\tabcolsep}{0pt}
\setlength{\tabcolsep}{6pt}

\begin{figure}
\epsscale{1.05}
\plotone{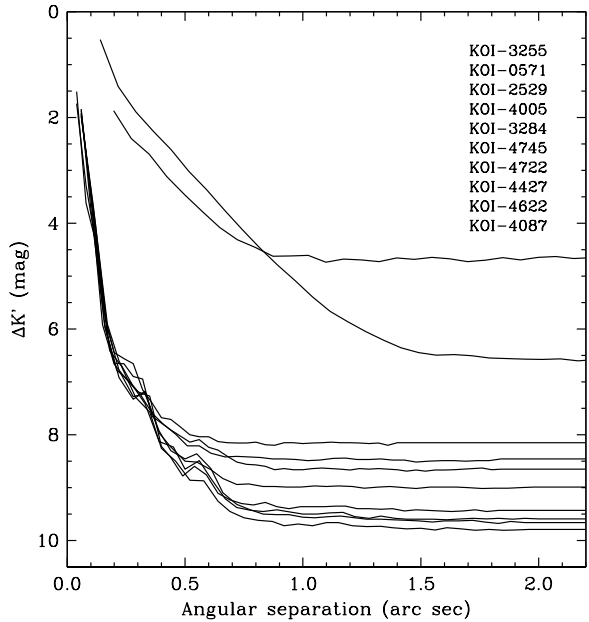}

\figcaption[]{Sensitivity curves from our Keck AO imaging observations
  in the (2.124\,$\mu$m) $K^\prime$ band. Curves correspond to the
  KOIs as labeled, top to bottom.\label{fig:sensitivity}}

\end{figure}

Point source detection limits were estimated in a series of concentric
annuli drawn around each target star.  The separation and widths of
the annuli were set to the full width at half maximum (FWHM) of the
target point spread function (PSF).  The standard deviation of the
background counts was calculated for each annulus, and the 5$\sigma$
limits were determined for each annular region \citep[see
  also][]{Adams:12}.  The PSF widths for our Keck images were
typically found to be 4--5 pixels, corresponding to
0\farcs04--0\farcs05 FWHM.  Typical contrast levels from the images
are 2--3 magnitudes at a separation of 1 FWHM, 7--8 magnitudes at 5+
FWHM, and deeper past 10 FWHM. Sensitivity curves in the $K^{\prime}$
band for each of the targets are shown in
Figure~\ref{fig:sensitivity}, and extend to angular separations of
approximately 3\arcsec, set by the dither pattern overlap. We
supplemented these measurements with similar sensitivity estimates
made from the UKIRT images available on the CFOP Web site, which reach
to much wider separations but are not as deep.

Additional high-resolution imaging observations for four of our
targets (KOI-0571, 1422, 3255, and 3284) were obtained with the
Differential Speckle Survey Instrument \citep[DSST;][]{Horch:09,
  Howell:11} mounted on the 8.1\,m Gemini-N telescope during the time
period UT 2012 July 25--28. The seeing conditions were generally
0\farcs5--0\farcs8 during those nights. The large aperture of the
Gemini-N primary mirror allows for high SNR observations at faint
magnitudes and achieves a very small diffraction limit for
visible-light observations. The speckle camera provides simultaneous
observations in two filters. For all Gemini-N observations discussed
here we used filters centered at 692\,nm and 880\,nm (corresponding
very roughly to the Cousins $R$ and Sloan $z$ bands), with
band widths of 40\,nm and 50\,nm, respectively.  Although full details
of the observing protocols have been given previously by
\cite{Horch:12}, we summarize them here for completeness.

For bright targets observed with this instrument ($V < 12$) a single
sequence of 1000 speckle frames of 60\,ms duration each has usually
been found to be sufficient, for a total of 3--4 minutes of observing
time. The much fainter sources discussed here ($15 < V < 17$) required
up to eight, 1000-frame sequences that were later combined, or about
30 minutes of on-source exposure time each. Calibration images and
sequences were taken as is usual to allow us to measure and define
dispersion effects, position angles, faint limit sensitivity,
photometric performance, and for point source reconstruction.

The speckle observations revealed no stars in the vicinity of
KOI-0571, but KOI-1422 was found to have a close companion (seen in
both filters) that was also subsequently detected in the \emph{HST}
observations by \cite{Gilliland:14}. Its brightness and position
relative to the primary are given in Table~\ref{tab:ao}. Close
companions were also found around KOI-3255 and KOI-3284, which are the
same as detected in our AO imaging with Keck.

\begin{figure}
\epsscale{1.05}
\plotone{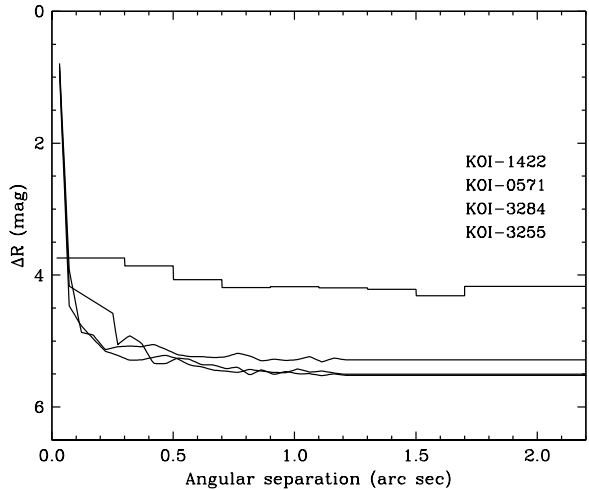}

\figcaption[]{Similar to Figure~\ref{fig:sensitivity} for the four
  targets observed with speckle interferometry at the Gemini North
  telescope. Curves are shown for the (692\,nm) $R$-band observations,
  and correspond to the KOIs as labeled, top to
  bottom.\label{fig:sensitivity_speckle}}

\end{figure}

For estimating our detection threshold for faint companions we
considered each local maximum in the reconstructed image as a
potential stellar source, and determined the statistics of these peaks
(i.e., their average values and standard deviation) as a function of
angular distance from the primary star. We then adopted a conservative
5$\sigma$ threshold for the detection of any companion stars.  The
details of the calibration procedures mentioned above and the
precision that may be obtained are described by \cite{Horch:11,
  Horch:12}. The sensitivity curves for the four \kepler\ targets
observed here with speckle imaging are shown in
Figure~\ref{fig:sensitivity_speckle}.

\subsection{Centroid analysis}
\label{sec:centroids}

One method to identify possible false positives due to background
eclipsing binaries measures the location of the transit signal
relative to each KOI host star via difference imaging.  For each
quarter, a difference image may be formed by subtracting an average of
in-transit pixel values from out-of-transit pixel values.  If the
transit signal is due to a stellar source, the difference image will
show that stellar source, whose location is determined by Pixel
Response Function centroiding \citep[PRF; see][]{Bryson:10}.  This
location may then be compared to the location of the host star,
determined from a PRF-fit centroid of an average out-of-transit image
(assuming the host is well isolated).  Repeating this for each
quarter in which transits occur gives a collection of offsets of the
transit source from the host star.  The average of these quarterly
offsets defines the final offset of the transit source from the host
star. This metric is reported for each KOI on the NASA Exoplanet
Archive. If the offset is greater than three times its uncertainty,
there is cause for concern that the signal may be coming from a
background source rather than the KOI itself. In that case we would
conclude that the likelihood that the transit signal is due to a
planet around the KOI host star is low \citep[for details,
  see][]{Bryson:13}.  Any transit source location less than 3$\sigma$
from the host star is considered statistically indistinguishable from
the host star.  This 3$\sigma$ exclusion radius depends on the KOI
rather than the host star, as it is primarily a function of the
signal-to-noise ratio of the KOI's transit signal.

Table~\ref{tab:centroids} lists the average offset of the transit
source for each KOI from the KOI host star, as reported on the
Exoplanet Archive, as well as the 3$\sigma$ exclusion radius for each
KOI.  All KOIs have offsets that are less than three times their
uncertainties, with the exception of KOI-4427.01. Therefore, on
centroid grounds alone KOI-4427.01 has a non-negligible chance of
being a false positive.

The companions at 4\arcsec\ to 5\arcsec\ from KOI-2529, 3284, and 4427
listed in Table~\ref{tab:ao} are significantly farther than 3$\sigma$
from the respective observed transit positions, so they are ruled out
as possible sources of the transit, as are any wider companions
detected in the UKIRT images. On the other hand, the close ($<
0\farcs5$) companions to KOI-1422, 2529, 3255, and 3284 cannot be
ruled out by centroid analysis alone. We address this issue later.
Additional closer (unresolved) companions may of course also be present.

\begin{deluxetable}{cccc}
\tablewidth{0pc}
\tablecaption{Centroid results for the candidates.\label{tab:centroids}}
\tablehead{
\colhead{} &
\colhead{Offset from} &
\colhead{Offset from} &
\colhead{3$\sigma$ exclusion}
\\
\colhead{Candidate} &
\colhead{star (\arcsec)} &
\colhead{star ($\sigma$)} &
\colhead{radius (\arcsec)}
}
\startdata
KOI-0571.05 & $0.74\pm0.82$ & 0.90 & 2.45  \\
KOI-1422.04 & $0.34\pm0.34$ & 1.00 & 1.03  \\
KOI-1422.05 & $0.54\pm0.33$ & 1.67 & 0.98  \\
KOI-2529.02 & $0.08\pm0.64$ & 0.13 & 1.92  \\
KOI-3255.01 & \phm{\tablenotemark{a}}$0.55\pm0.19$\tablenotemark{a} & 2.82 & \phm{\tablenotemark{a}}0.58\tablenotemark{a}  \\
KOI-3284.01 & \phm{\tablenotemark{a}}$0.60\pm0.94$\tablenotemark{a} & 0.64 & \phm{\tablenotemark{a}}2.82\tablenotemark{a}  \\
KOI-4005.01 & $0.23\pm0.61$ & 0.38 & 1.84  \\
KOI-4087.01 & $0.11\pm0.41$ & 0.26 & 1.22  \\
KOI-4427.01 & \phm{\tablenotemark{b}}$0.83\pm0.24$\tablenotemark{b} & 3.47 & \phm{\tablenotemark{b}}0.72\tablenotemark{b}  \\
KOI-4622.01 & $0.92\pm0.52$ & 1.76 & 1.57  \\
KOI-4742.01 & $1.19\pm1.17$ & 1.01 & 3.52  \\
KOI-4745.01 & $0.76\pm0.36$ & 2.09 & 1.08 
\enddata

\tablenotetext{a}{The presence of relatively bright close companions to KOI-3255 and
  KOI-3284 can cause biases in the centroid measurements.  These
  biases are expected to be smaller than the separation of the close
  companion, but we have nevertheless conservatively doubled the uncertainty found
  on the Exoplanet Archive. }

\tablenotetext{b}{The Exoplanet Archive reports the offset of
  KOI-4427.01 as $0\farcs94\pm1\farcs57$, but there is a single large
  outlier due to a quarter with a noisy difference image.  The values
  reported here are after removing the outlier.}

\end{deluxetable}

The centroid offsets in Table~\ref{tab:centroids}, on the basis of
which we rule out the 4--5\arcsec\ companions, were computed on the
assumption that the host stars do not have closer ($<4$\arcsec)
companions of comparable brightness.  However, as described in
Sect.~\ref{sec:imaging}, this is not the case for KOI-3255 and
KOI-3284.  The relatively bright close companions for these two KOIs will
introduce error in the centroid measurement of the KOI star position.
(Such error does not arise for the difference-image centroids
measuring the transit source position because, assuming negligible
variability, the difference image shows only the star hosting the
transit signal.)  To account for this unknown error, we have chosen to
conservatively double the offset uncertainties for KOI-3255 and
KOI-3284, although we do not expect the error to be larger than the
separation of the companion. The uncertainties in the table already
include this conservative doubling. The companion for KOI-2529 is
sufficiently dim that the centroid error due to this companion should
be negligible.

\subsection{Optical spectroscopy}
\label{sec:opticalspectroscopy}

In order to characterize the target stars and search for additional
companions, we acquired high-resolution spectra for the eleven stars
in our sample with the Keck~I Telescope on Mauna Kea (HI) and the
HIRES spectrometer \citep{Vogt:94}, using the standard setup of the
California Planet Search \citep[CPS; see][]{Howard:10}. The wavelength
coverage is approximately 3800--8000\,\AA, recorded on three
contiguous CCD detectors.  Observations were made over the course of
four nights (UT 2012 July 26 and September 22, and 2013 June 7 and 28)
under clear skies, with a typical seeing of $\sim$1\arcsec.  Exposure
times of about 15 minutes resulted in signal-to-noise ratios of 12--25
per pixel at a mean wavelength near the \ion{Mg}{1}\,b triplet
(5150--5200\,\AA).  The exception was KOI-3284, which was observed for
45 minutes yielding a SNR of 45 per pixel.  All spectra were taken
using the C2 decker ($0\farcs86 \times 14\arcsec$), giving a resolving
power of $R \approx 60,\!000$. The 14\arcsec\ long slit allows for
subtraction of the sky background that is superimposed on the stellar
spectrum. Sky subtraction was performed by measuring the sky
background light on a wavelength-by-wavelength basis.

For the reduction of the raw spectra we used the standard CPS
pipeline. Images from each of the three HIRES CCDs were independently
reduced and subjected to sky subtraction, flat fielding, and cosmic
ray removal. The pixel columns at each wavelength were then summed,
providing photon counts as a function of wavelength for each
pixel. Consistent wavelength solutions were ensured by aligning a
carefully chosen set of Thorium-Argon emission lines onto the same
pixels at the beginning of each night's observations.

The radial velocity (RV) of each star was measured using the A-band and
B-band telluric line features as wavelength fiducials. We measured the
placement of the stellar absorption lines relative to these features,
and then referenced them to stars of known radial velocity
\citep{Chubak:12}. The final radial velocity measurements in the frame
of the solar system barycenter are accurate to $\pm$0.1\,\kms, and are
given in Table~\ref{tab:keck} with other details of the observations.

To aid in eliminating some of the false positives that might be
causing the transit-like signals in our targets, we searched each of
our spectra for evidence of a second set of lines from another star
possibly falling on the spectrometer slit during our observations
(i.e., at angular separations smaller than about 0\farcs43, or half of
the slit width).  We first cross-correlated each spectrum against a
large library of stellar spectra obtained with the same instrument and
setup (see Sect.~\ref{sec:stellarproperties}) in order to identify the
best match. After flux normalization and placement of the target star
and library stars onto a common wavelength solution, we subtracted
this best match from the observed spectra, and the residuals were
subsequently cross-correlated against the same set of library stars
\citep{Kolbl:15}. A resulting peak in the cross-correlation
function would be evidence of a second star contaminating the
spectrum. For our targets this method is sensitive to physically
associated as well as unrelated (background/foreground) companion
stars so long as the relative RV of the two objects is greater than
about 10\,\kms. For smaller velocity differences the technique loses
sensitivity due to imperfect subtraction of the primary star.

We detected no companions to any of the 11 stars in our sample down to
about 1\% of the brightness of the primary (corresponding to a
magnitude difference $\Delta K\!p = 5$) for targets brighter than
$K\!p = 15$, and down to 2\% ($\Delta K\!p \approx 4.2$) for our
fainter targets. This indicates that at least two of the close
companions found in our Keck AO imaging that are bright enough to have
been detected spectroscopically, those around KOI-3255 and KOI-3284,
must have radial velocities within 10\,\kms\ of the main star, which
strongly suggests they are physically associated with the target.

\begin{deluxetable}{lcccc}
\tablewidth{0pc}
\tablecaption{Spectroscopic observations (Keck/HIRES).\label{tab:keck}}
\tablehead{
\colhead{} &
\colhead{BJD} &
\colhead{Exposure} &
\colhead{SNR} &
\colhead{RV}
\\
\colhead{Star} &
\colhead{(2,400,000+)} &
\colhead{(sec)} &
\colhead{(pix$^{-1}$)} &
\colhead{(\kms)}
}
\startdata
KOI-0571 &  56135.11242 &    900 &  18 &  $-$61.7 \\
KOI-1422 &  56192.77304 &    900 &  13 &  $-$24.8 \\
KOI-2529 &  56472.07615 &    900 &  13 &  $-$12.2 \\
KOI-3255 &  56472.06602 &    729 &  24 &  $-$15.3 \\
KOI-3284 &  56450.88292 &   2700 &  45 &  $-$90.9 \\
KOI-4005 &  56472.05630 &    785 &  25 &  $-$41.0 \\
KOI-4087 &  56472.04532 &    900 &  15 &  $-$20.1 \\
KOI-4427 &  56472.03348 &    900 &  12 &  $-$11.6 \\
KOI-4622 &  56472.08746 &    900 &  15 &  $-$98.6 \\
KOI-4742 &  56472.02224 &    900 &  20 &  $-$76.2 \\
KOI-4745 &  56472.09849 &    900 &  14 &  $-$24.0 
\enddata
\end{deluxetable}

\subsection{Near-infrared spectroscopy}
\label{sec:nirspectroscopy}

For two of our cooler targets (KOI-4087 and KOI-4622) we gathered
near-infrared spectra with the TripleSpec instrument \citep{Herter:08}
on the Apache Point Observatory 2.5\,m telescope to supplement the
material described above and to aid in the determination of the
stellar properties (Sect.~\ref{sec:stellarproperties}). For both
observations, taken on UT 2014 June 26, we used the 1\farcs1 slit
giving a resolving power of $R \approx 3500$. We alternated exposures
in an `ABBA' fashion, integrating for 4 minutes at each position until
we achieved a SNR of at least 80 per pixel throughout the $H$ band
(our SNR was slightly higher, about 100, for KOI-4622). Our total
integrations were 48 minutes and 64 minutes for KOI-4087 and KOI-4622,
respectively. We took care to gather spectra of the \ion{A0}{5}
telluric standard star 51\,Dra (using an exposure time of 10\,s per
image in the same ABBA nod strategy) at intervals separated by no more
than 30 minutes in time from the KOI observations, corresponding to
changes in airmass of 0.1 from the positions of the KOIs. We also
gathered flat field images using the bright quartz lamp at the Apache
Point Observatory. We used the \verb+spextool+ pipeline
\citep{Cushing:04} to co-add flat fields and science frames and to
extract the spectra from the full-frame images. The \verb+xtellcor+
package was then employed to compare the spectra of our standard
\ion{A0}{5} star 51\,Dra to that of Vega, in order to identify and
remove telluric absorption lines in both KOI spectra.

\section{Stellar properties}
\label{sec:stellarproperties}

We used our Keck/HIRES spectra to estimate the effective temperature
($T_{\rm eff}$), surface gravity ($\log g$), and metallicity ([Fe/H])
of our stars using a procedure referred to as {\tt SpecMatch}
(Petigura et al., in prep.). {\tt SpecMatch} compares an observed
target spectrum against a set of approximately 800 library spectra
obtained with the same instrument that span a wide range of parameters
($T_{\rm eff} = 3500$--7500\,K, $\log g = 2.0$--5.0 (cgs), and [Fe/H]
values from $-1.0$ to $+0.5$\,dex). For each comparison {\tt
  SpecMatch} places the target spectrum onto the same wavelength scale
as the library spectrum and computes $\chi^2$, the sum of the squares
of the pixel-by-pixel differences in normalized intensity. We adopted
the mean $T_{\rm eff}$, $\log g$, and [Fe/H] of the ten closest
matching library spectra (weighted by the $\chi^2$) as the final set
of stellar parameters for each target spectrum. They are reported in
Table~\ref{tab:specmatch} for about half of our sample. We estimate
these parameters to be accurate to about 100\,K in $T_{\rm eff}$,
0.10\,dex in $\log g$, and 0.10\,dex in [Fe/H] based on comparisons
with benchmark stars having parameters measured using detailed LTE
modeling.

\begin{deluxetable}{lcccc}
\tablewidth{0pc}
\tablecaption{Spectroscopic parameters.\label{tab:specmatch}}
\tablehead{
\colhead{} &
\colhead{$T_{\rm eff}$} &
\colhead{$\log g$} &
\colhead{[Fe/H]} &
\colhead{}
\\
\colhead{Star} &
\colhead{(K)} &
\colhead{(cm\,s$^{-2}$)} &
\colhead{(dex)} &
\colhead{Source}
}
\startdata
KOI-0571  & $3755 \pm \phn90$  &     \nodata      &  $-0.26 \pm 0.12$    &  1,2   \\
KOI-1422  & $3572 \pm \phn80$  &     \nodata      &  $-0.12 \pm 0.12$    &  1,2   \\
KOI-2529  & $4651 \pm 100$     & $4.64 \pm 0.10$  &  $+0.01 \pm 0.10$    &  3     \\
KOI-3255  & $4551 \pm 100$     & $4.67 \pm 0.10$  & \phs$0.00 \pm 0.10$  &  3     \\
KOI-3284  & $3748 \pm 112$     &     \nodata      &  $+0.16 \pm 0.14$    &  2     \\
KOI-4005  & $5431 \pm 100$     & $4.50 \pm 0.10$  &  $+0.02 \pm 0.10$    &  3     \\
KOI-4087  & $4134 \pm 154$     &     \nodata      &  $-0.30 \pm 0.15$    &  4     \\ 
KOI-4427  & $3813 \pm 112$     &     \nodata      &  $-0.07 \pm 0.14$    &  2     \\
KOI-4622  & $4340 \pm 177$     &     \nodata      &  $-0.57 \pm 0.18$    &  4     \\ 
KOI-4742  & $4402 \pm 100$     & $4.71 \pm 0.10$  &  $-0.37 \pm 0.10$    &  3     \\
KOI-4745  & $4723 \pm 100$     & $4.62 \pm 0.10$  &  $-0.01 \pm 0.10$    &  3    
\enddata
\tablecomments{Sources are: 1. \cite{Mann:13a, Mann:13b};
  2. \cite{Muirhead:14}; 3. {\tt SpecMatch}; 4. New estimates from
  this paper.}
\end{deluxetable}

For stars cooler than about 4300\,K this technique becomes less
reliable.  In that temperature range the library of reference spectra
is quite sparse, and it is possible for the {\tt SpecMatch} $T_{\rm
  eff}$ value to be biased high simply because the best matches are
hotter. Similar biases are likely in $\log g$ and
[Fe/H].\footnote{Even though {\tt SpecMatch} has difficulty obtaining
  accurate parameters for cool stars, it is still easily capable of
  distinguishing dwarfs from giants as the pressure-sensitive lines
  such as the \ion{Mg}{1}\,b triplet are very much narrower in giants
  than in our dwarf library spectra. All of the targets in this paper
  are verified to be dwarfs.}  Six stars in our sample are in this
regime: KOI-0571, 1422, 3284, 4087, 4427, and 4622. Spectroscopic
parameters (except for $\log g$) for the first three and for KOI-4427
were reported by \cite{Mann:13a, Mann:13b} and/or
\cite{Muirhead:14}\footnote{The \cite{Muirhead:14} values supersede
  earlier ones reported by \cite{Muirhead:12} that used the same
  spectra.} based on the measurement of temperature- and
metallicity-sensitive features in near infrared $H$-band or $K$-band
spectra. We adopted averages of those determinations here.

For KOI-4087 and KOI-4622 we derived new parameters based on the NIR
spectra described in Sect.~\ref{sec:nirspectroscopy}. Effective
temperatures were inferred from the empirical $H$-band calibrations of
\cite{Newton:14b}, which were established using stars with bolometric
luminosities and temperatures from long-baseline interferometry
\citep{Boyajian:13}. They are valid between 3100\,K and 4800\,K, and
have a typical scatter of 72\,K. Spectral indices based on the
\ion{Mg}{1} doublet at 1.50\,$\mu$m and the \ion{Al}{1} doublet at
1.67\,$\mu$m were measured as described by \cite{Newton:14b}, and
temperature uncertainties were obtained from numerical simulations
following the same work.  Metallicities for KOI-4087 and KOI-4622 were
determined using the NIR calibrations of \cite{Mann:13c}, which have
the advantage over other calibrations developed in recent years
\citep{Rojas:10, Rojas:12, Terrien:12, Newton:14a} that they were
established including K dwarfs with effective temperatures greater
than 4000\,K. They are thus the most appropriate calibrations for
these two KOIs, which lie on the K/M-dwarf boundary.  \cite{Mann:13c}
showed that the $K$-band relations display the strongest correlation
with [Fe/H]. We adopted their calibration in Eq.(16), which has a
scatter of 0.11\,dex.  The analysis of the spectra, the measurement of
the spectral features (equivalent widths of the 2.21\,$\mu$m
\ion{Na}{1} doublet, the CO\,(2-0) band-head in the $K$ band, and the
H$_2$O-K2 index from \citealt{Rojas:12}), as well as the uncertainty
estimation followed the procedures described recently by
\cite{Muirhead:14}, who applied the same methodology to a similar set
of spectra of about 100 M-dwarf KOIs.  The new temperatures and
metallicities for KOI-4087 and KOI-4622 may be found in
Table~\ref{tab:specmatch}.

The spectroscopic parameters for our targets were used to estimate the
stellar properties (primarily the mass and radius) by appealing to
stellar evolution models from the Dartmouth series \citep{Dotter:08},
following a Monte Carlo procedure similar to that described by
\cite{Torres:08}. While this is straightforward for the stars with
complete information ($T_{\rm eff}$, [Fe/H], and $\log g$), we lack a
$\log g$ estimate for the six cooler stars because the spectroscopic
techniques applied to them do not constrain that property. Thus, we
cannot establish their precise ages, or equivalently, their
sizes. This uncertainty is relatively unimportant, however, because
the radii of cool main-sequence stars change little with age. Most of
those stars show periodic brightness variations that may be
interpreted as rotational modulation, and in principle knowledge of
the rotation periods ($P_{\rm rot}$) enables one to infer a rough age
using gyrochronology relations.  The age, in turn, may be used in
place of $\log g$ as a constraint for the stellar evolution modeling,
although in practice the constraint is weak because the radius is not
very sensitive to age, as just stated. Our determination of $P_{\rm
  rot}$ for KOI-3284, 4087, 4427, and 4622 is explained in the
following section; for KOI-0571 and KOI-1422 we adopted the periods
measured previously by others (see below).  Our age estimates relied
on the gyrochronology relations of \cite{Epstein:14}, which are
claimed to provide values good to $\sim$1\,Gyr for stars with masses
above 0.55\,$M_{\sun}$, degrading rapidly below that.  The $B-V$ color
indices required by these calibrations were taken from the $UBV$
survey of the \kepler\ field by \cite{Everett:12}, along with
reddening values from the KIC. Generous uncertainties of 0.1 mag were
assigned to the de-reddened color indices to account for possible
systematic errors. For KOI-1422 and KOI-3284 the colors were
additionally corrected for the presence of the close companions
described in Sect.~\ref{sec:imaging}, assuming they are physically
associated.  The formal ages inferred for KOI-0571, 1422, 3284, 4087,
and 4622 are $3.89_{-0.51}^{+0.62}$, $3.57_{-3.21}^{+3.25}$,
$4.36_{-0.61}^{+0.76}$, $1.20_{-0.52}^{+0.39}$, and
$1.82_{-0.31}^{+0.44}$\,Gyr, respectively. As described below,
KOI-4427 does not display a clear signature of rotation, so we have
conservatively adopted a broad interval of possible rotation periods
(10--45\,days) based on the full range of periods observed for M stars
of similar temperature, as reported by \cite{McQuillan:13} (see their
Fig.~10). The corresponding age range from the gyrochronology
relations of \cite{Epstein:14} is 1--6\,Gyr, which we then used as a
constraint for the isochrone modeling.

KOI-0571 and KOI-1422 are systems with multiple transiting planet
candidates (five each). In such cases, if the candidates can be
assumed to transit the same star in near-circular orbits (required for
stability), the transit light curve modeling can provide a much
stronger constraint on the mean stellar density ($\rho_{\star}$) than
for single-planet candidates because the density estimates from the
different candidates within the same system can be averaged together
(see Sect.~\ref{sec:ap_multi}). For these two KOIs we have therefore
made use of the photometric $\rho_{\star}$ in our Monte Carlo
procedure simultaneously with the age constraint to strengthen the
determination of the stellar characteristics.

The final properties from our stellar evolution modeling are listed in
Table~\ref{tab:stellar}, computed from the mode of the corresponding
posterior distributions. Uncertainties correspond to the 68.3\%
(1$\sigma$) credible intervals from the same distributions. Distances
were computed from the absolute $K_s$ magnitudes inferred from the
isochrones along with the apparent brightness from the Two Micron All-Sky Survey \citep[2MASS;][]{Cutri:03} in the same
passband, which is the one least affected by interstellar
extinction. We nevertheless made corrections for extinction that we
inferred from reddening values for each KOI adopted from the KIC, as
above. For the four targets with close companions (KOI-1422, 2529,
3255, and 3284) the apparent magnitudes were corrected for the light
contribution of the neighbors. On the assumption that those stars are
physically bound to the targets, their typical orbital semimajor axes
would be approximately 52, 248, 75, and 64\,AU, and their orbital
periods roughly 550, 4600, 770, and 690\,yr, respectively.

\begin{deluxetable*}{lccccccccc}
\tablewidth{0pc}
\tablecaption{Stellar properties.\label{tab:stellar}}
\tablehead{
\colhead{} &
\colhead{Age} &
\colhead{$M_{\star}$} &
\colhead{$R_{\star}$} &
\colhead{$\log g$} &
\colhead{$\rho_{\star}$} &
\colhead{$L_{\star}$} &
\colhead{$M_V$} &
\colhead{$M_{K_s}$} &
\colhead{Distance}
\\
\colhead{Star} &
\colhead{(Gyr)} &
\colhead{($M_{\sun}$)} &
\colhead{($R_{\sun}$)} &
\colhead{(cm\,s$^{-2}$)} &
\colhead{(g\,cm$^{-3}$)} &
\colhead{($L_{\sun}$)} &
\colhead{(mag)} &
\colhead{(mag)} &
\colhead{(pc)}
}
\startdata
KOI-0571  & $4.0_{-0.6}^{+0.6}$ & $0.544_{-0.021}^{+0.024}$ & $0.523_{-0.021}^{+0.023}$ & $4.736_{-0.019}^{+0.020}$ & $5.29_{-0.39}^{+0.54}$ & $0.055_{-0.006}^{+0.011}$  & $9.01_{-0.24}^{+0.24}$  & $5.41_{-0.15}^{+0.14}$  & $172_{-10}^{+13}$ \\ [+1ex]
KOI-1422  & $4.2_{-1.6}^{+3.4}$ & $0.454_{-0.035}^{+0.033}$ & $0.426_{-0.027}^{+0.038}$ & $4.833_{-0.041}^{+0.025}$ & $7.94_{-1.08}^{+1.34}$ & $0.027_{-0.004}^{+0.008}$  & $10.12_{-0.35}^{+0.25}$  & $6.07_{-0.23}^{+0.20}$  & $226_{-18}^{+28}$ \\ [+1ex]
KOI-2529  & $3.0_{-0.3}^{+7.7}$ & $0.729_{-0.029}^{+0.033}$ & $0.697_{-0.023}^{+0.028}$ & $4.619_{-0.028}^{+0.015}$ & $3.03_{-0.25}^{+0.24}$ & $0.199_{-0.025}^{+0.039}$  & $6.98_{-0.24}^{+0.25}$  & $4.36_{-0.11}^{+0.11}$  & $618_{-30}^{+34}$ \\ [+1ex]
KOI-3255  & $2.9_{-0.3}^{+7.5}$ & $0.707_{-0.027}^{+0.033}$ & $0.680_{-0.024}^{+0.026}$ & $4.629_{-0.026}^{+0.015}$ & $3.18_{-0.25}^{+0.25}$ & $0.173_{-0.022}^{+0.035}$  & $7.20_{-0.24}^{+0.26}$  & $4.45_{-0.11}^{+0.11}$  & $417_{-21}^{+24}$ \\ [+1ex]
KOI-3284  & $4.4_{-0.7}^{+0.8}$ & $0.544_{-0.061}^{+0.041}$ & $0.520_{-0.061}^{+0.038}$ & $4.740_{-0.029}^{+0.059}$ & $5.52_{-0.77}^{+1.53}$ & $0.044_{-0.012}^{+0.017}$  & $9.55_{-0.44}^{+0.54}$  & $5.50_{-0.25}^{+0.41}$  & $145_{-23}^{+20}$ \\ [+1ex]
KOI-4005  & $7.2_{-3.9}^{+3.6}$ & $0.884_{-0.038}^{+0.044}$ & $0.866_{-0.040}^{+0.076}$ & $4.514_{-0.073}^{+0.035}$ & $1.89_{-0.38}^{+0.30}$ & $0.581_{-0.079}^{+0.153}$  & $5.46_{-0.25}^{+0.21}$  & $3.69_{-0.19}^{+0.12}$  & $693_{-38}^{+66}$ \\ [+1ex]
KOI-4087  & $1.3_{-0.2}^{+0.6}$ & $0.575_{-0.047}^{+0.043}$ & $0.559_{-0.054}^{+0.029}$ & $4.706_{-0.016}^{+0.049}$ & $4.76_{-0.48}^{+1.03}$ & $0.079_{-0.022}^{+0.023}$  & $8.33_{-0.32}^{+0.55}$  & $5.02_{-0.13}^{+0.42}$  & $261_{-46}^{+16}$ \\ [+1ex]
KOI-4427  & $3.6_{-1.3}^{+2.6}$ & $0.526_{-0.062}^{+0.040}$ & $0.505_{-0.065}^{+0.038}$ & $4.751_{-0.030}^{+0.067}$ & $5.79_{-0.82}^{+1.87}$ & $0.043_{-0.012}^{+0.017}$  & $9.34_{-0.41}^{+0.57}$  & $5.54_{-0.24}^{+0.42}$  & $240_{-39}^{+32}$ \\ [+1ex]
KOI-4622  & $1.9_{-0.4}^{+0.5}$ & $0.572_{-0.053}^{+0.049}$ & $0.550_{-0.054}^{+0.038}$ & $4.715_{-0.024}^{+0.047}$ & $4.94_{-0.66}^{+1.03}$ & $0.089_{-0.026}^{+0.038}$  & $8.02_{-0.44}^{+0.62}$  & $5.00_{-0.20}^{+0.41}$  & $284_{-48}^{+28}$ \\ [+1ex]
KOI-4742  & $2.9_{-0.2}^{+8.1}$ & $0.609_{-0.026}^{+0.030}$ & $0.598_{-0.024}^{+0.023}$ & $4.673_{-0.021}^{+0.018}$ & $4.01_{-0.30}^{+0.37}$ & $0.117_{-0.016}^{+0.024}$  & $7.73_{-0.25}^{+0.28}$  & $4.79_{-0.11}^{+0.15}$  & $342_{-22}^{+19}$ \\ [+1ex]
KOI-4745  & $3.2_{-0.4}^{+7.5}$ & $0.738_{-0.029}^{+0.033}$ & $0.706_{-0.024}^{+0.028}$ & $4.614_{-0.029}^{+0.016}$ & $2.96_{-0.25}^{+0.24}$ & $0.217_{-0.027}^{+0.043}$  & $6.83_{-0.23}^{+0.25}$  & $4.32_{-0.11}^{+0.11}$  & $779_{-38}^{+45}$ 
\enddata

\tablecomments{Ages, $\log g$, and $\rho_{\star}$ are best-fit values
    from the Dartmouth models, constrained by either a spectroscopic
    $\log g$, a light-curve-derived $\rho_{\star}$, and/or a
    gyrochronology age based on the rotation period (see text).}

\end{deluxetable*}

\subsection{Stellar rotational periods}
\label{sec:rotation}

Here we give the details of our determination of the rotation periods
used above for four of the cooler targets in our sample that have no
spectroscopic estimate of $\log g$: KOI-3284, 4087, 4427, and
4622. For this application the processing of the raw photometry was
somewhat different than that described earlier, because we wished to
retain the astrophysical variations present in the light curve while
at the same time removing the instrumental effects. We therefore used
the Presearch Data Conditioning Maximum A-Posteriori (PDC-MAP) data
from \kepler\ \citep{Smith:12}, which is designed to meet those goals.

Since the data are unevenly sampled and each quarter has a unique
offset, we elected to use a Lomb-Scargle style
periodogram.\footnote{Stars with active regions have a non-uniform
  surface brightness distribution, leading to brightness variations as
  the star rotates \citep{Budding:77}.  These active regions tend to
  evolve in location and amplitude over timescales of days to years,
  which can cause the periodicities to change as well due to
  differential rotation \citep{Reinhold:13}. Despite the complex
  nature of individual spots, the ensemble population tends to imprint
  the rotation period as a dominant peak in the Fourier domain,
  allowing for an estimate of the rotation period using photometry
  alone \citep{Basri:11,Nielsen:12}.}  The light curve model is a
simple sinusoid and thus is linear with respect to the model
parameters for any trial rotation period, $P_{\rm rot}$.  Using
weighted linear least squares we are guaranteed to find the global
maximum likelihood solution at each trial $P_{\rm rot}$. We scanned in
frequency space from twice the cadence up to twice the total baseline
of observations, taking $10^5$ uniform steps in frequency. At each
realization, we defined the ``power'' as $(BIC_{\rm null}-BIC_{\rm
  trial})/BIC_{\rm null}$, where $BIC$ is the Bayesian Information
Criterion and ``null'' and ``trial'' refer to the two models under
comparison. We also performed a second periodogram analysis with a
finer grid step around any prominent peaks found in the original
periodograms.

\begin{figure}
\centering
\begin{tabular}{c}

\includegraphics[trim = 0pt 0pt 0pt 30pt, clip, width=8.0cm]{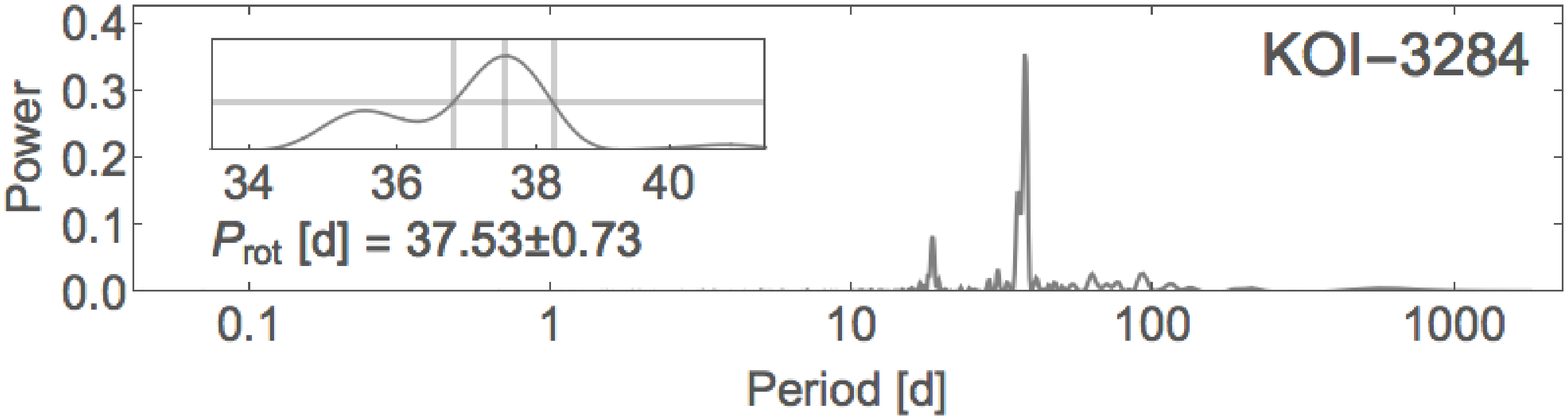} \\[-1ex]
\includegraphics[trim = 0pt 0pt 0pt 30pt, clip, width=8.0cm]{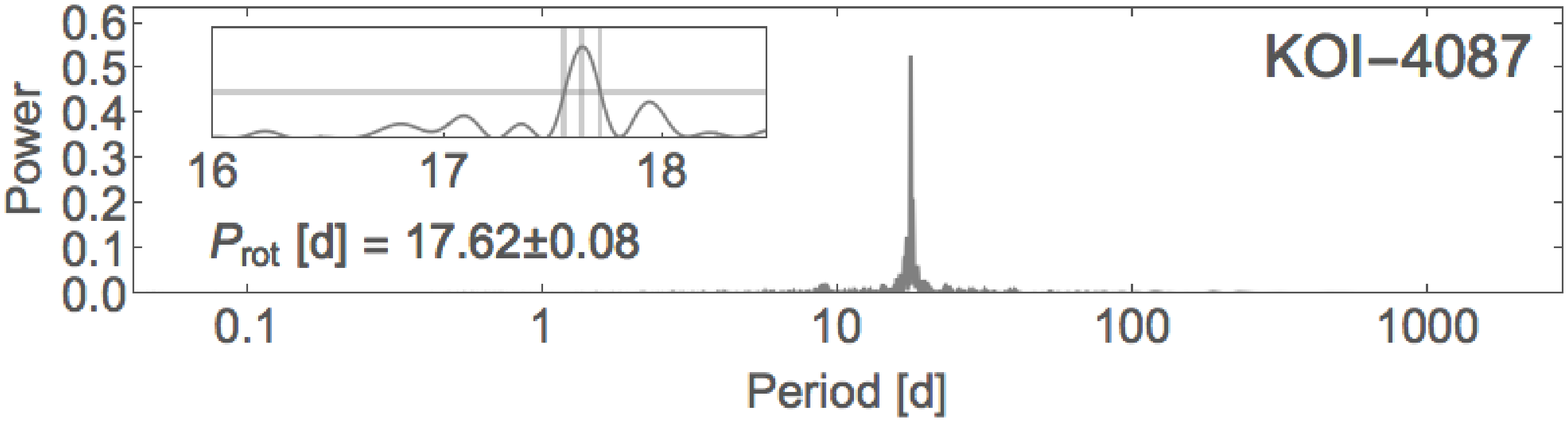} \\[-1ex]
\includegraphics[trim = 0pt 0pt 0pt 30pt, clip, width=8.0cm]{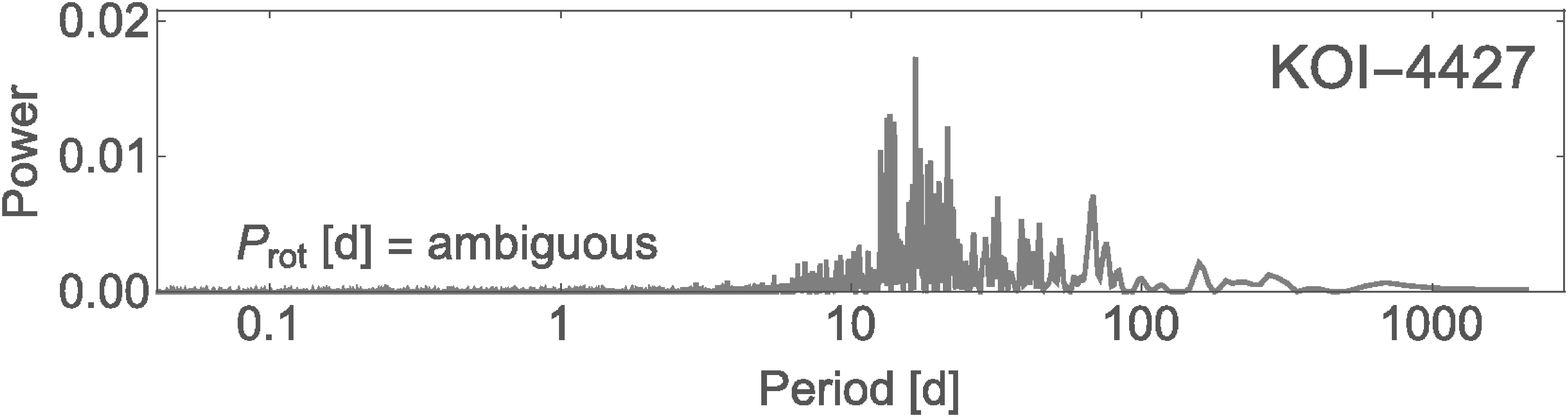} \\[-1ex]
\includegraphics[trim = 0pt 0pt 0pt 30pt, clip, width=8.0cm]{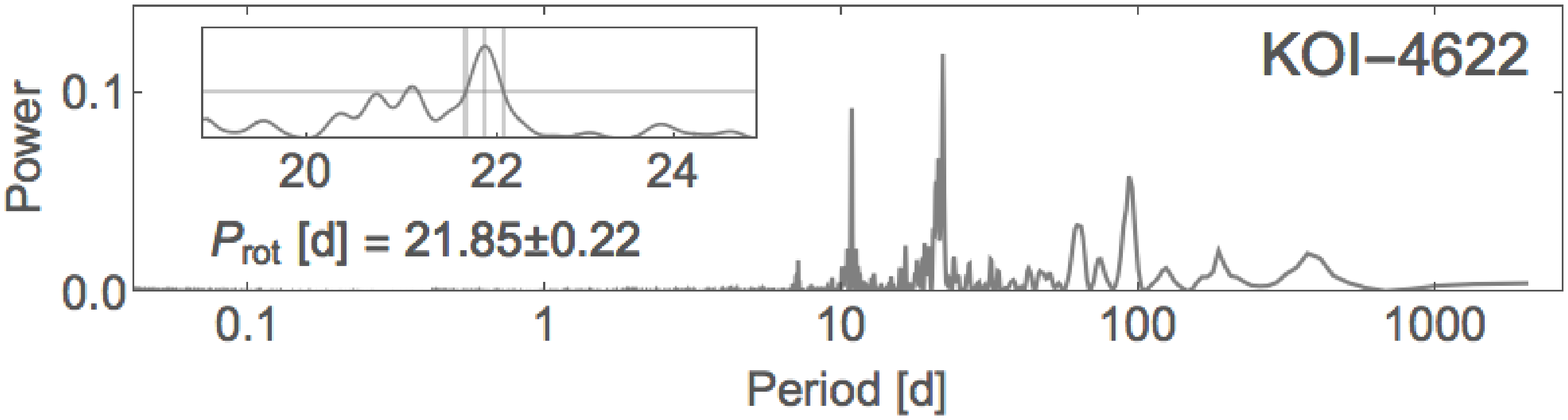} \\

\end{tabular}

\figcaption[]{Periodograms of the \kepler\ PDC-MAP photometry for four of the
  KOIs for which a spectroscopic surface gravity constraint is
  lacking. The insets show a close-up of the main peak, with an
  indication of the full width at half maximum that we assign as the
  error in $P_{\rm rot}$. No clear result is obtained for
  KOI-4427.\label{fig:periodograms}}

\end{figure}

The resulting periodograms for KOI-3284, 4087, 4427 and 4622 are shown
in Figure~\ref{fig:periodograms}. KOI-3284 and KOI-4087 show clear
uni-modal peaks at $37.53 \pm 0.73$\,d and $17.62 \pm 0.08$\,d,
respectively, which we attribute to their rotation periods. Here we
have taken the 1$\sigma$ confidence intervals to be the full width of
the peak in the periodogram at half maximum. For comparison, we note
that \cite{McQuillan:13} used an autocorrelation periodogram and found
peaks at similar periods (37.026\,d and 17.542\,d). Additionally,
KOI-4087 is in the sample studied by \cite{Nielsen:13}, who obtained
$P_{\rm rot} = 17.496$\,d, in good agreement with the other
results. KOI-4427 shows no clear peaks above the noise level, in
agreement with the findings of \cite{McQuillan:13}, so we were unable
to assign a rotation period in this case. A possible reason for
complex structure of the periodogram in this case is the presence of
other unresolved stars in the aperture.

KOI-4622 exhibits two peaks and is not included in either the
\cite{McQuillan:13} or the \cite{Nielsen:13} samples. The dominant
peak occurs at $21.85 \pm 0.22$\,d, with a second peak at $\sim$25\%
lower power with $P = 10.72 \pm 0.06$\,d. We note that this second
peak occurs not only at a lower power but also at nearly exactly one
half the period of the higher peak. We argue that the longer period is
that of the stellar rotation, and the shorter period is an alias
caused by more than one active region on the stellar surface.

Two other cool KOIs have had their rotation periods determined
previously \citep[$33.70 \pm 0.31$\,d for KOI-0571 and $36.5 \pm
  17.6$\,d for KOI-1422;][]{McQuillan:13}, and we adopt those values
here as published.

\section{Candidate validation}
\label{sec:validation}

The Doppler signals (velocity semi-amplitudes) expected of the
candidates in our sample, if due to planets around the targets, range
from 0.4 to 1.4\,\ms\ based on rough estimates of their masses
using the preliminary radius measures and the mass-radius relation of
\cite{Weiss:14}. Given the faintness of the parent stars ($K\!p =
14.3$--15.9), the detection of such small variations would be very
challenging, and therefore the planetary nature of these objects
cannot presently be ``confirmed'' in the usual way, i.e., by
establishing that the orbiting objects are of planetary mass. Instead
they must be ``validated'' statistically, by showing that the
likelihood of a true planet (which we refer to hereafter as the
`planet prior') is orders of magnitude larger than that of a false
positive. We describe this process below.

\subsection{Overview}
\label{sec:overview}

The types of astrophysical false positives we consider in our analysis
involve other unseen stars in the photometric aperture of
\kepler\ that are eclipsed by an orbiting object, and have their
eclipses attenuated by the light of the target such that they mimic
shallow planetary transits.  Examples of these situations include
background or foreground eclipsing binaries (`BEB'), background or
foreground stars transited by a (larger) planet (`BP'), and physically
associated stars transited by a smaller star or by a planet.  Such
physically associated stars will typically be close enough to the
target that they are generally unresolved in high-resolution
imaging. We refer to these hierarchical triple configurations as `HTS'
or `HTP', depending on whether the object orbiting the physical
companion is a star or a planet.\footnote{We point out that other
  statistical studies using \kepler\ data \citep[e.g.,][]{Morton:11, Morton:12,
    Lissauer:14} have not considered the HTP scenario as a false
  positive, arguing that the exact location of the planet (whether on
  the intended target or a bound companion) is inconsequential
  although its size may be larger if around the companion. Validation
  without considering these kinds of false positives is significantly
  easier, particularly since they tend to dominate the blend
  frequency, as we show below. For the present work, however, planet
  size is critical because it affects habitability: a planet that
  transits an unseen companion instead of the target may be too large
  to be rocky, and is thus less interesting for our purposes
  \citep[see, e.g.,][]{Alibert:14, Rogers:14}. We therefore count HTP
  configurations as blends.}

The procedure we used for validation is \blender\ \citep{Torres:04,
  Torres:11, Fressin:12}, which has been applied successfully in the
past to many of the most interesting candidates revealed by
\kepler\ \citep[see, e.g.,][]{Borucki:13, Barclay:13, Meibom:13,
  Ballard:13, Kippingetal:14b}. \blender\ makes full use of the detailed
shape of the transits to limit the pool of viable blends. It does this
by simulating large numbers of blend scenarios and comparing each of
them with the \kepler\ photometry in a $\chi^2$ sense. Fits that give
the wrong shape for the transit are considered to be ruled out. This
enables us to place useful constraints on the properties of the
objects that make up the blend, including their sizes and masses,
overall color and brightness, the linear distance between the
background/foreground eclipsing pair and the target, and even the
eccentricities of the orbits. Those constraints are then used to
estimate the frequencies of blends of different kinds. We note that
the simulated light curves generated by \blender\ take full account of
any known extra light in the aperture, such as that coming from the
close companions reported earlier (see Table~\ref{tab:ao}). We also
point out that here we only consider blends involving main-sequence
stars, as BEBs with a giant component produce light curves with a
shape that does not mimic a true planetary transit when observed with
such high precision as delivered by \kepler.

The photometric data we use here are the long-cadence PDC-MAP time
series, as in Sect.~\ref{sec:rotation}, detrended 
to remove signals at time scales long compared to
the transit durations.  Throughout this analysis we followed the
nomenclature established in previous \blender\ studies, designating
the objects in the eclipsing pair as the ``secondary'' and
``tertiary'', and the target itself as the ``primary''. We drew
stellar properties for the primary, secondary, and tertiary (masses,
radii, and absolute brightness in the \kepler\ and other passbands)
from model isochrones from the Dartmouth series \citep{Dotter:08}; the
selection of the isochrone for the primary was based on the
spectroscopic properties given earlier.  The \blender\ studies cited
above may be consulted for further technical details.

In summary, our validations proceed in two stages. First, we use
\blender\ proper to derive constraints on blend scenarios from their
light curve shapes. Then, we incorporate other constraints from
follow-up observations to compute the blend frequencies and the planet
prior by means of Monte Carlo simulations. We now address each of
these steps in turn.

\subsection{\blender\ constraints}
\label{sec:blender}

\begin{figure}
\epsscale{1.15}
\plotone{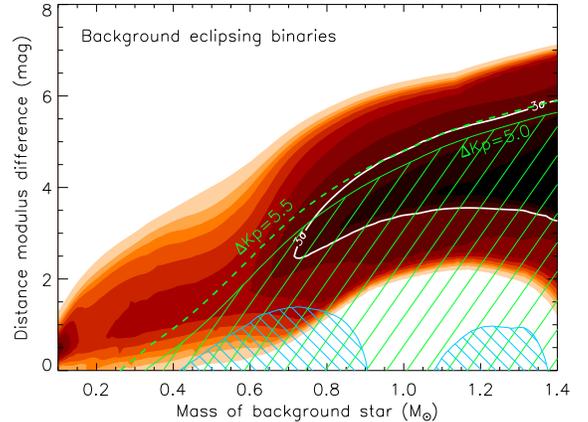}

\figcaption[]{Map of the $\chi^2$ surface (goodness of fit) for
  KOI-4005.01 corresponding to blends involving background eclipsing
  binaries. On the vertical axis we represent the linear distance
  between the BEB and the target ($D_{\rm BEB} - D_{\rm targ}$),
  expressed for convenience in terms of the difference in distance
  modulus, $\Delta\delta = 5\log(D_{\rm BEB}/D_{\rm targ}$). Only
  blends within the solid white contour (darker colors) provide fits
  to the \kepler\ light curve that are within acceptable limits
  \citep[3$\sigma$, where $\sigma$ is the significance level of the
    $\chi^2$ difference compared to a transiting planet model fit;
    see][]{Fressin:12}. Other concentric colored areas (lighter
  colors) represent fits that are increasingly worse (4$\sigma$,
  5$\sigma$, etc.), which we consider to be ruled out. The blue
  cross-hatched areas correspond to regions of parameter space where
  the blends are either too red (left) or too blue (right) compared to
  the measured $r-K_s$ color of the target, by more than three times
  the measurement uncertainty. The dashed green line labeled $\Delta
  K\!p = 5.5$ is tangent to the white contour from above and
  corresponds to the faintest viable blends. The green line labeled
  $\Delta K\!p = 5.0$ represents the spectroscopic limit on faint
  background stars. All simulated blends below this line (green
  hatched region) are brighter and are generally excluded if the BEB
  is angularly close enough to the target to fall within the slit of
  the spectrograph. Thus, very few blends remain viable.
\label{fig:bs_4005}}

\end{figure}

To illustrate the constraints derived from the shape of the transit,
we focus here on one of our candidates,
KOI-4005.01. \blender\ indicates that background eclipsing binaries
are only able to produce viable false positives if the primary star of
the eclipsing pair is restricted to a narrow range of main-sequence
masses\footnote{Stars more massive than about 1.4\,$M_{\sun}$ will
  generally have evolved to become giants for ages typical of the
  field ($\sim$3\,Gyr), and such stars yield light curves that have a
  very different shape than a transit, as stated earlier, so they do
  not constitute viable blends.} between about 0.7\,$M_{\sun}$ and
1.4\,$M_{\sun}$, and to a limited interval in brightness ($K\!p$
magnitude) relative to the target corresponding to $\Delta K\!p \leq
5.5$.  Figure~\ref{fig:bs_4005} shows the $\chi^2$ landscape for all
blends of this kind (BEB scenario) in a representative cross-section
of parameter space. Regions outside of the 3$\sigma$ contour
correspond to configurations with light curves giving a poor fit to
the observations, i.e., much worse than a true planet fit. These
blends are therefore excluded. The figure also illustrates some of the
additional constraints available from our follow-up observations for
this candidate. For example, analysis of our Keck/HIRES spectra of
KOI-4005 generally rules out companions within 5 magnitudes of the
primary if they are angularly close enough to fall within the
spectrograph slit.\footnote{For typical seeing conditions at the Keck
  telescope stars that are beyond 0\farcs43 can still imprint their
  lines on the target spectrum, and be detected, as can stars at wider
  separations that happen to be aligned along the slit. Nevertheless,
  to be conservative we assume here that these companions are
  spectroscopically undetectable.}  This can eliminate much of
parameter space (see green hatched area in the figure\footnote{Note,
  however, that not all blends in the green hatched area are
  excluded. This is only the case if the background star falls on the
  slit, and its RV is more than 10\,\kms\ different from that of the
  target. In all other cases we consider the blends to be viable, even
  if they have $\Delta K\!p < 5$.}).  Additionally, by comparing the
$r-K_s$ colors of the simulated blends with the measured color index
of the target \citep[$r-K_s = 1.580 \pm 0.029$;][]{Brown:11}, we find
that some of the BEB scenarios we have simulated are either too blue
or too red by more than 3$\sigma$ (blue hatched areas in the figure),
and are therefore also excluded. However, for this particular KOI all
BEB blends with the wrong color are already excluded by \blender\ for
giving poor fits (i.e., they are outside of the 3$\sigma$
contour). \cite{Santerne:13} have pointed out that an additional
source of blends involves eclipsing binaries with eccentric orbits
that are oriented so that they show only a secondary eclipse, as
viewed by \kepler. More generally, we note that they could also show
only a primary eclipse. We find, though, that while the depth and
shape of the diluted eclipses may indeed match the transit signal in
some cases, the combined color and brightness of these blends are such
that they are generally ruled out by the spectroscopic constraint
and/or the measured color of the KOI.

\begin{figure}
\epsscale{1.15}
\plotone{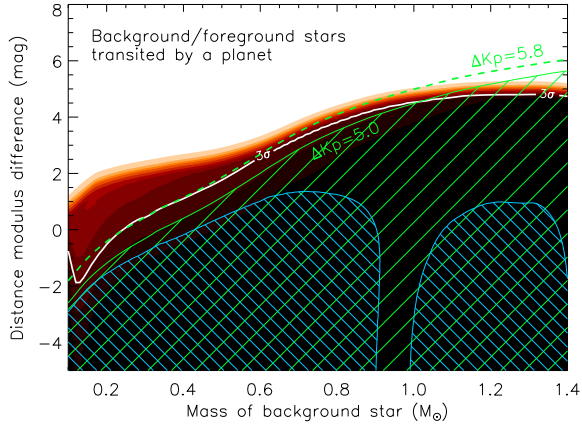}

\figcaption[]{Similar to Figure~\ref{fig:bs_4005} (and with the same
  color scheme) for blends involving background or foreground stars
  transited by a planet (BP scenario). The faintest blends giving
  acceptable fits have $\Delta K\!p = 5.8$ relative to the target
  (dashed green line). Blends below the solid green ($\Delta K\!p =
  5.0$, hatched green area) are generally excluded by the
  spectroscopic constraint unless the intruding star is more than
  0\farcs43 from the target, or closer than 0\farcs43 but with $\Delta
  RV < 10$\,\kms\ (see text).\label{fig:bp_4005}}

\end{figure}

For blends involving a background or foreground star transited by a
planet (BP scenario) there is a wide range of secondary masses that
produce good fits to the \kepler\ photometry of KOI-4005, as shown in
Figure~\ref{fig:bp_4005}. \blender\ indicates that the faintest of
these blends are about 5.8 magnitudes fainter than the target in the
$K\!p$ passband. In this case, however, the spectroscopic and color
constraints drastically reduce the pool of viable false positives.

\begin{figure}
\epsscale{1.15}
\plotone{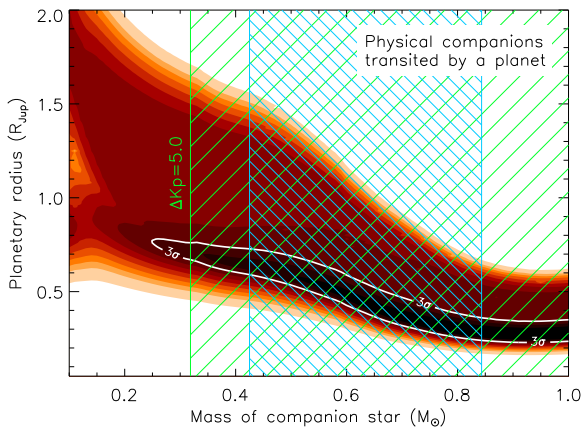}

\figcaption[]{Similar to Figure~\ref{fig:bs_4005} for the case of
  physical companions to KOI-4005 that are transited by a planet
  (HTP). Only companion stars with masses larger than about
  0.85\,$M_{\sun}$ or smaller than about 0.42\,$M_{\sun}$ yield blend
  colors that are consistent with the measured $r-K_s$ index of
  KOI-4005. However, many of these blends are eliminated by the
  spectroscopic constraint, as they are brighter than $\Delta K\!p =
  5.0$.\label{fig:htp_4005}}

\end{figure}

The $\chi^2$ map for blends involving a physically associated
companion to KOI-4005 transited by a larger planet (HTP scenario) is
seen in Figure~\ref{fig:htp_4005}, and shows the size of the tertiary
as a function of the mass of the companion star (secondary). In this
case \blender\ restricts the false positives to be in a narrow strip
of parameter space corresponding to secondary masses larger than about
0.25\,$M_{\sun}$, and planetary sizes between 0.25 and about 0.8\,$R_{\rm
  Jup}$ (2.8--9.0 \,$R_{\earth}$). As in the other scenarios, color
and brightness constraints allow us to reject many of these blends.

Finally, \blender\ indicates that pairs of eclipsing stars orbiting
the target (HTS scenario) invariably produce light curves with the
wrong shape for a transit, or feature noticeable secondary eclipses
that are not observed in the photometry of KOI-4005.01, or, if
  they show only a single eclipse due to a high eccentricity and
  special orientation \citep{Santerne:13}, the overall brightness
  would make the eclipsing binary detectable and/or its color
  inconsistent with the measurement.  These types of configurations
are therefore easily excluded. A similar result was found for each of
the other candidates in our sample, and indeed all previous
\blender\ analyses of KOIs have also found that HTS blends are always
ruled out with photometry of the quality delivered by the
\kepler\ instrument, when combined with the observational
  constraints.  Consequently, we do not consider HTS scenarios any
further.

\begin{figure*}
\centering
\begin{tabular}{ccc}

\includegraphics[width=6.0cm]{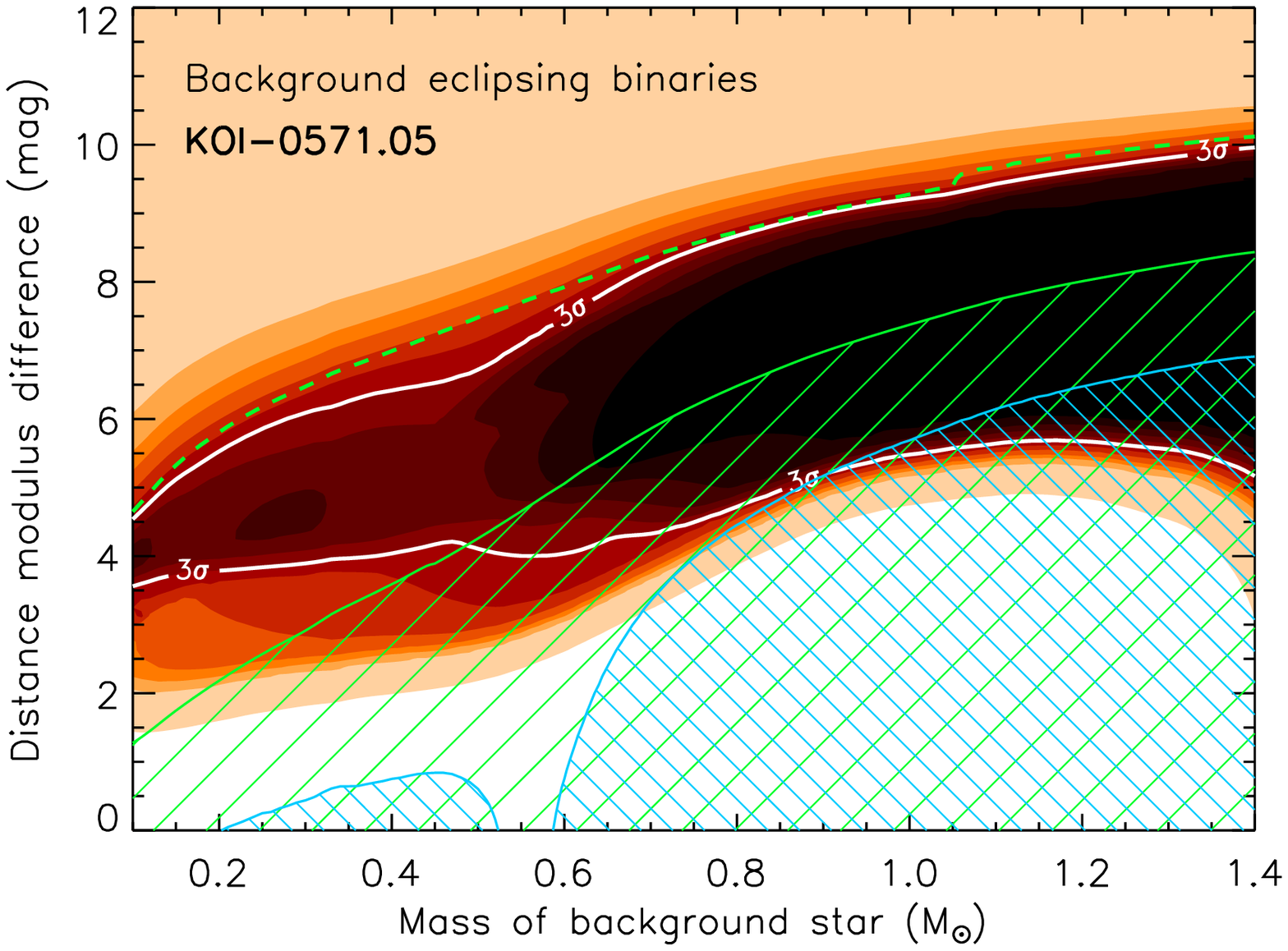} &
\includegraphics[width=6.0cm]{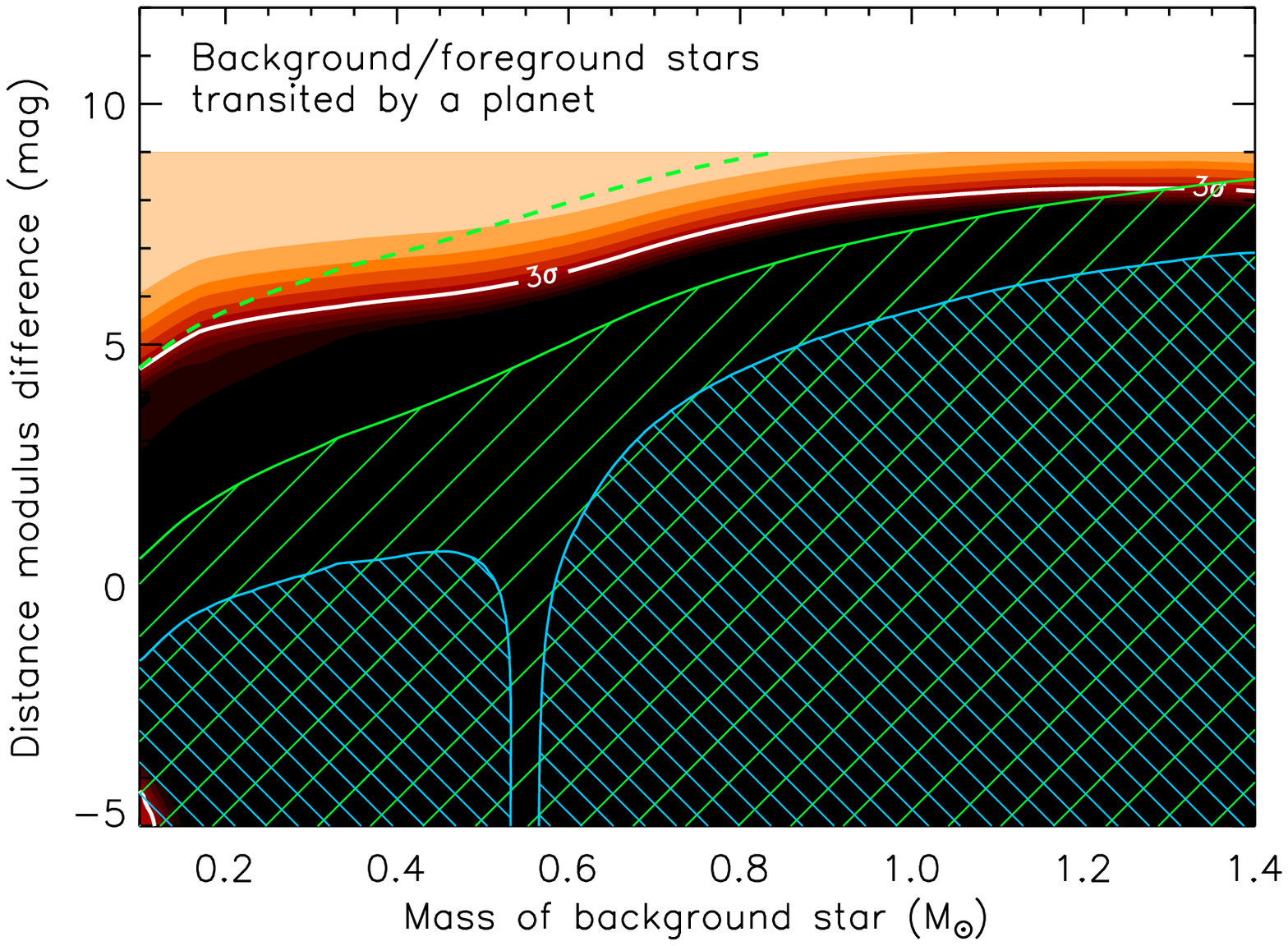} &
\includegraphics[width=6.0cm]{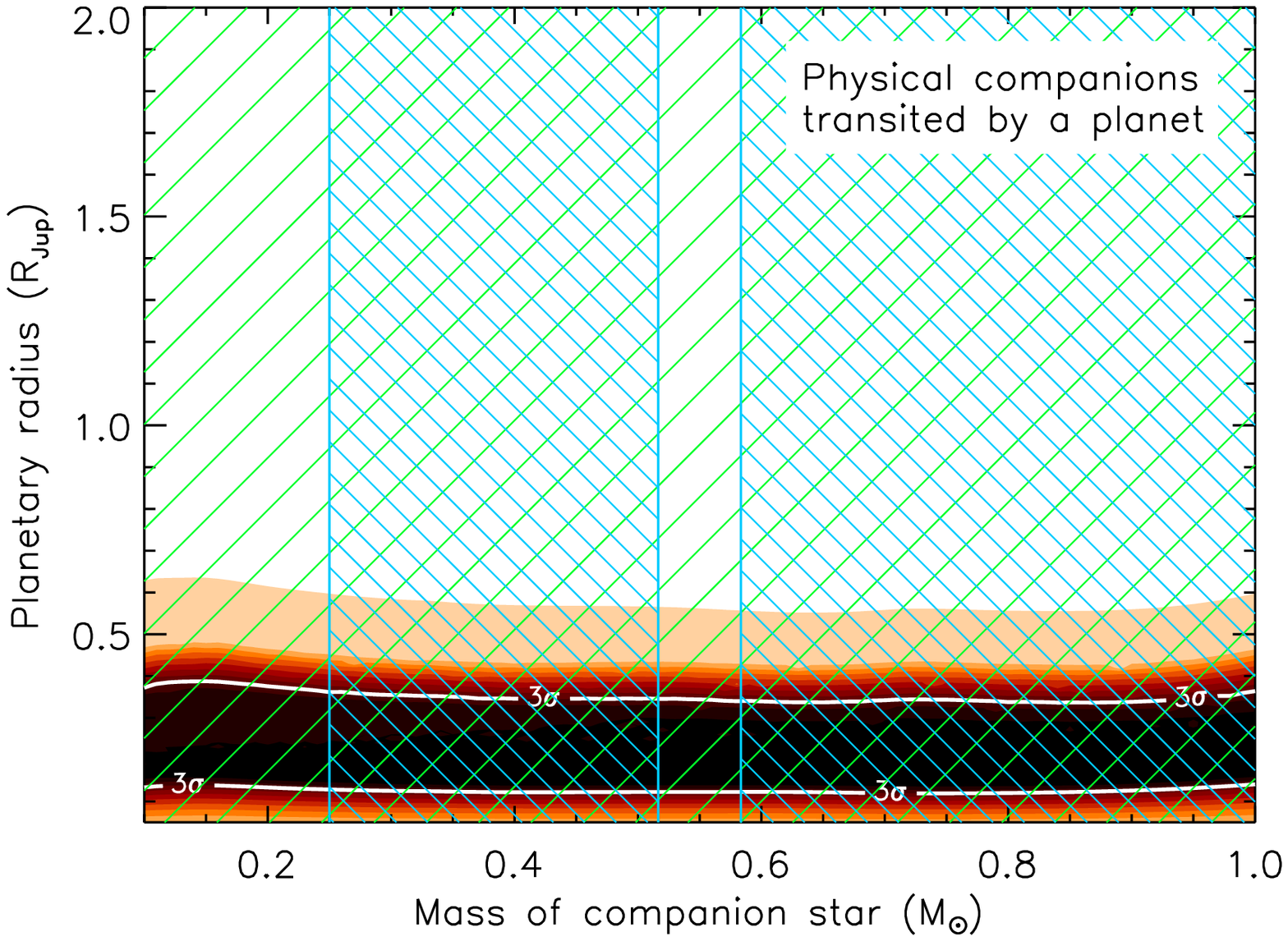} \\

\includegraphics[width=6.0cm]{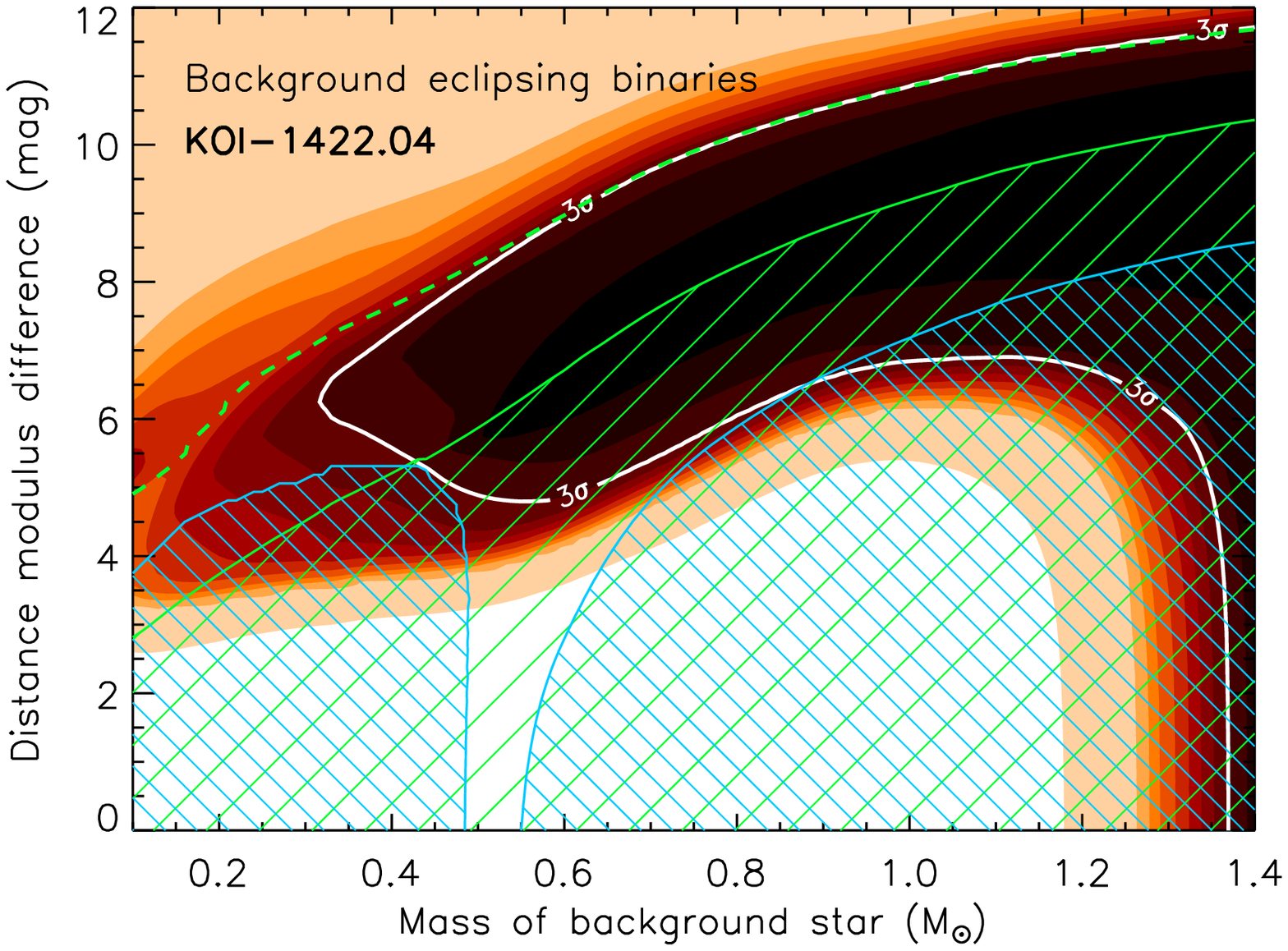} &
\includegraphics[width=6.0cm]{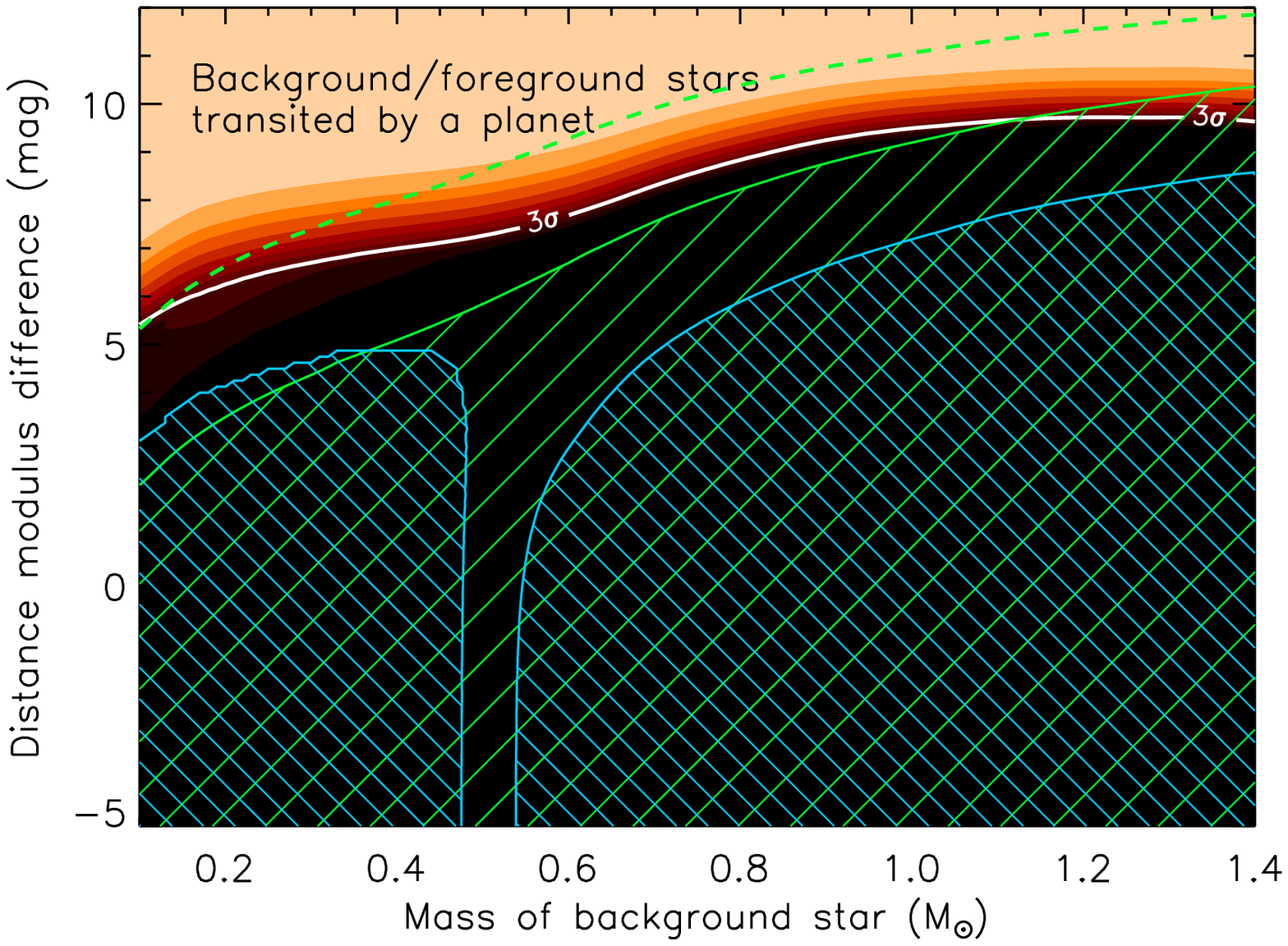} &
\includegraphics[width=6.0cm]{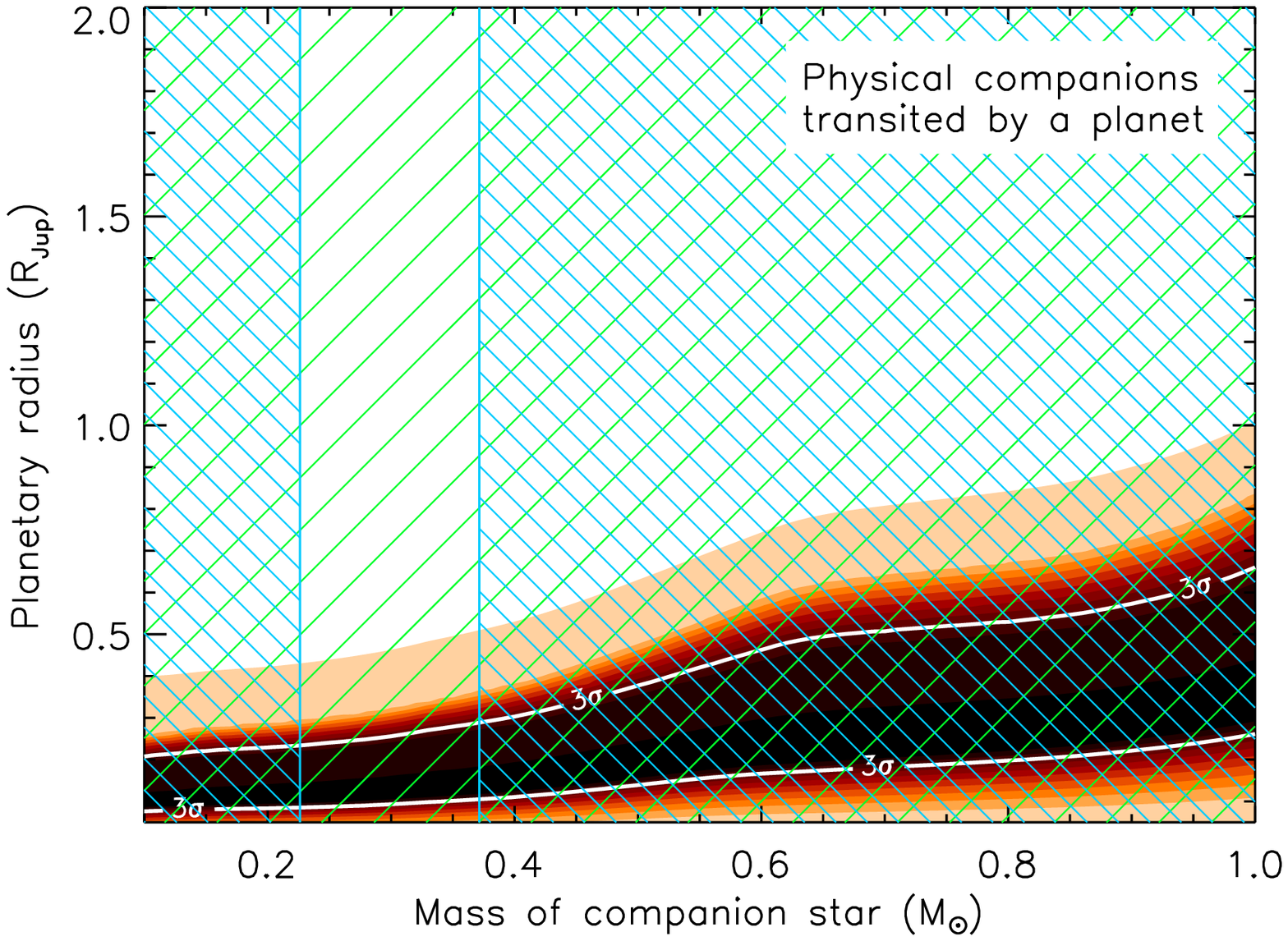} \\

\includegraphics[width=6.0cm]{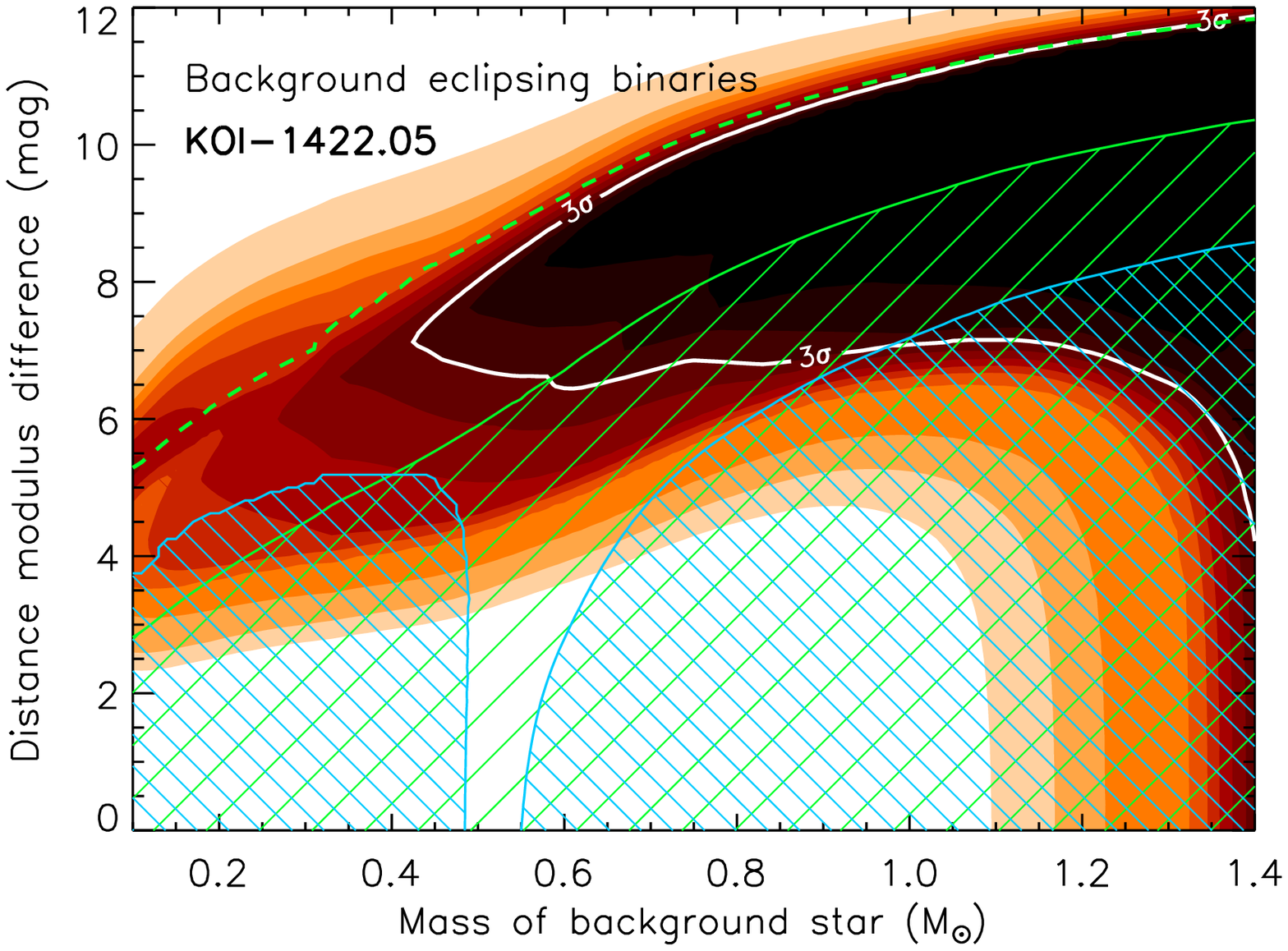} &
\includegraphics[width=6.0cm]{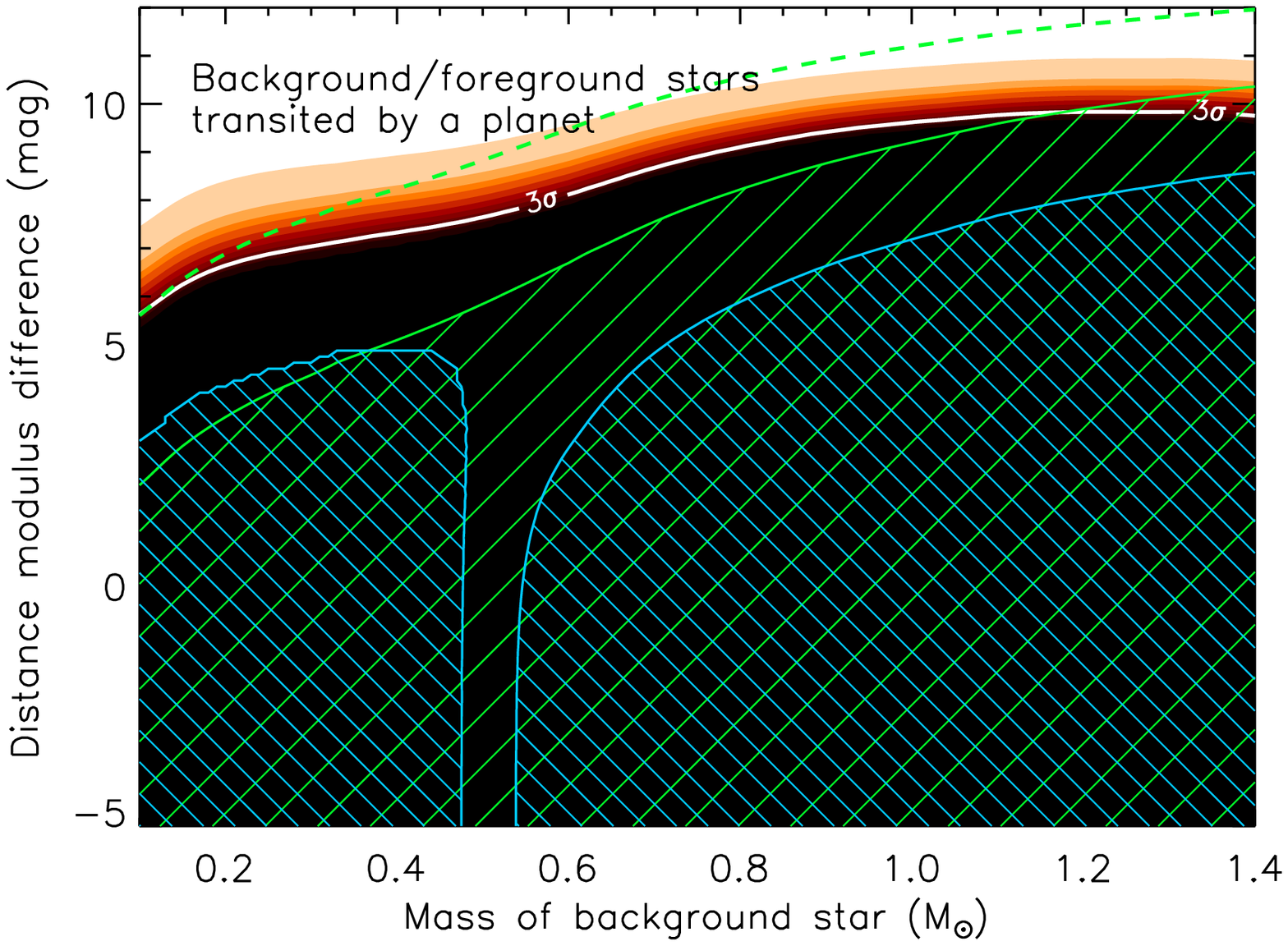} &
\includegraphics[width=6.0cm]{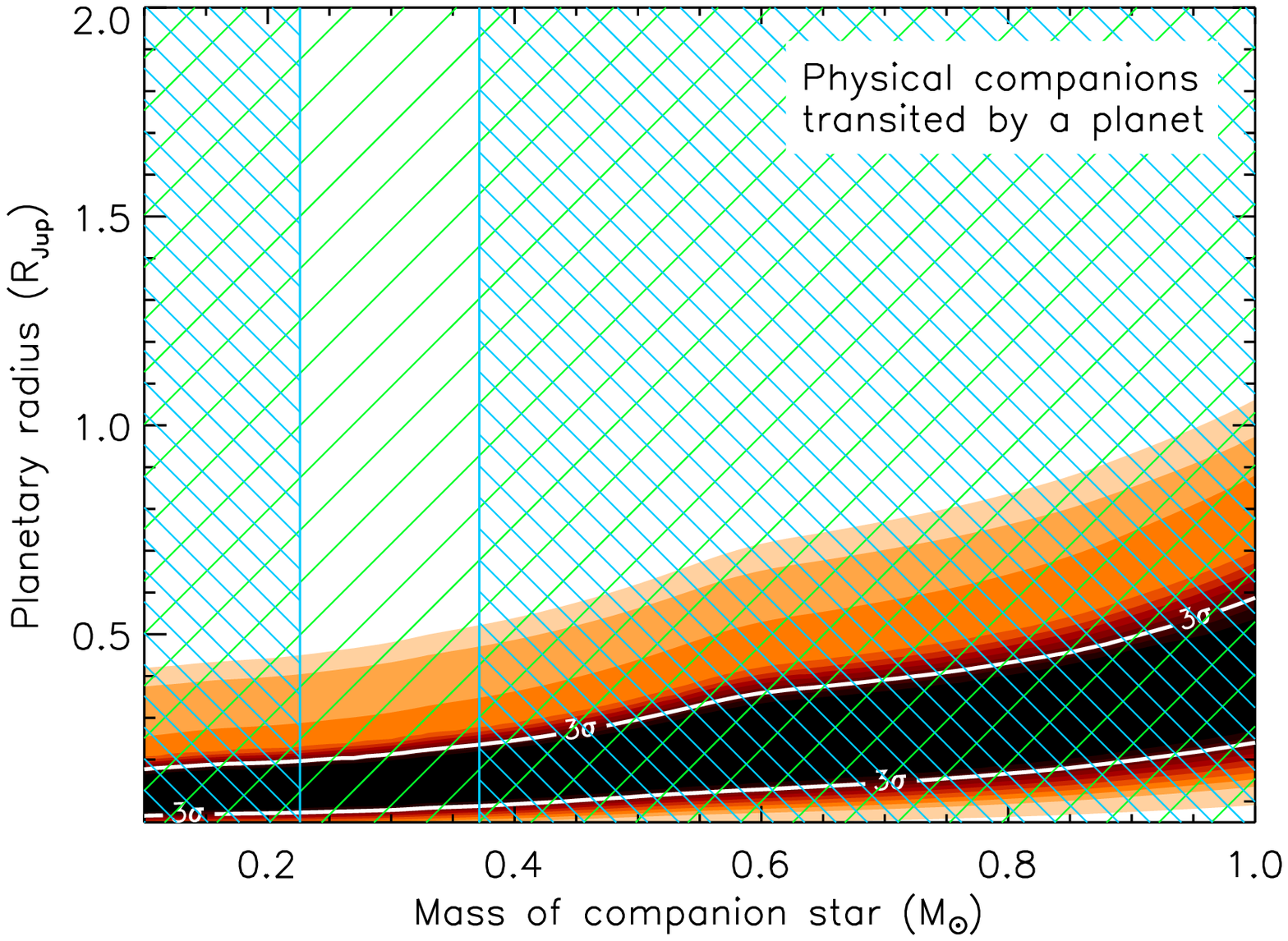} \\

\includegraphics[width=6.0cm]{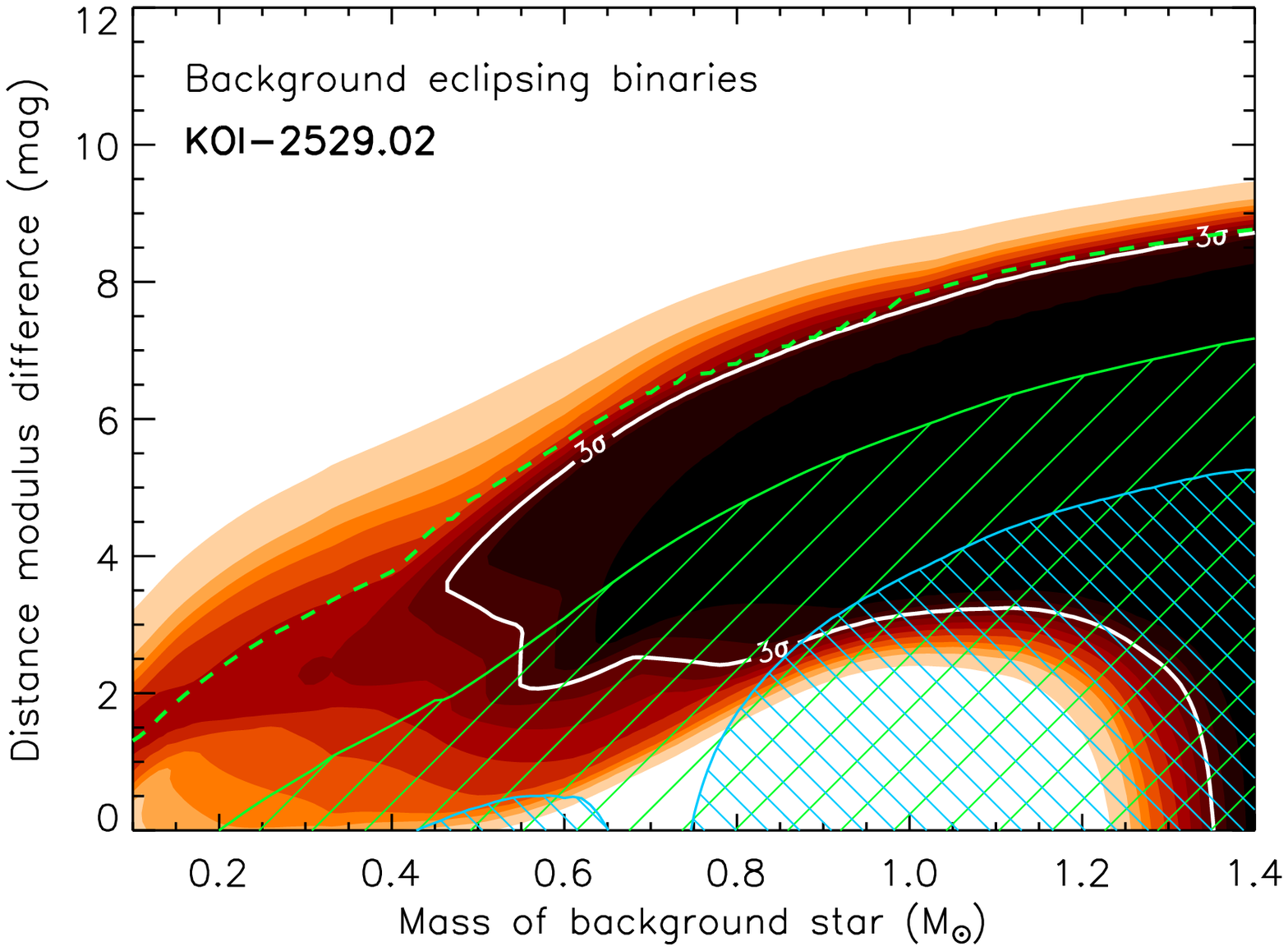} &
\includegraphics[width=6.0cm]{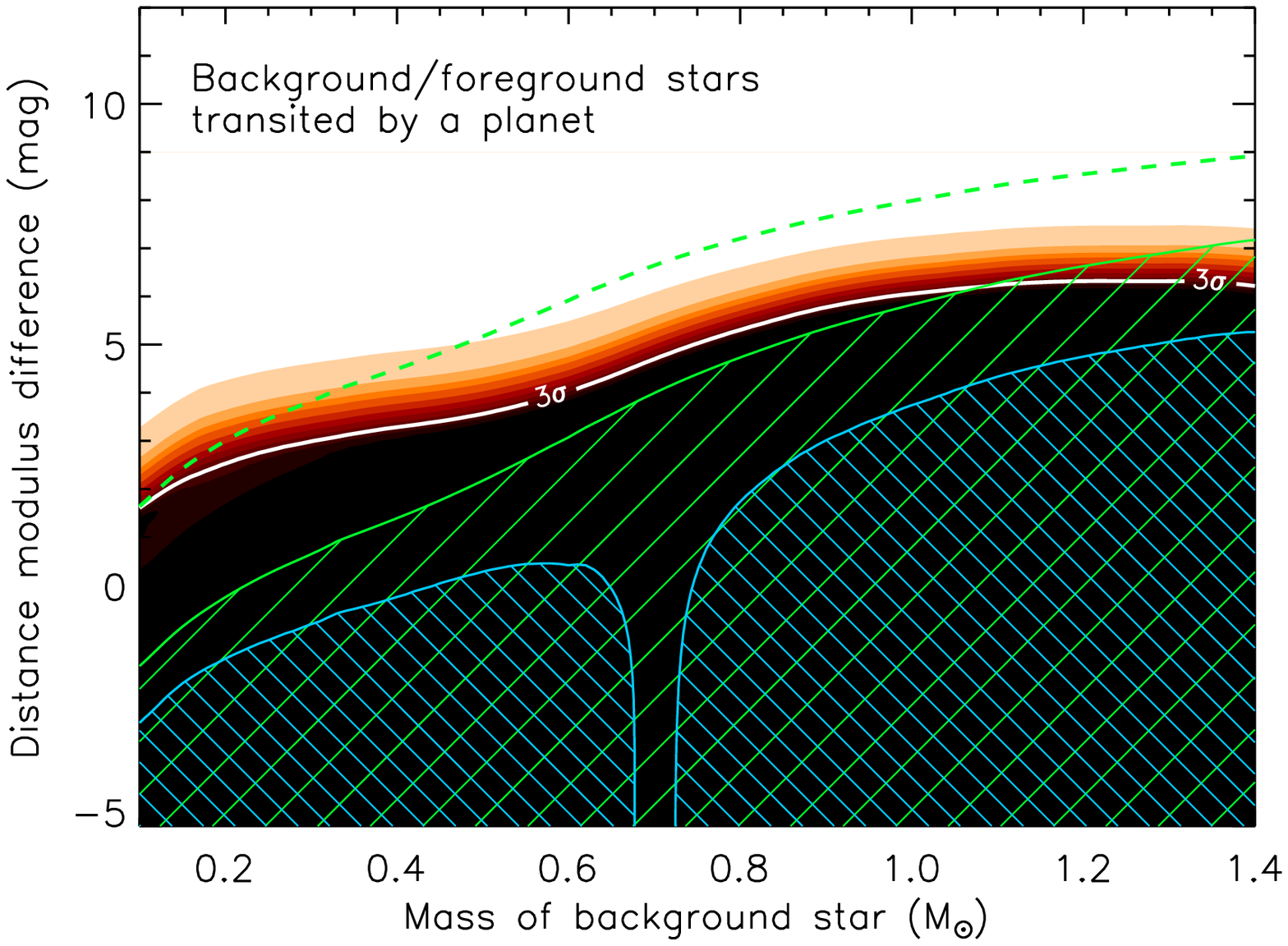} &
\includegraphics[width=6.0cm]{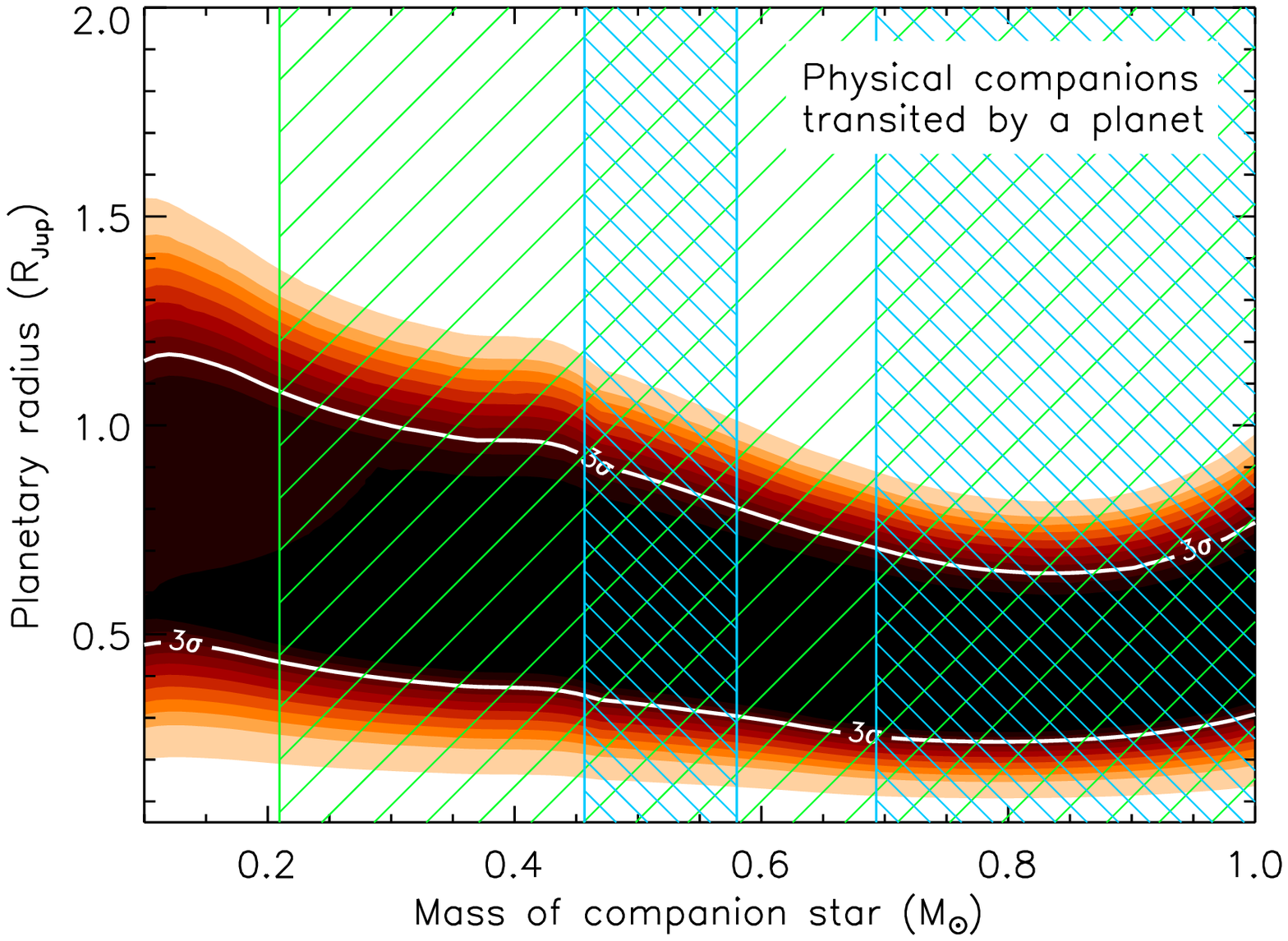} \\

\includegraphics[width=6.0cm]{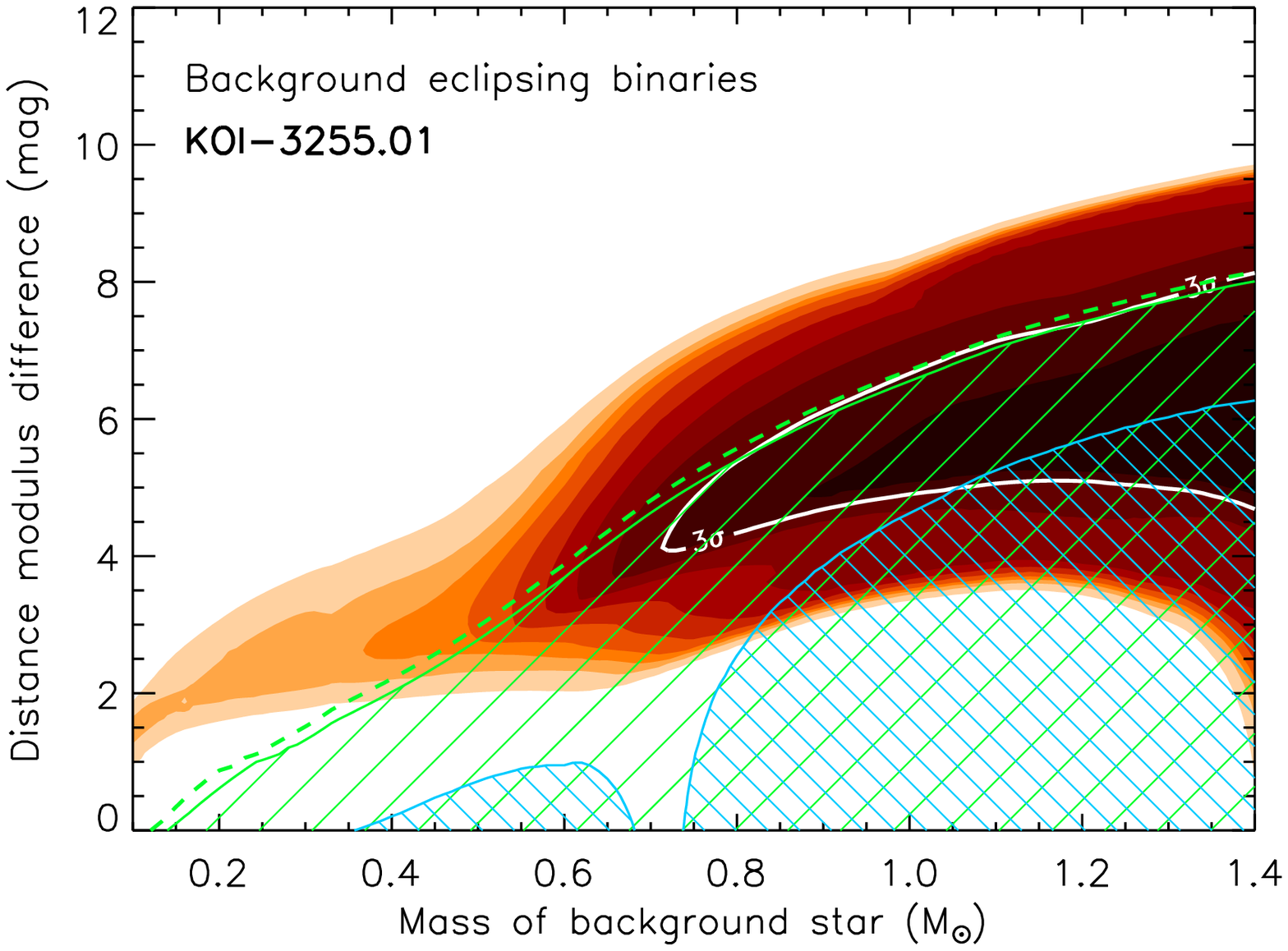} &
\includegraphics[width=6.0cm]{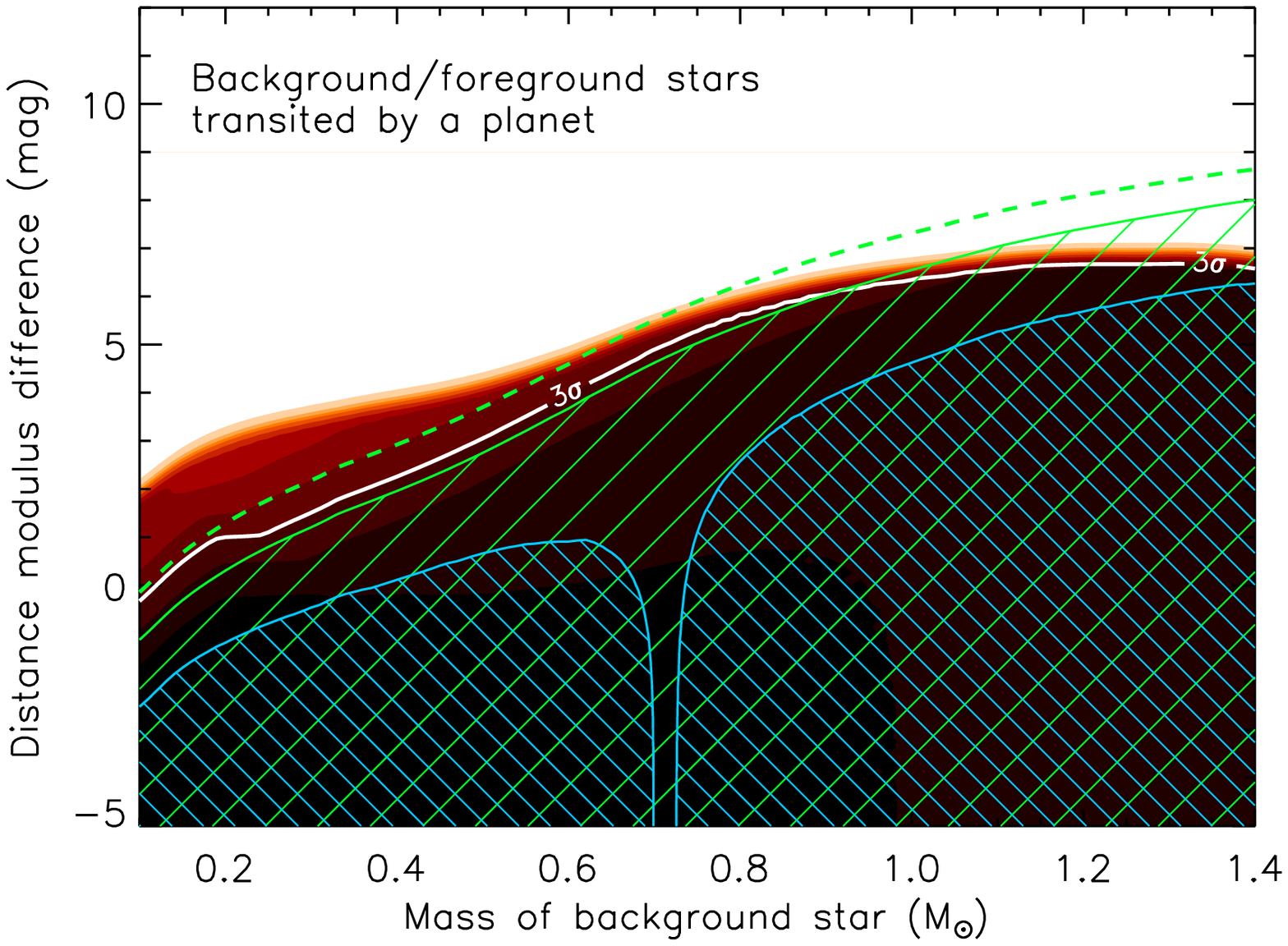} &
\includegraphics[width=6.0cm]{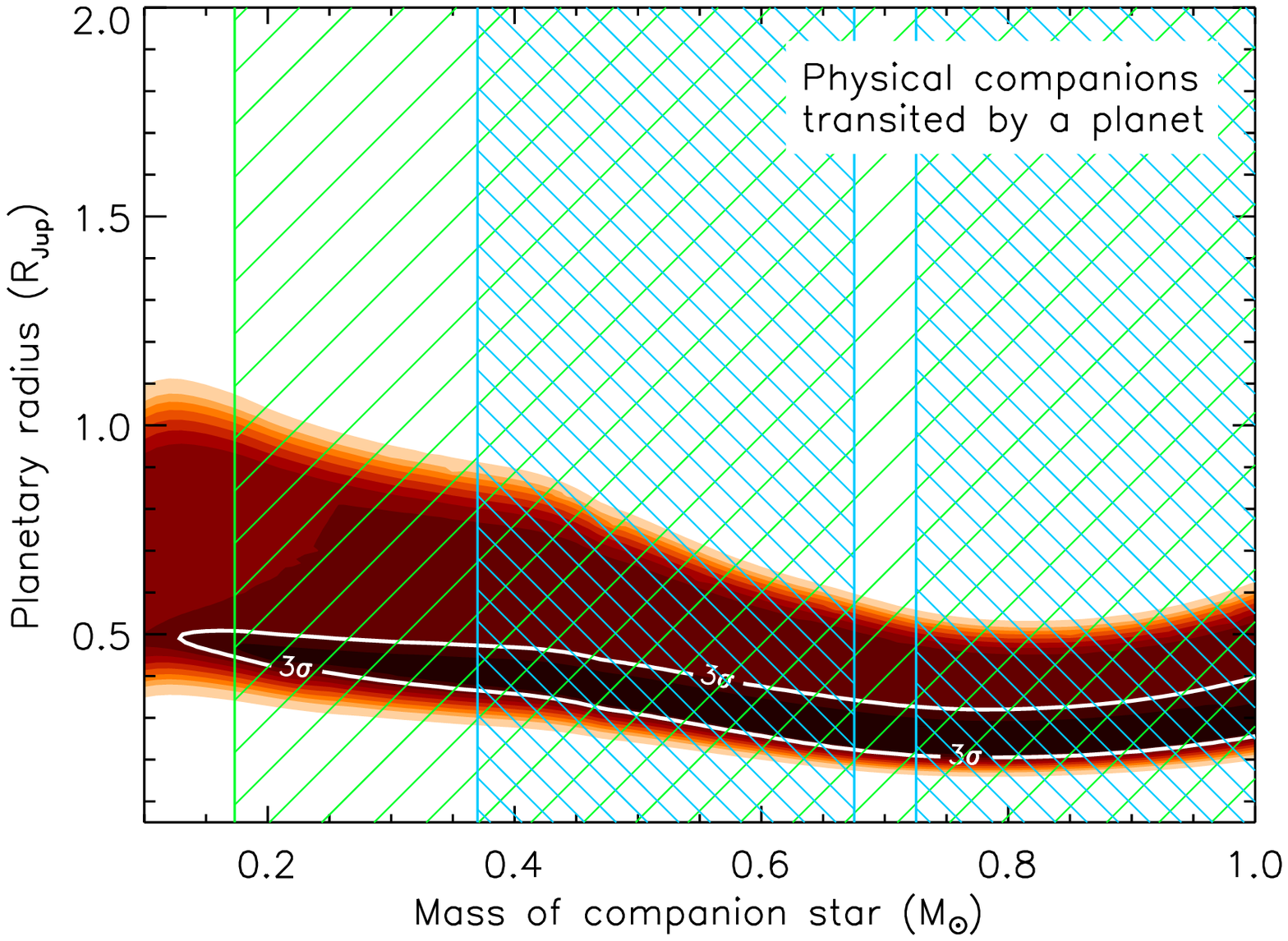} \\

\end{tabular}

\figcaption[]{\blender\ $\chi^2$ landscapes for five KOI targets.
  Each row corresponds to the KOI labeled in the left panel, and shows
  the three blend scenarios. Observational constraints from
  spectroscopy and color are also shown. See
  Figures~\ref{fig:bs_4005}--\ref{fig:htp_4005} for a detailed
  description.\label{fig:blender1}}

\end{figure*}

\begin{figure*}
\centering
\begin{tabular}{ccc}

\includegraphics[width=6.0cm]{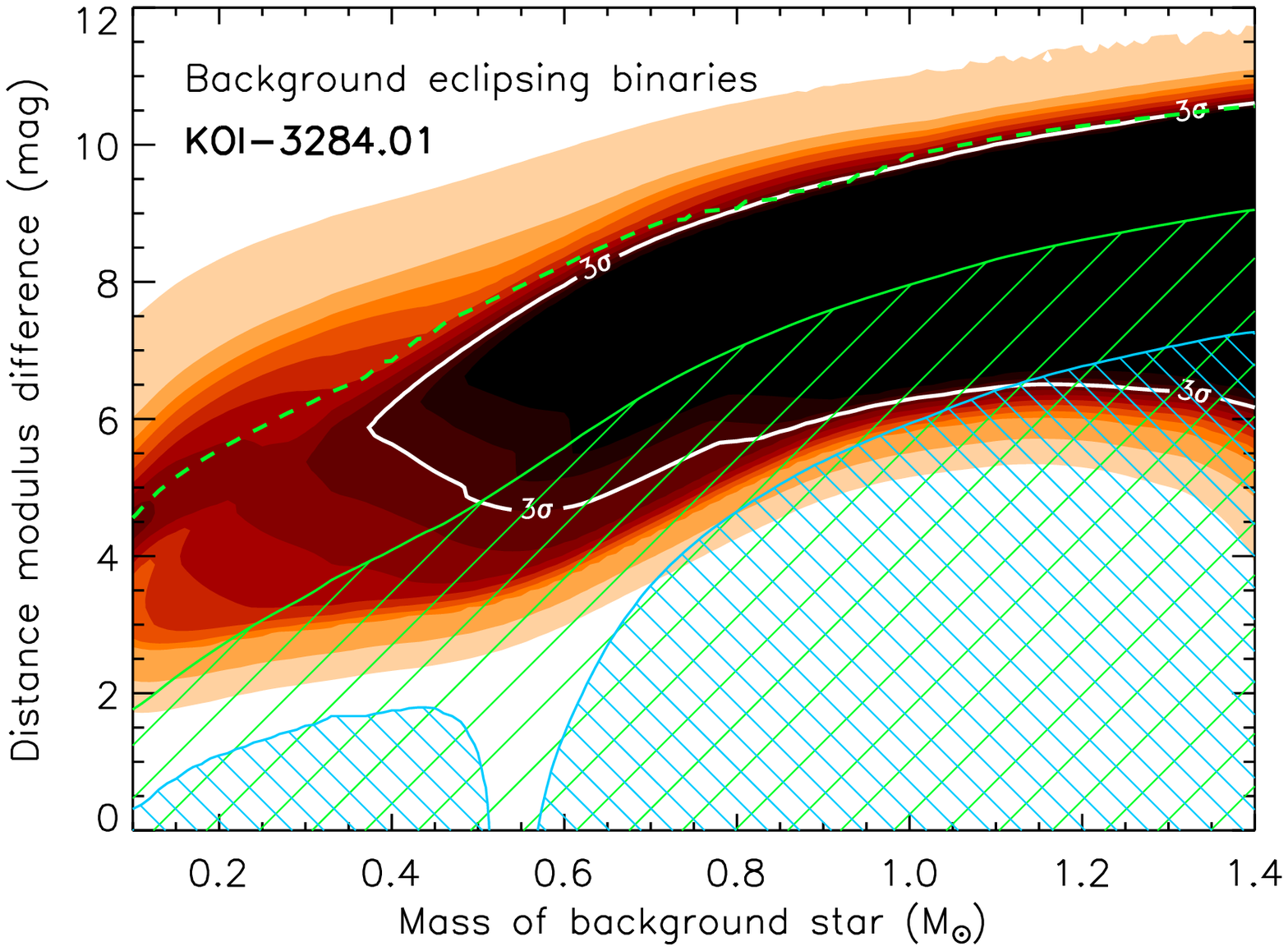} &
\includegraphics[width=6.0cm]{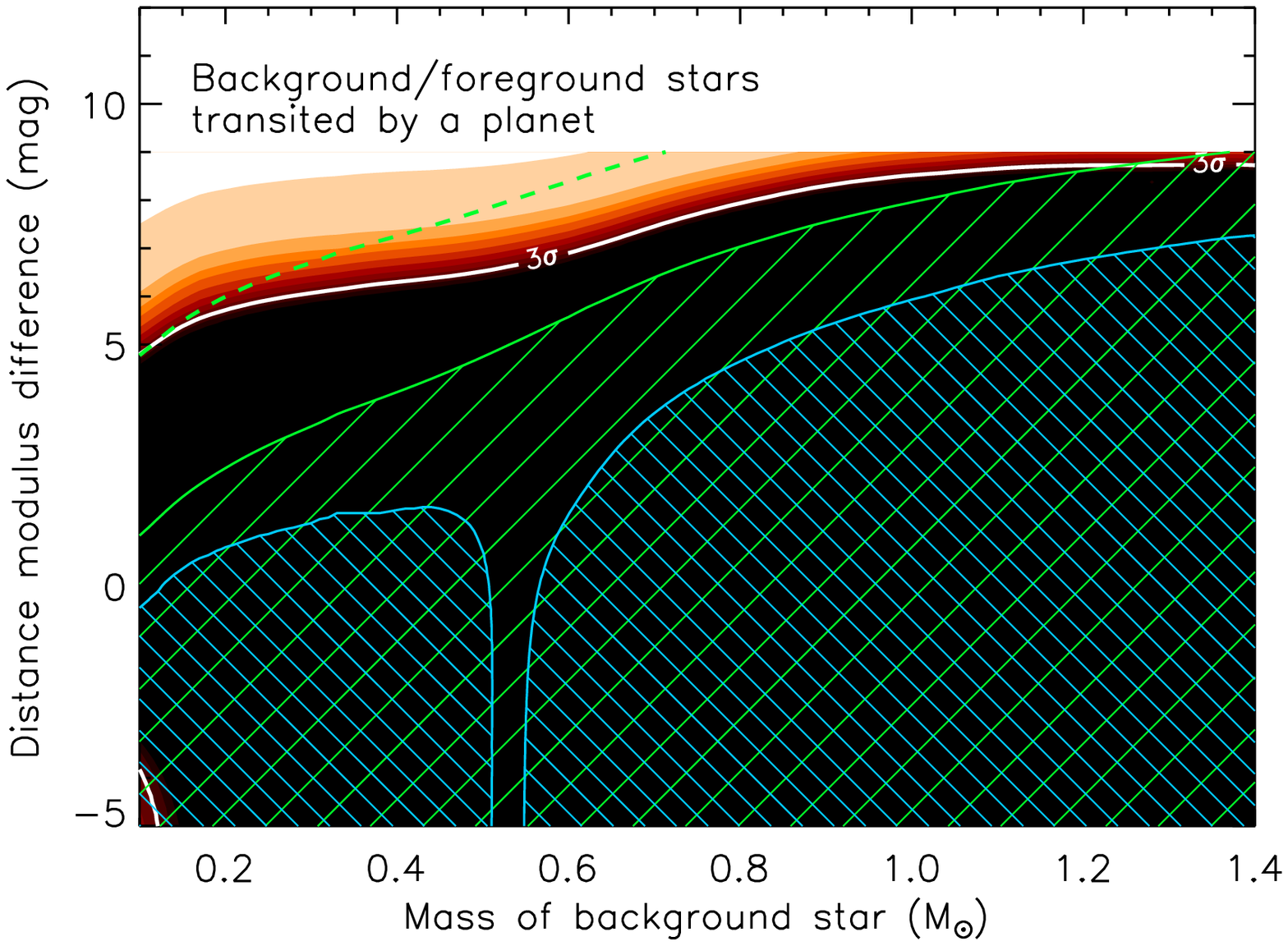} &
\includegraphics[width=6.0cm]{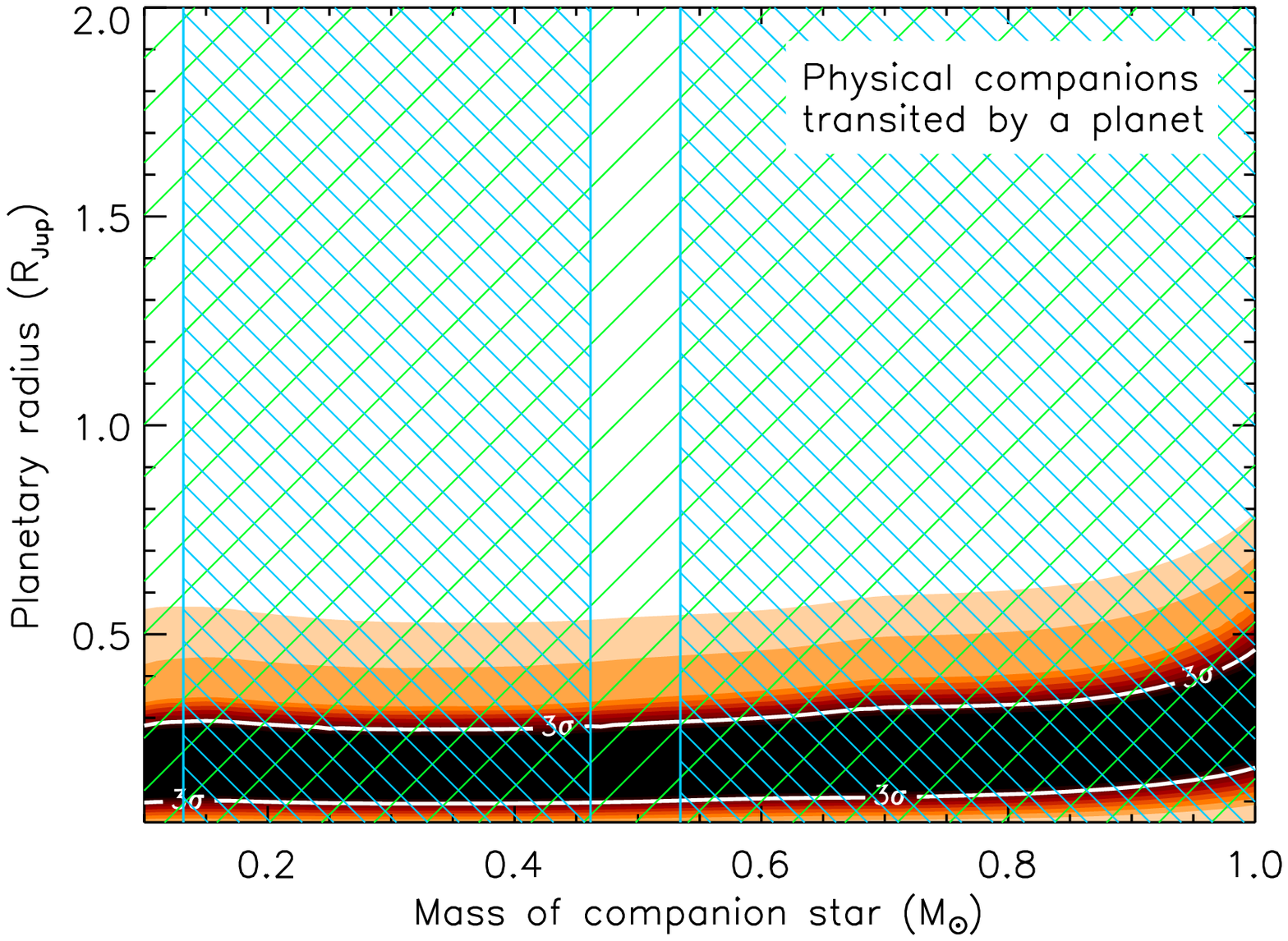} \\

\includegraphics[width=6.0cm]{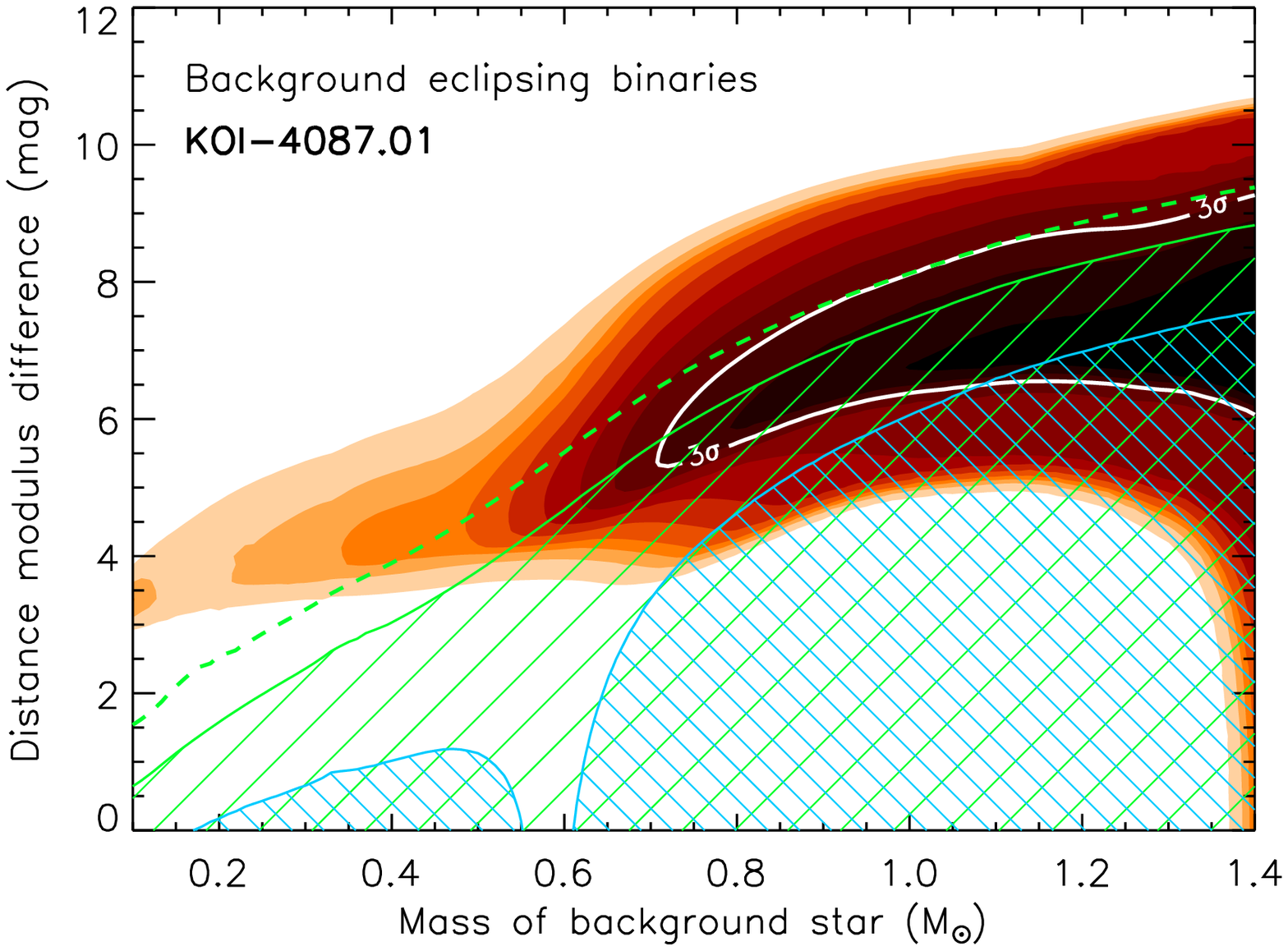} &
\includegraphics[width=6.0cm]{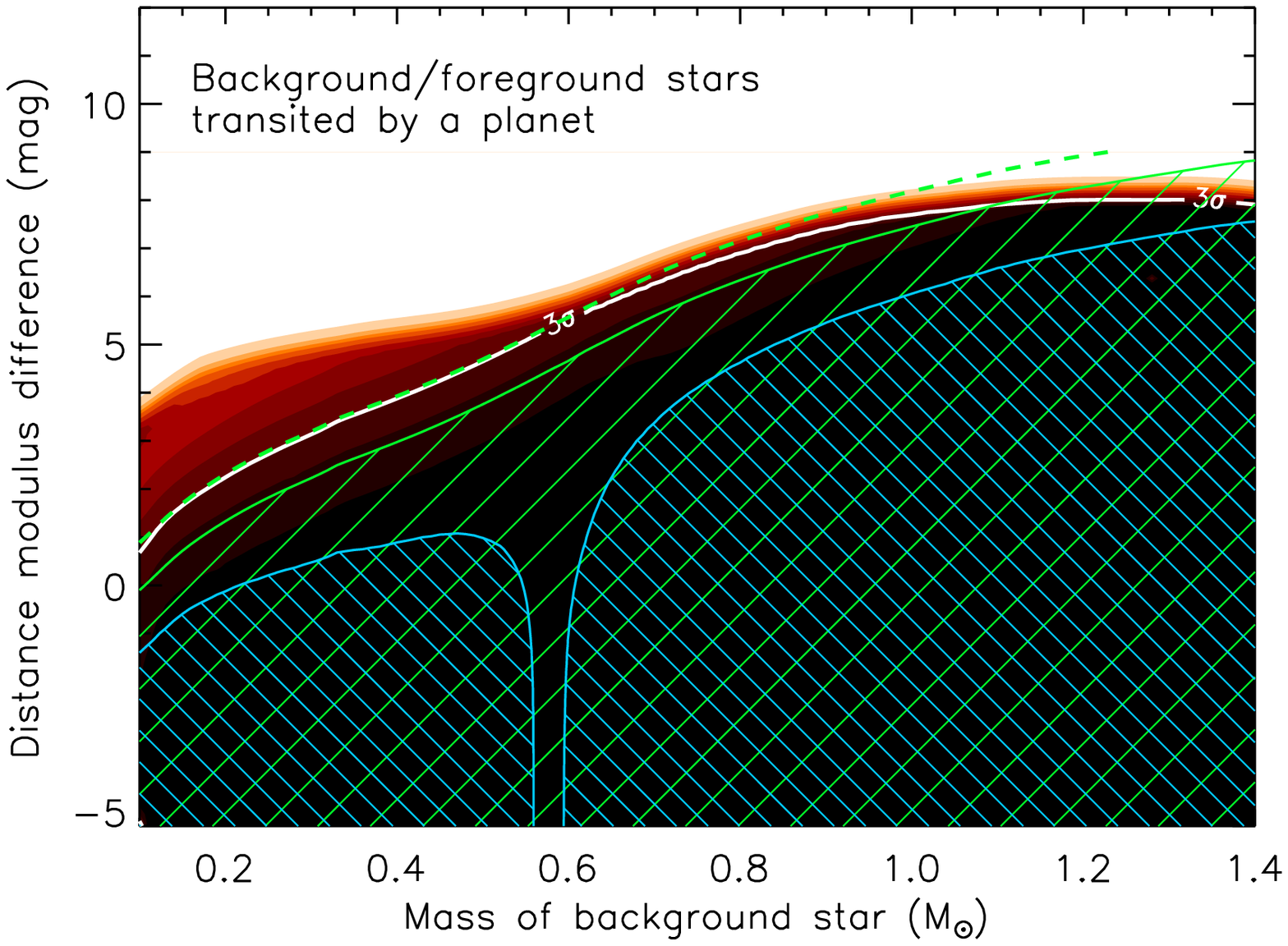} &
\includegraphics[width=6.0cm]{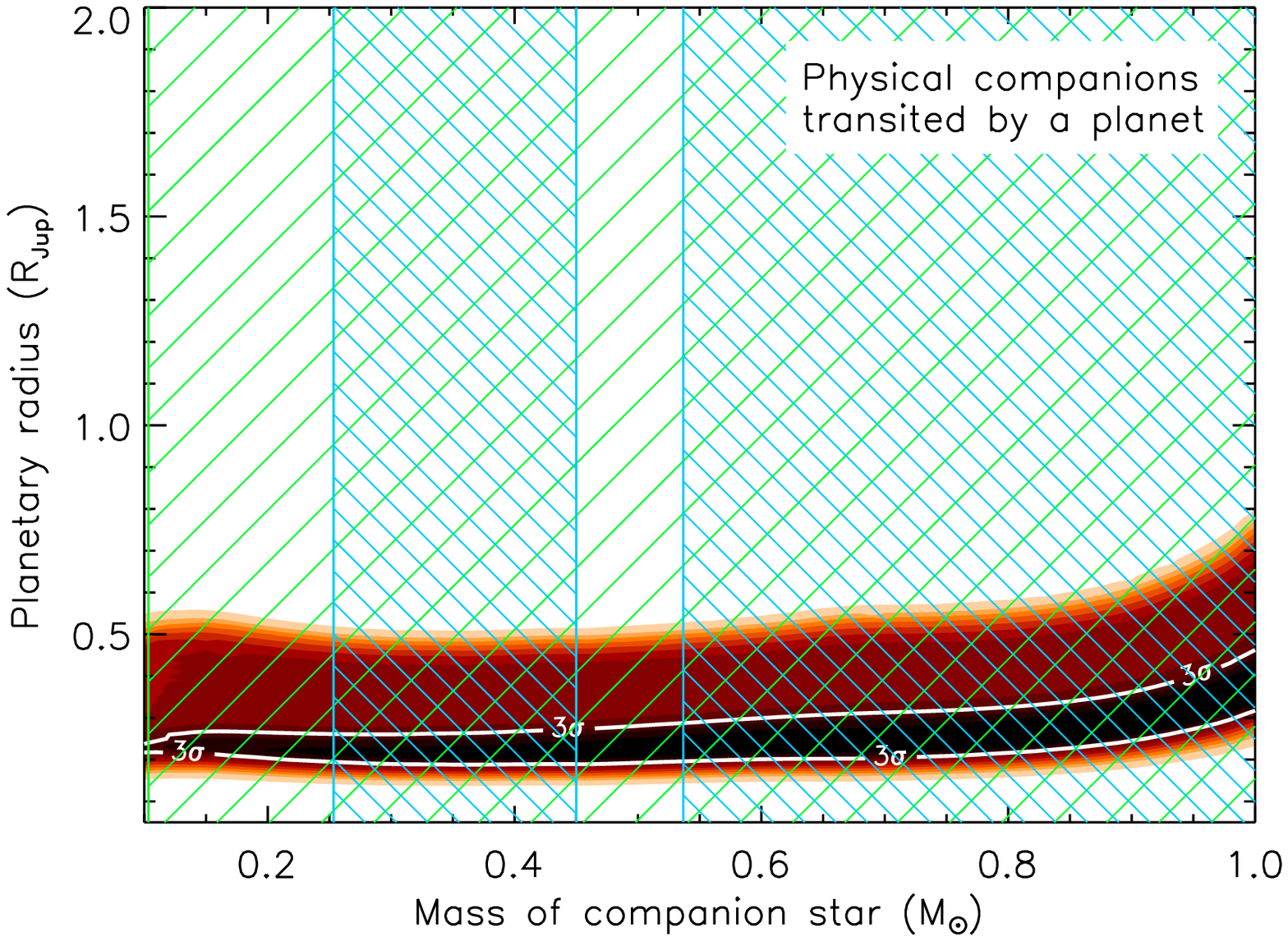} \\

\includegraphics[width=6.0cm]{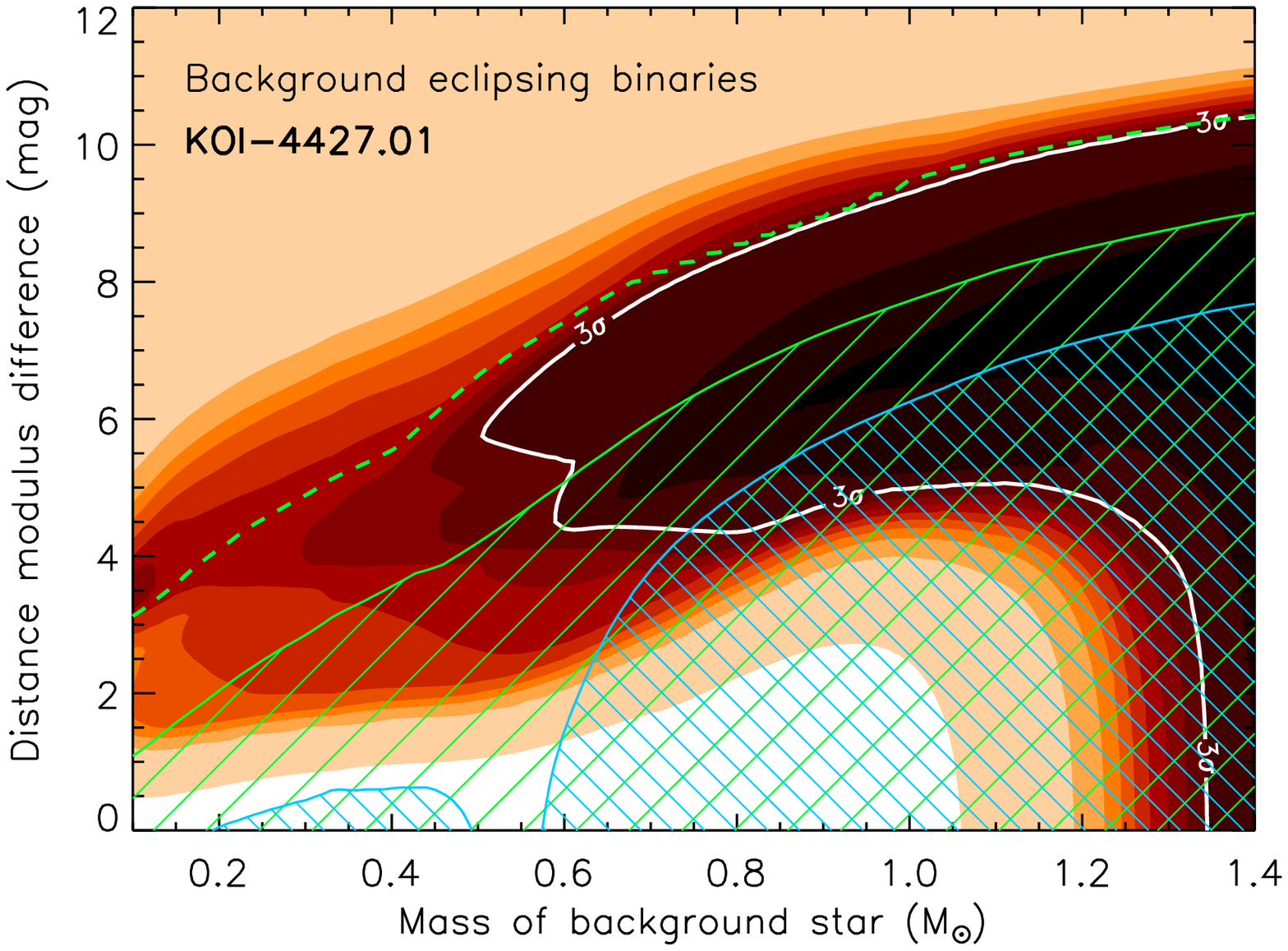} &
\includegraphics[width=6.0cm]{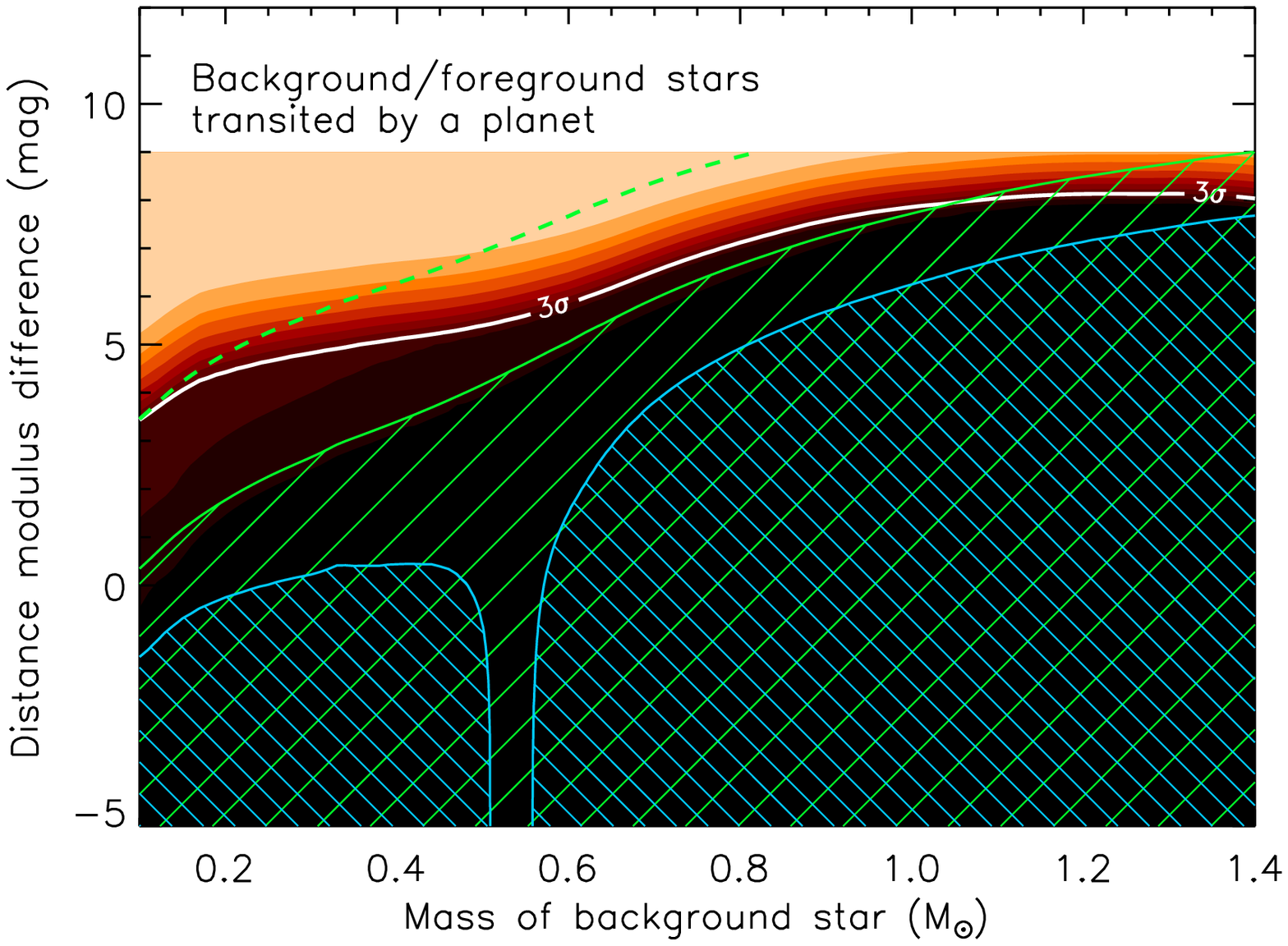} &
\includegraphics[width=6.0cm]{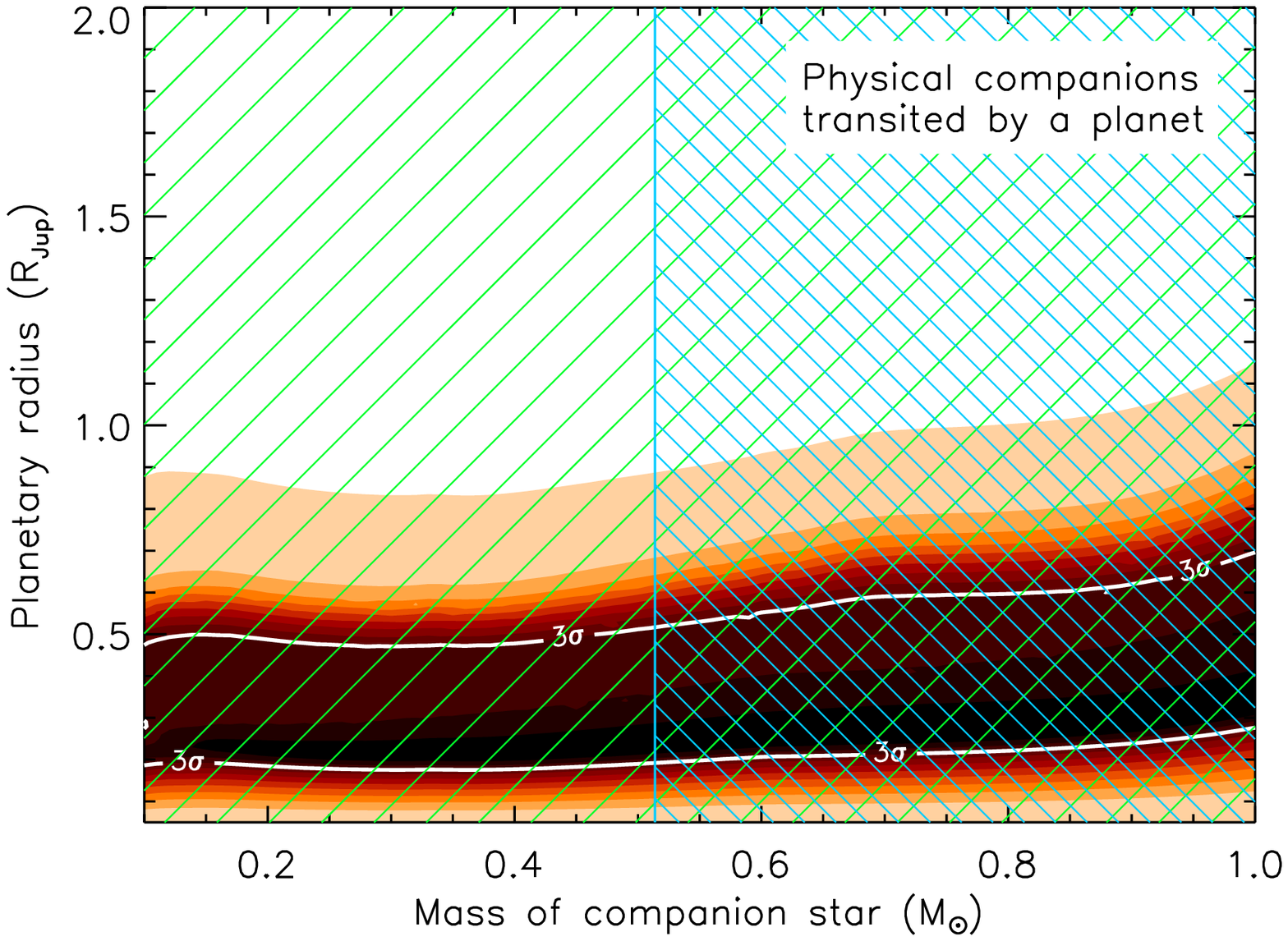} \\

\includegraphics[width=6.0cm]{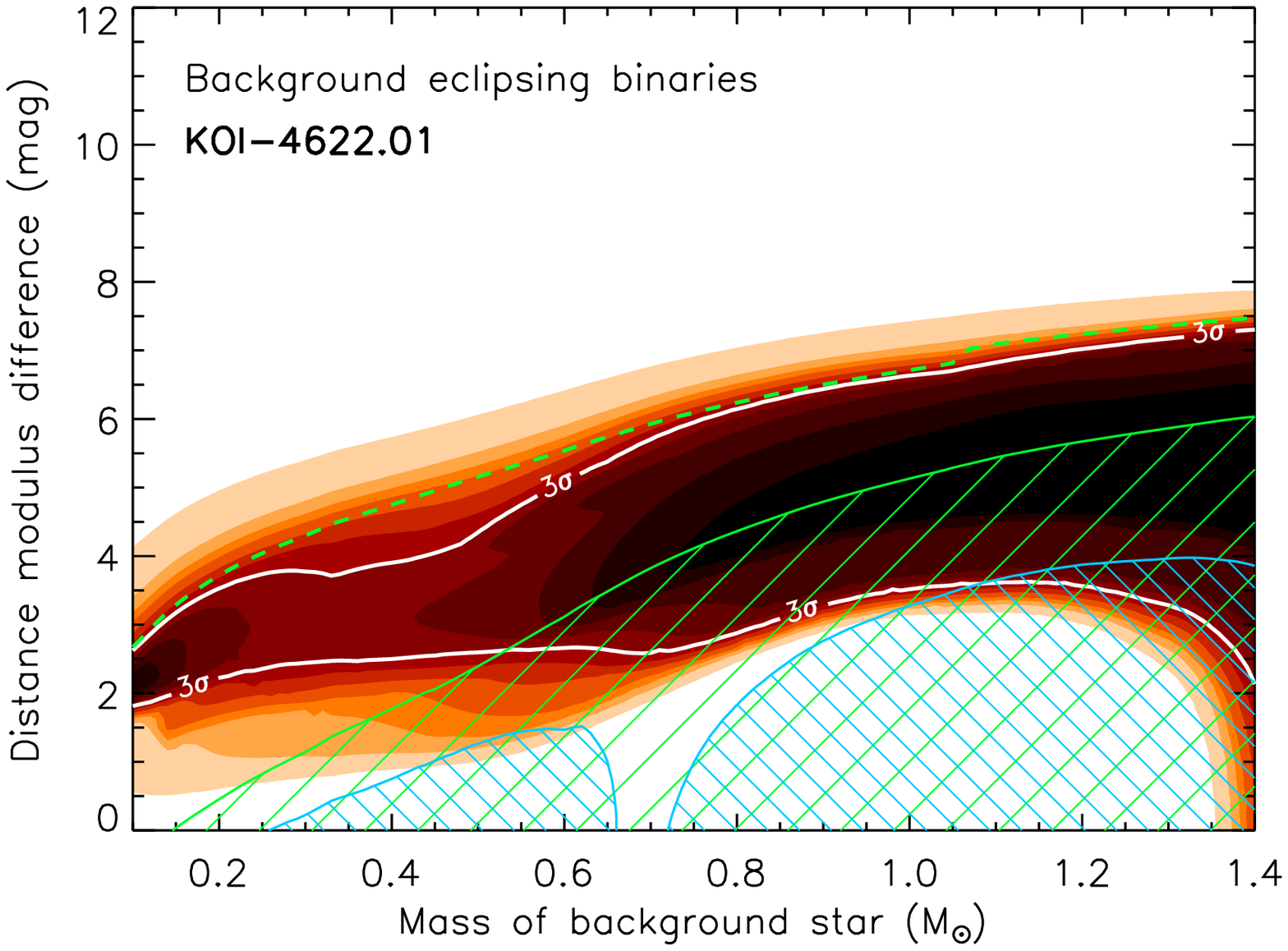} &
\includegraphics[width=6.0cm]{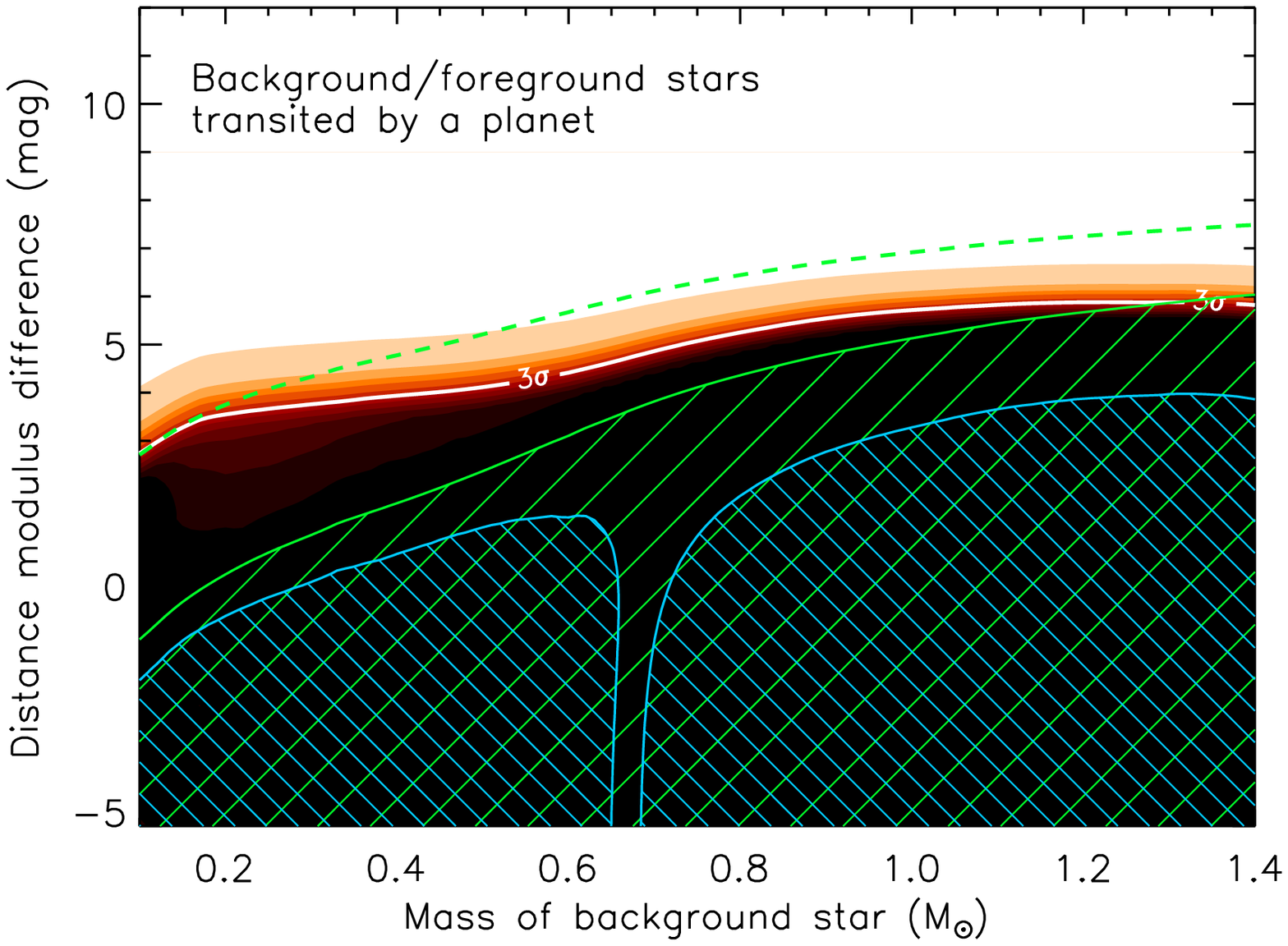} &
\includegraphics[width=6.0cm]{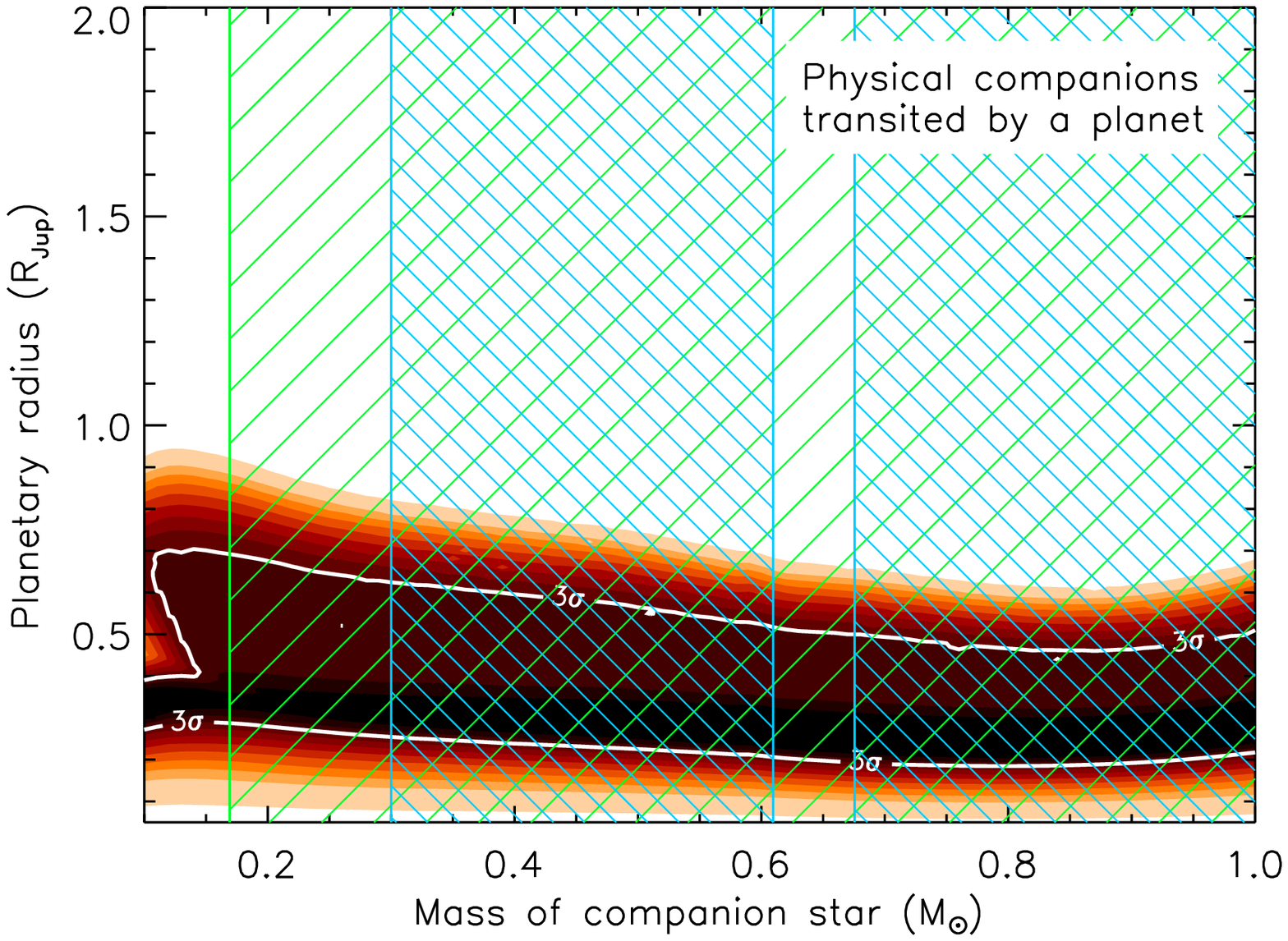} \\

\includegraphics[width=6.0cm]{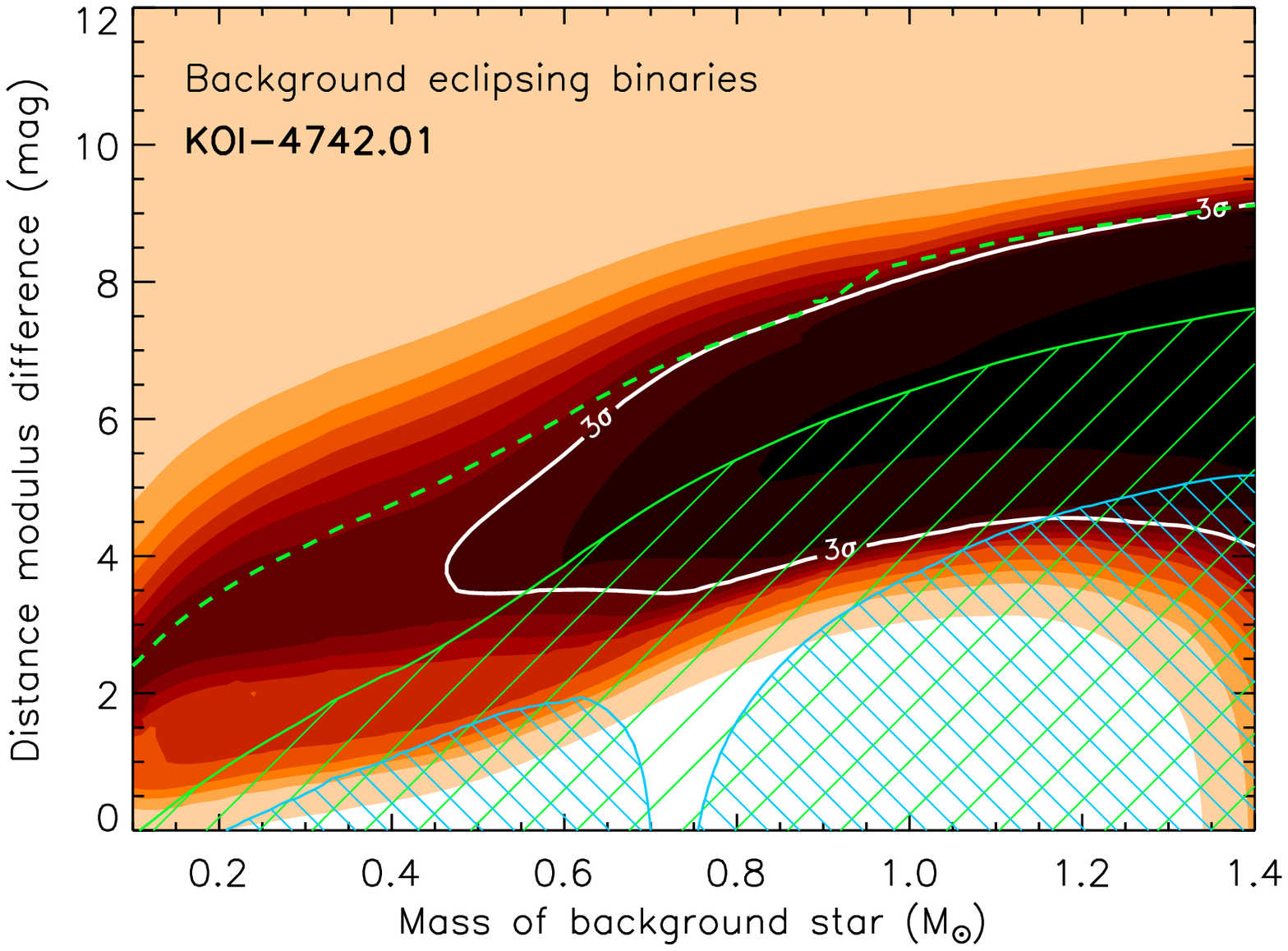} &
\includegraphics[width=6.0cm]{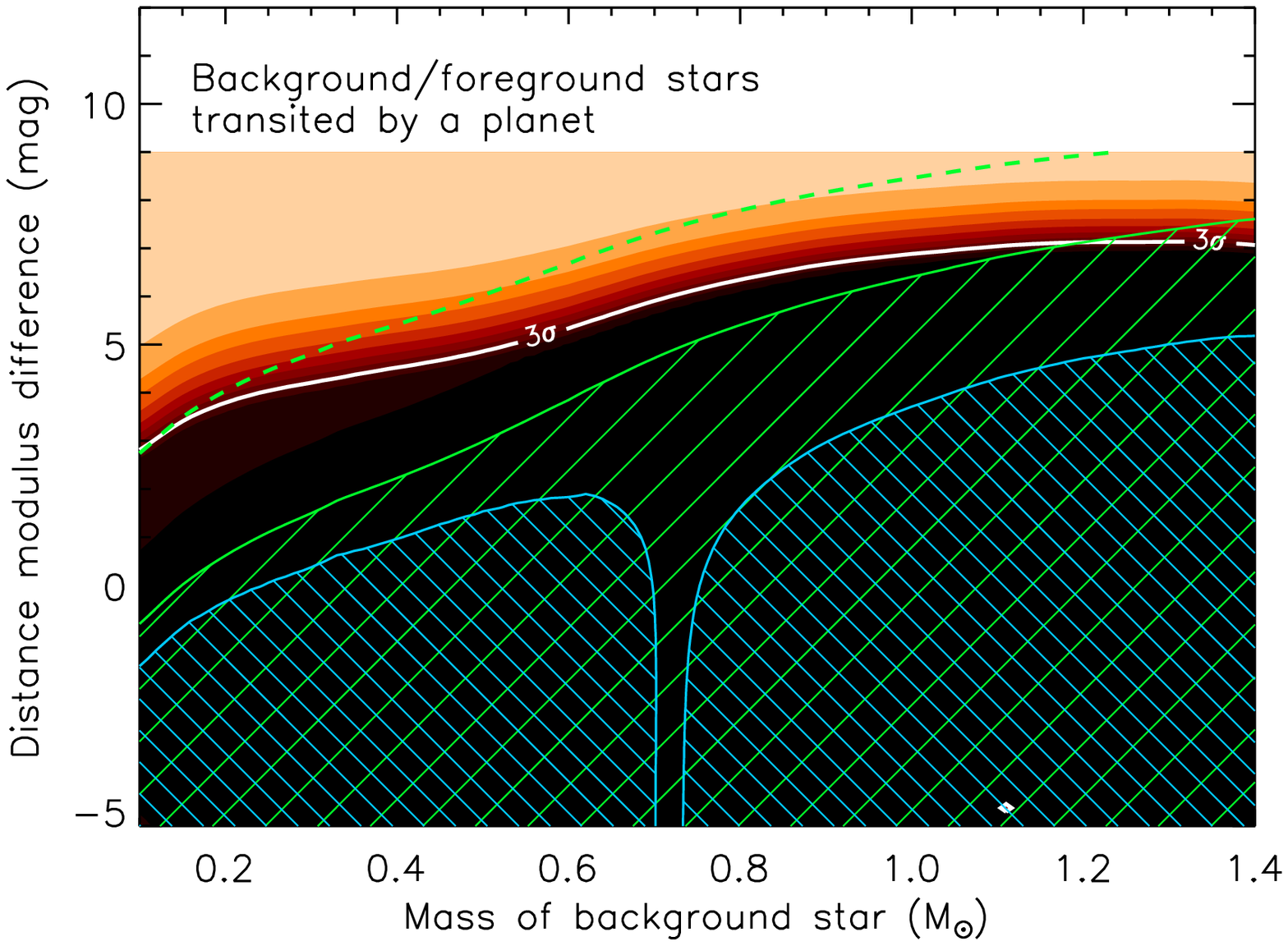} &
\includegraphics[width=6.0cm]{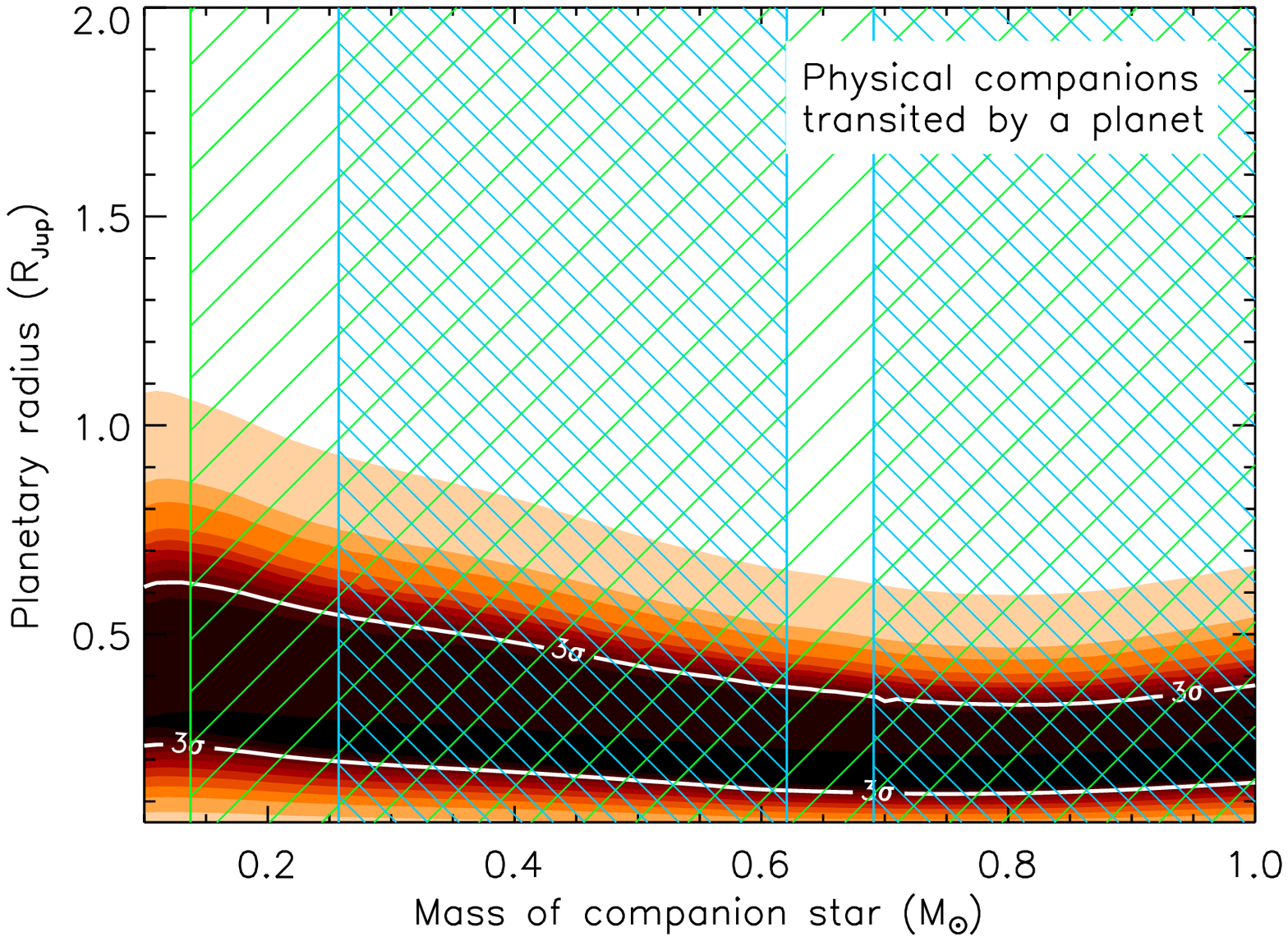} \\

\end{tabular}

\figcaption[]{Same as Figure~\ref{fig:blender1} for a second group of five KOIs
  in the sample.\label{fig:blender2}}

\end{figure*}

\begin{figure*}
\centering
\begin{tabular}{ccc}

\includegraphics[width=6.0cm]{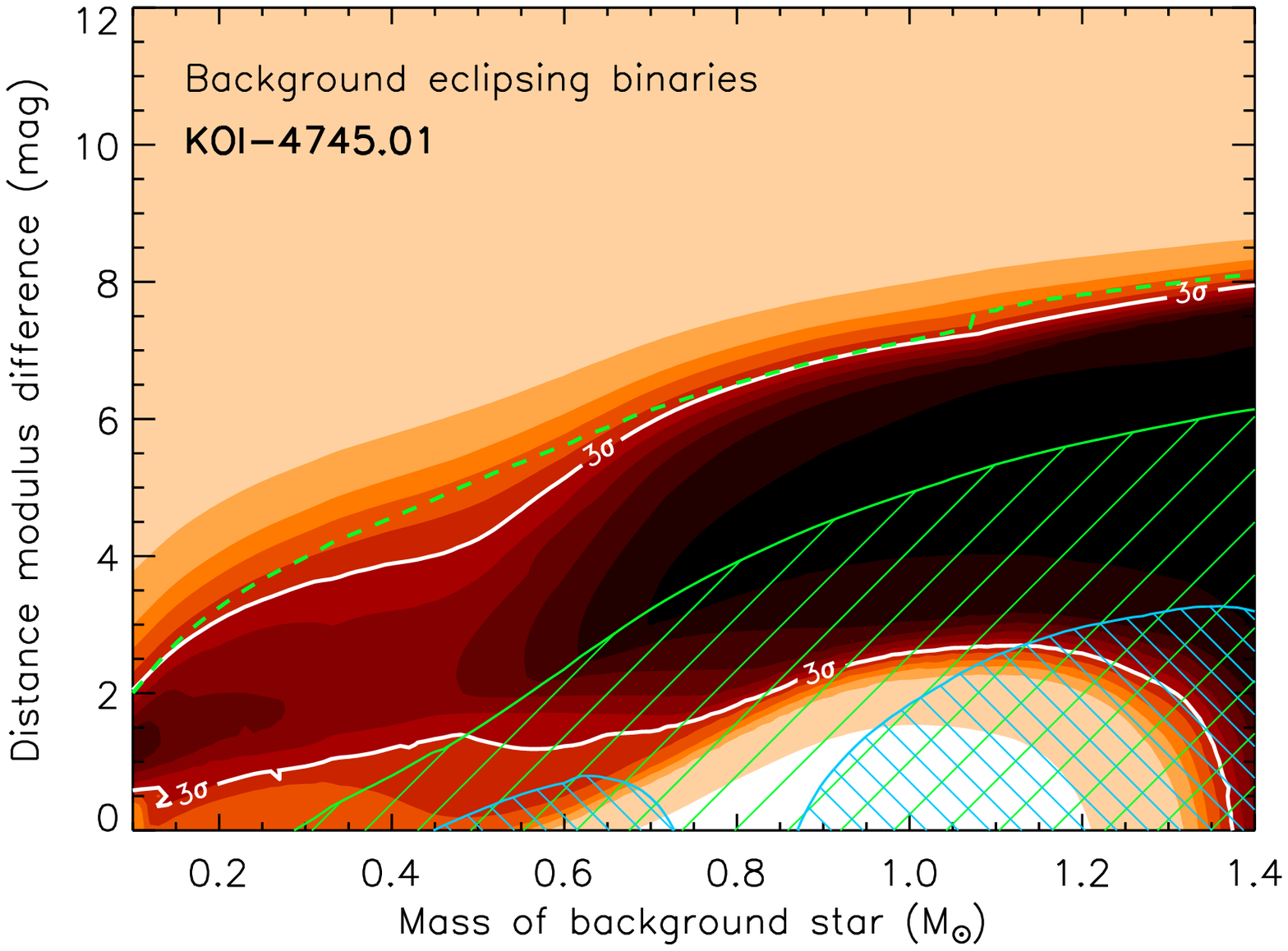} &
\includegraphics[width=6.0cm]{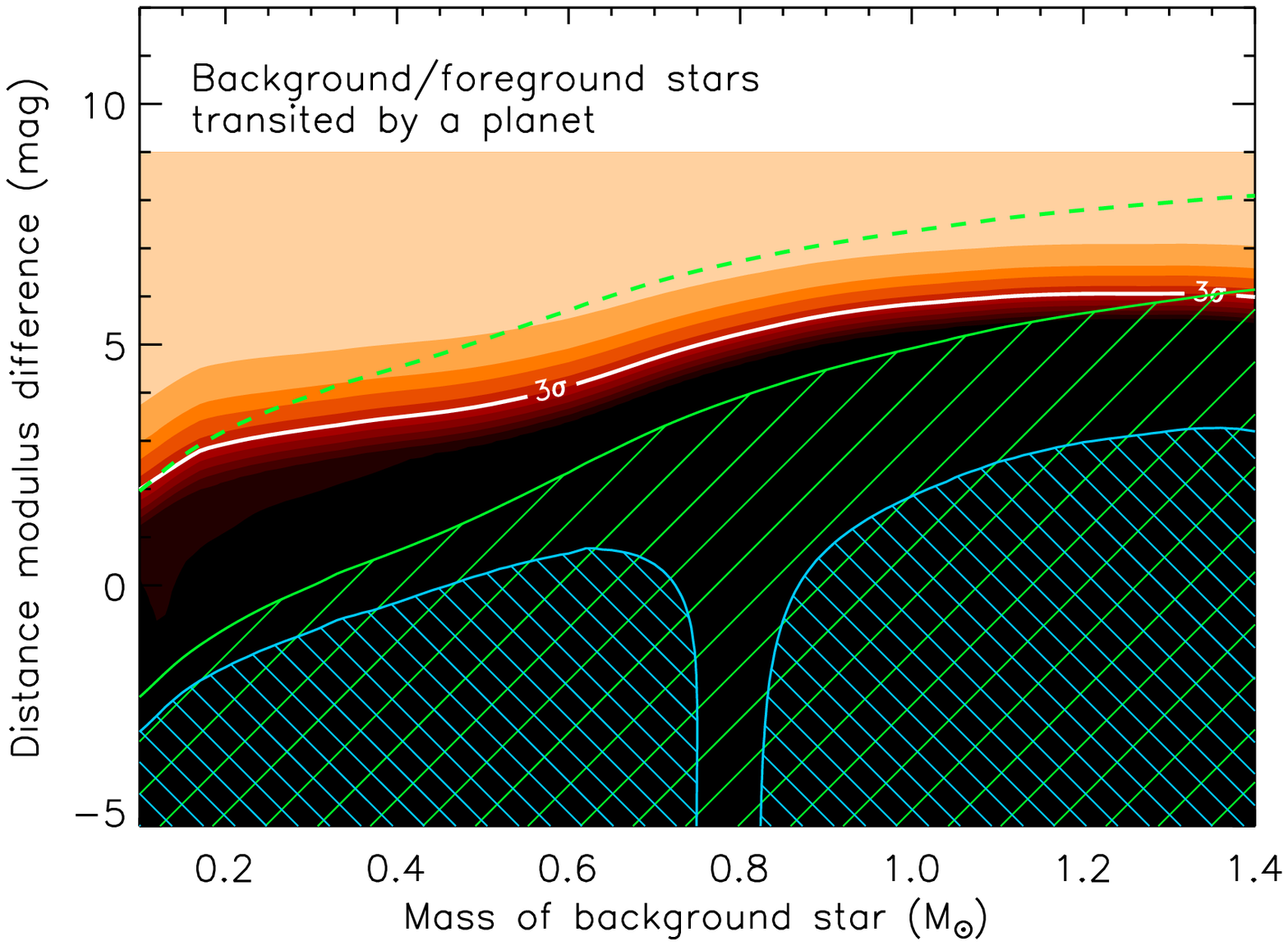} &
\includegraphics[width=6.0cm]{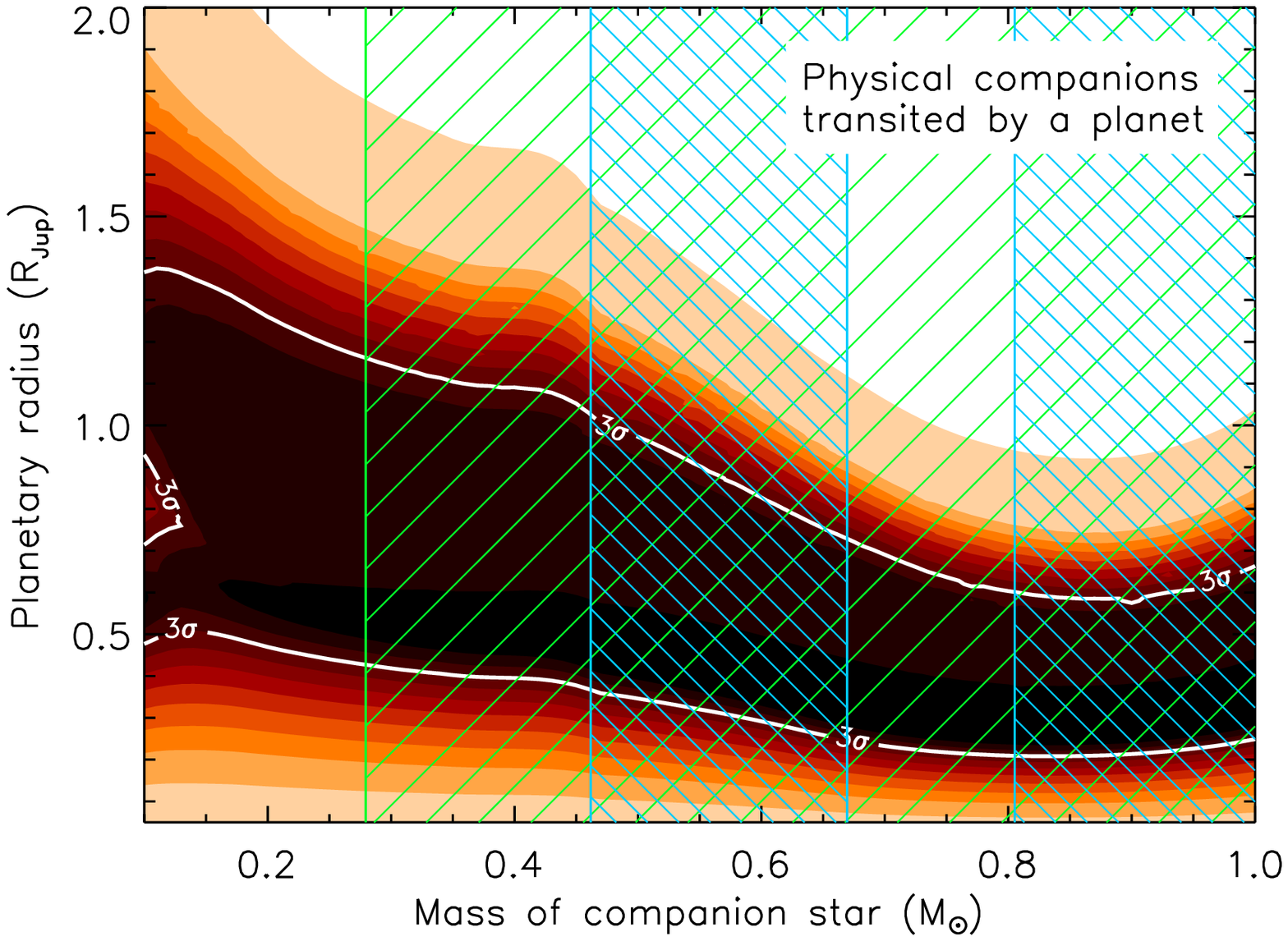} \\

\end{tabular}

\figcaption[]{Same as Figure~\ref{fig:blender1} for the remaining KOI
  in the sample.\label{fig:blender3}}

\end{figure*}

Maps of the $\chi^2$ landscapes for the BEB, BP, and HTP scenarios
analogous to those described above, along with the additional
restrictions based on the color and the spectroscopic brightness
limits, are shown for each of our other
candidates in Figure set~\ref{fig:blender1}--\ref{fig:blender3}.

\setlength{\tabcolsep}{-4pt}

\setlength{\tabcolsep}{6pt}   

\subsection{Estimating blend frequencies and the planet prior}
\label{sec:frequencies}

The \blender\ constraints just described significantly reduce the
number of viable BEB, BP, and HTP blends, but not all of the ones that
remain can be ruled out by our follow-up observations. The next step
of the analysis is therefore to quantify their expected frequencies,
which we did via Monte Carlo simulations. Specifically, we generated
large numbers of realistic false positives according to each of the
blend scenarios, we eliminated those that \blender\ tells us would not
match the \kepler\ light curve or that would have been detected in our
follow-up observations, and we counted the survivors.  These numerical
experiments rely on the known distributions of binary star properties,
the number density of stars near each target, and estimates of the
rates of occurrence of transiting planets and of eclipsing binaries
from the \kepler\ Mission itself.  Those rates of occurrence (as well
as any dependence they may have on orbital period or other properties)
are implicit in the lists of KOIs and eclipsing binaries generated by
the \kepler\ team \citep{Burke:14, Slawson:11} that we used in our
analysis, when normalized by the total number of targets observed by
the spacecraft.  Since we have recently made significant enhancements
in these Monte Carlo procedures compared to previous applications, we
take the opportunity to describe them in some detail here.

For the HTP scenario we simulated companion stars to each target
following the distributions of binary properties (mass ratios, orbital
periods, eccentricities) proposed by \cite{Raghavan:10} and
\cite{Tokovinin:14}. We placed the companions in random orbits around
the target assuming isotropically distributed inclination angles ($i$)
and uniformly distributed longitudes of periastron ($\omega$), as well
as random orbital phases. Other relevant properties of these simulated
companions (size, absolute magnitudes, and colors corresponding to
their simulated masses) were inferred from the same isochrone used for
the target, as the two stars are assumed to be coeval in this
scenario. In particular, we computed the distance to the system by
imposing the condition that the total apparent brightness of the two
stars (accounting for extinction) must be the same as the $K\!p$
magnitude associated with the KOI, and this in turn enables us to
compute the angular separation. We then assigned to each of these
secondary stars a random transiting planet drawn from the actual list
of KOIs hosted at the NASA Exoplanet Archive (list downloaded on 2014
March 26), which ensures that their properties (including any
correlations among them) are as realistic as possible.\footnote{As
  pointed out by \cite{Lissauer:14}, when attempting to validate
  targets with a single transiting planet candidate it is more
  appropriate to exclude from the list of known KOIs all stars with
  multiple candidates, as this unduly inflates the rates of occurrence
  of planets (particularly small ones that are more common in multiple
  systems). We have followed this recommendation here.}  We accepted
only planets with periods similar to the KOI under investigation
(within a factor of two). The rationale for this is that the relevant
HTP blend frequency is that of configurations that involve planets
with periods near that of the candidate, as those frequencies depend
strongly on orbital period.  We then examined the properties of the
stellar companions and their planets, and rejected scenarios that do
not satisfy the \blender\ restrictions on companion mass, planetary
size, and orbital eccentricity of the planet (related to the transit
duration), as they yield poor fits to the transit light curve. We also
rejected configurations with companions bright enough that they would
have been detected from our high-resolution imaging observations, our
spectroscopic observations, by our centroid motion analysis, or that
would yield an overall color for the blend inconsistent with the
measured color of the target \citep{Brown:11}. In applying the
spectroscopic constraint we excluded companions brighter than the
limits described in Sect.~\ref{sec:opticalspectroscopy} only if they
are within 0\farcs43 of the target (half width of the spectrometer
slit), unless their radial velocity computed from the simulated orbit
around the target is within 10\,\kms\ of that of the primary. In that
case, blending of the spectral lines of the secondary with those of
the primary would prevent detection, so we considered those blends
still viable. Finally, we retained only the false positive
configurations that would be dynamically stable according to the
criterion of \cite{Holman:99}. Since not every stellar companion will
necessarily have a transiting planet and therefore act as a blend, as
we assumed above, we adjusted the blend frequencies by accounting for
the rate of occurrence of planets of each size and period as inferred
from the KOI list itself. Minor corrections were applied to those
rates for incompleteness and for the incidence of false positives in
the KOI list, by means of simulations as described by
\cite{Fressin:13}. An additional adjustment was made to account for
the dependence of the transit probability on the size of the secondary
relative to the average size of the host stars in the KOI list. The
final blend frequencies were obtained by multiplying by the occurrence
rate of binary stars with planets as a function of binary semimajor
axis, following \cite{Wang:14}.

The calculation of the frequency of background or foreground stars
transited by a planet (BP) proceeded in a similar fashion, and scales
with the number density of stars in the vicinity of each target, for
which we relied on predictions from the Galactic structure model of
\cite{Robin:03, Robin:12} (Besan\c{c}on model).\footnote{We note that the
  current online version of the Besan\c{c}on simulator (available at
  {\tt http://model.obs-besancon.fr}) relies on stellar evolution
  models by \cite{Haywood:94} to generate some of the stellar
  properties returned, including absolute magnitudes, colors,
  temperatures, and radii. Many of the simulated main-sequence stars
  of interest for our work are significantly cooler than the Sun
  (spectral type K and M), and the \cite{Haywood:94} models are not
  specifically designed for such stars as they adopt gray boundary
  conditions and a rather simple equation of state, both of which
  result in increasingly biased predictions for cool stars
  \citep[e.g.,][]{Chabrier:01}.  Because of this, we have preferred to
  recompute the relevant properties of all simulated stars adopting
  from the output of the Besan\c{c}on simulator only the basic
  characteristics that derive from the initial mass function and the
  Galactic structure model, which are the stellar mass, distance, age,
  and metallicity. We then used model isochrones from the Dartmouth
  series \citep{Dotter:08} to generate all other stellar quantities
  for our purposes, as the physical ingredients of these isochrones
  give more realistic predictions for the lower main-sequence.}  We
generated a list of simulated stars in a 5 square-degree area around
each of our targets, including their kinematic properties (radial
velocity), and drew stars randomly from this list assigning them a
random angular separation from the target within the corresponding
3$\sigma$ exclusion region derived from our centroid motion analysis
(since stars outside this area would have been detected; see
Sect.~\ref{sec:centroids}). We assigned a random planet to each of
these stars that we drew from the KOI list, keeping only those with
orbital periods within a factor of two of that of the candidate, as
done previously.  False positive scenarios that do not meet the
constraints from \blender\ were rejected, along with those that would
have been flagged by our high-resolution imaging observations or color
information. Further blends were excluded by the spectroscopic limits
on the brightness of any unseen companions, as before, except for
background/foreground stars whose spectral lines would be blended with
those of the primary (i.e., those with simulated radial velocities
within 10\,\kms\ of the measured heliocentric velocity of the target),
which we regarded as viable.  Surviving blends were weighted by the
corresponding rate of occurrence of transiting planets of that size
and period, corrected as in the HTP case for incompleteness and false
positives and adjusted for the dependence of the transit probability
on the secondary radius.  The final blend frequencies were then
normalized by the ratio of the areas between the centroid exclusion
region and 5 square degrees.

To estimate the frequency of BEBs acting as blends we again drew stars
randomly from a list generated with the Besan\c{c}on Galactic
structure model, and paired them with an orbiting stellar companion
(tertiary) generated from the distributions of binary properties by
\cite{Raghavan:10} and \cite{Tokovinin:14}. We assigned orbital
periods to these secondary-tertiary pairs drawn randomly from the
catalog of \kepler\ eclipsing binaries of \cite{Slawson:11}, retaining
only those within a factor of two of the period of the candidate. Each
blend was also assigned a random angular separation from the target
within the 3$\sigma$ exclusion limit from our centroid motion
analysis. After rejecting scenarios that do not meet the constraints
from \blender\ for acceptable fits to the transit light curve, we
applied the observational constraints (imaging, spectroscopy, color
information) as in the BP scenario, and tallied the remaining viable
blends. For computing the final BEB frequency, each surviving blend
was assigned a weight given by the ratio of the eclipse probability of
the particular blend (a function of the secondary and tertiary sizes)
to the average eclipse probability for the \cite{Slawson:11} sample as
a whole.

The probability that the candidate is a true planet as opposed to a
false positive (`planet prior', PL) was estimated by simply counting
the number of known KOIs with periods within a factor of two of each
candidate, radii within 3$\sigma$ of the measured planetary radius,
and with similar transit durations.  Incompleteness and false positive
corrections were applied in the same way as above.

\subsection{Results}
\label{sec:results}

Blend frequencies for the HTP, BP, and BEB scenarios are presented for
each of our candidates in Table~\ref{tab:frequencies}, along with the
planet priors, PL. For the present work we have set a threshold for
validation equivalent to a 3$\sigma$ confidence level, consistent with
previous applications of \blender.  The statistical significance may
be expressed as ${\rm PL}/({\rm PL}+{\rm HTP}+{\rm BP}+{\rm
  BEB})$. For a 3$\sigma$ validation we therefore require a planet
prior that is at least $1/(1/99.73\%-1) \approx 370$ times larger than
the total blend frequency. We list these `odds ratios' in
Table~\ref{tab:frequencies}, with the corresponding significance
levels.  For eleven of the twelve candidates we have achieved very
robust validations at the 3$\sigma$ level or higher, strongly
supporting their true planetary nature. On this basis, the eight that
have not been previously validated by others are given the new planet
designations Kepler-436\,b to Kepler-443\,b (see Table~\ref{tab:targets}).
KOI-4427.01 has a lower odds ratio of only $\sim$100, corresponding to
a significance level slightly over 99\%. To avoid confusion in the
rest of the paper we will continue to refer to the newly validated
planets by their KOI numbers.

\begin{deluxetable*}{lcccccccc}
\tablewidth{0pc}
\tablecaption{Blend frequencies, planet priors, odds ratios, and
  significance level of the validation for our
  targets.\label{tab:frequencies}}
\tablehead{
\colhead{Candidate} &
\colhead{HTP} &
\colhead{BP} &
\colhead{BEB} &
\colhead{PL} &
\colhead{Odds ratio} &
\colhead{Significance} &
\colhead{PL$_{comp}$} &
\colhead{$\mathcal{P}[{\rm targ}]$}
}
\startdata
 KOI-0571.05  &  $1.15 \times 10^{-6}$  &  $1.32 \times 10^{-7}$  &  $6.60 \times 10^{-8}$  &  $2.15 \times 10^{-3}$  & 1595  &   99.94\%  & \nodata & \nodata \\
 KOI-1422.04  &  $7.75 \times 10^{-7}$  &  $4.74 \times 10^{-8}$  &  $1.28 \times 10^{-8}$  &  $3.19 \times 10^{-4}$  &  382  &   99.74\%  & $5.94 \times 10^{-4}$ & 78.4\% \\
 KOI-1422.05  &  $4.49 \times 10^{-7}$  &  $4.84 \times 10^{-8}$  &  $1.29 \times 10^{-8}$  &  $4.63 \times 10^{-4}$  &  907  &   99.89\%  & $7.25 \times 10^{-5}$ & 86.5\% \\
 KOI-2529.02  &  $3.24 \times 10^{-6}$  &  $2.17 \times 10^{-9}$  &  $5.55 \times 10^{-9}$  &  $1.31 \times 10^{-3}$  &  403  &   99.75\%  & $5.61 \times 10^{-6}$ & 99.6\% \\
 KOI-3255.01  &  $4.29 \times 10^{-7}$  &  \phm{$^1$}$4.79 \times 10^{-11}$ & \nodata       &  $4.91 \times 10^{-4}$  & 1144  &   99.91\%  & $4.39 \times 10^{-4}$ & 52.8\% \\
 KOI-3284.01  &  $6.71 \times 10^{-7}$  &  $1.10 \times 10^{-9}$  &  \phm{$^1$}$9.27 \times 10^{-10}$  &  $1.57 \times 10^{-3}$  & 2333  &   99.96\%  & $8.02 \times 10^{-4}$ & 66.2\% \\
 KOI-4005.01  &  $3.71 \times 10^{-7}$  &     \nodata             & \nodata                 &  $6.61 \times 10^{-4}$  & 1782  &   99.94\%  & \nodata & \nodata \\
 KOI-4087.01  &  $1.34 \times 10^{-7}$  &        \nodata          &        \nodata          &  $1.32 \times 10^{-4}$  &  985  &   99.90\%  & \nodata & \nodata \\
 KOI-4427.01  &  $2.23 \times 10^{-6}$  &  $1.07 \times 10^{-9}$  &  \phm{$^1$}$1.39 \times 10^{-10}$ &  $2.62 \times 10^{-4}$  & 117   &   99.16\% & \nodata & \nodata \\
 KOI-4622.01  &  $9.98 \times 10^{-8}$  &  \phm{$^1$}$3.82 \times 10^{-10}$  & \phm{$^1$}$6.68 \times 10^{-12}$ &  $2.08 \times 10^{-3}$  & 20761 &   99.99\% & \nodata & \nodata \\
 KOI-4742.01  &  $1.37 \times 10^{-6}$  &  \phm{$^1$}$4.89 \times 10^{-10}$  & \phm{$^1$}$5.45 \times 10^{-12}$ &  $1.26 \times 10^{-3}$  & 919   &   99.89\% & \nodata & \nodata \\
 KOI-4745.01  &  $1.53 \times 10^{-6}$  &  $1.41 \times 10^{-9}$  &  \phm{$^1$}$9.19 \times 10^{-12}$ &  $2.85 \times 10^{-3}$  & 1861  &   99.95\% & \nodata & \nodata 
\enddata

\tablecomments{For several of the candidates the blend frequencies for
  the BP and/or BEB scenarios are negligible and are not listed. For
  the four targets with close companions PL$_{comp}$ is the planet
  prior assuming the planet transits the close companion, and
  $\mathcal{P}[{\rm targ}]$ the probability that the planet transits
  the target rather than the companion.}

\end{deluxetable*}

Strictly speaking, the above results compare the likelihood that the
transit signals originate on the target itself with the likelihood
that they come from eclipses associated with some other \emph{unseen}
star in the photometric aperture. However, four of our target stars
(KOI-1422, 2529, 3255, and 3284) have \emph{known} close companions
from our high-resolution imaging observations, none of which can be
ruled out as the source of the transit signal by our centroid motion
analysis in Sect.~\ref{sec:centroids}.  Thus, the possibility remains
that the planets causing the signals orbit the companions rather than
the targets.  In that case, the inferred radius of the planets would
be somewhat larger than the values derived in the next section, the
exact amount depending on the brightness difference (dilution effect)
and the physical size of the companion stars.

An estimate of the likelihood that the planets transit the companions
was derived as follows. In each of the four cases we made the
assumption that the companions are physically associated with the
target. This is reasonable, as the probability of having an unrelated
background star as close to the primary stars as observed is very
small compared to that of a physical companion \citep[see
  also][]{Horch:14}, based on the number density of stars around each
KOI from the Besan\c{c}on model and the known rate of occurrence of
binaries \citep{Raghavan:10, Wang:14}.  Furthermore, no signs of any
of these companions are seen in our Keck/HIRES spectra even though in
most cases they are bright enough to have been detected. This implies
that their radial velocities must be similar to those of the primary
stars so that the spectral lines are blended, preventing
detection. This again argues for physical association.\footnote{In an
  independent study of KOI-1422, \cite{Star:14} also reached the
  conclusion that the companion is physically bound, based on
  numerical experiments with model isochrones.}  In each case we then
used the same isochrone as for the primaries to infer the companion
properties, and we computed the `planet prior' in the same way as for
the targets, but this time assuming the planet transits the companion
(i.e., increasing the dilution factor by the appropriate amount). We
represent this planet prior for the companions as PL$_{comp}$, and we
report these values in Table~\ref{tab:frequencies}. Also given in the
table are the probabilities that the planets transit the target rather
than the companion, which we computed as $\mathcal{P}[{\rm targ}] = {\rm
  PL}/({\rm PL} + {\rm PL}_{comp})$. For KOI-2529.02 the results allow
us to state with high confidence that the planet orbits the
target. For KOI-1422.04, KOI-1422.05, and KOI-3284.01 the calculations
indicate a slight or modest preference for the planet being around the
target rather than the companion, whereas for KOI-3255.01 the result
is inconclusive. We return to this issue later in
Sect.~\ref{sec:APblends}.

\section{Light curve fits and planetary parameters}
\label{sec:fits}

Having validated eleven of our candidates as bona-fide planets with
very high confidence, and KOI-4427.01 to a lower degree of confidence,
we subjected their light curves to a detailed modeling to infer the
properties of the transiting objects. For this we used the detrended
SAP time series from \kepler\ described earlier in
Sect.~\ref{sec:photometry}.  The transits were modeled using the
standard \citet{Mandel:02} algorithm employing the quadratic
limb-darkening law. This simple model assumes a spherical, opaque
planet transiting a spherically symmetric luminous star on a circular
Keplerian orbit. We re-sampled the long-cadence data onto
short-cadence sampling following the method described by
\citet{Kipping:10}, to avoid smearing effects. Our model has eight
free parameters in total. These are the orbital period, $P$, the time
of transit center, $\tau$, the planet-to-star radius ratio,
$R_p/R_{\star}$, the mean stellar density, $\rho_{\star,\rm phot}$,
the impact parameter, $b$, the logarithm of fractional light
contamination, $\log \beta$, and the quadratic limb-darkening
coefficients $q_1$ and $q_2$.  All of these parameters have uniform
priors in our fits, except $\rho_{\star,\rm phot}$, for which we
employed a Jeffrey's prior between $10^{-3}$ and $10^3$\,g\,cm$^{-3}$,
and $\log\beta$, for which we used an informative Gaussian prior where
appropriate.

In our model, $\beta$ is the flux of any contaminating sources in the
aperture divided by the flux of the target star. The purpose of this
term is to account for dilution of the transit light curve due to
close companions identified by the high spatial resolution imaging
observations.  We computed these ratios here based on the magnitude
differences of the companions listed in Table~\ref{tab:ao},
transformed to the $K\!p$ band, and assuming that all their flux is
included in the aperture. For the wider companions
(4--5\arcsec\ separations) that are also in the aperture the
transformed magnitudes were taken from the CFOP Web site, and for the
closer ones we carried out the conversion using the Dartmouth
isochrone for each target, on the assumption that the companions are
physically associated. In the case of KOI-1422 we adopted the
magnitude difference in the $K\!p$ band as reported by \cite{Star:14}.
The $\beta$ factors range from $0.047 \pm 0.009$ for KOI-2529 to $0.83
\pm 0.18$ for KOI-3255, which has a very bright companion.
Additionally, other nearby \kepler\ sources may contaminate the light
curve (e.g., bleeding into the photometric aperture) and were
corrected for using the reported contamination factors from the
\kepler\ pipeline, treating them as fixed parameters.

It is worth noting that we did not directly fit for the standard
quadratic limb darkening coefficients $u_1$ and $u_2$, but instead we
used the transformed parameters $q_1$ and $q_2$ as advocated by
\citet{Kipping:13b}, in order to impose efficient, uninformative, and
physical priors for the limb darkening profile.  We note also that the
out-of-transit baseline flux for each transit epoch was fitted as
well. However, in this case we used a linear minimization to derive
the baseline flux, similar to that described by \citet{Kundurthy:11}.
This treats the baseline flux simply as a nuisance parameter that is
not marginalized against, but rather minimized at each Monte Carlo
realization, reducing the number of free parameters yet at the same
time allowing us to account for any residual offsets in the detrended
light curves.

\begin{figure*}
\centering
\begin{tabular}{ccc}
\includegraphics[width=5.5cm]{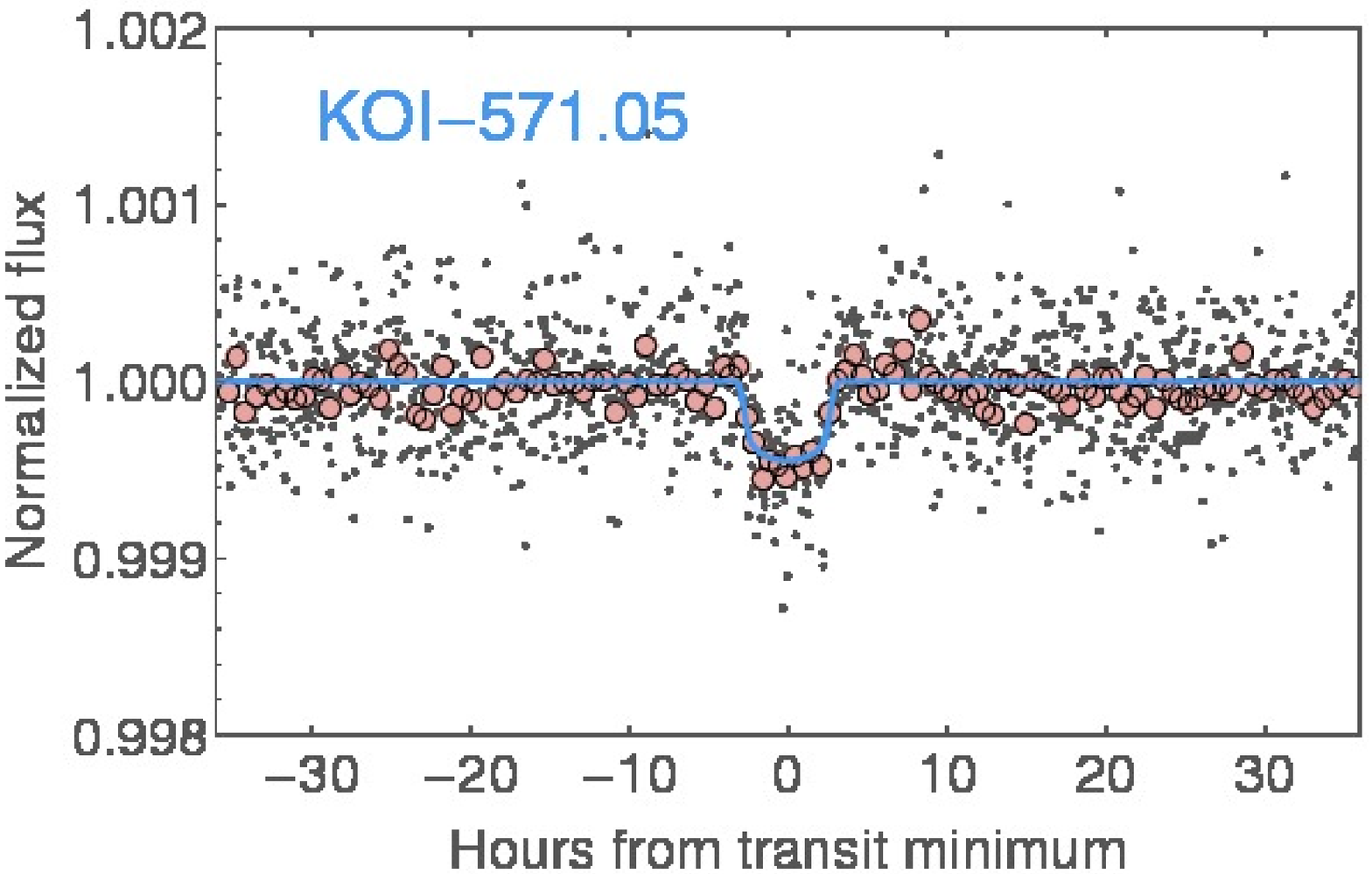} &
\includegraphics[width=5.5cm]{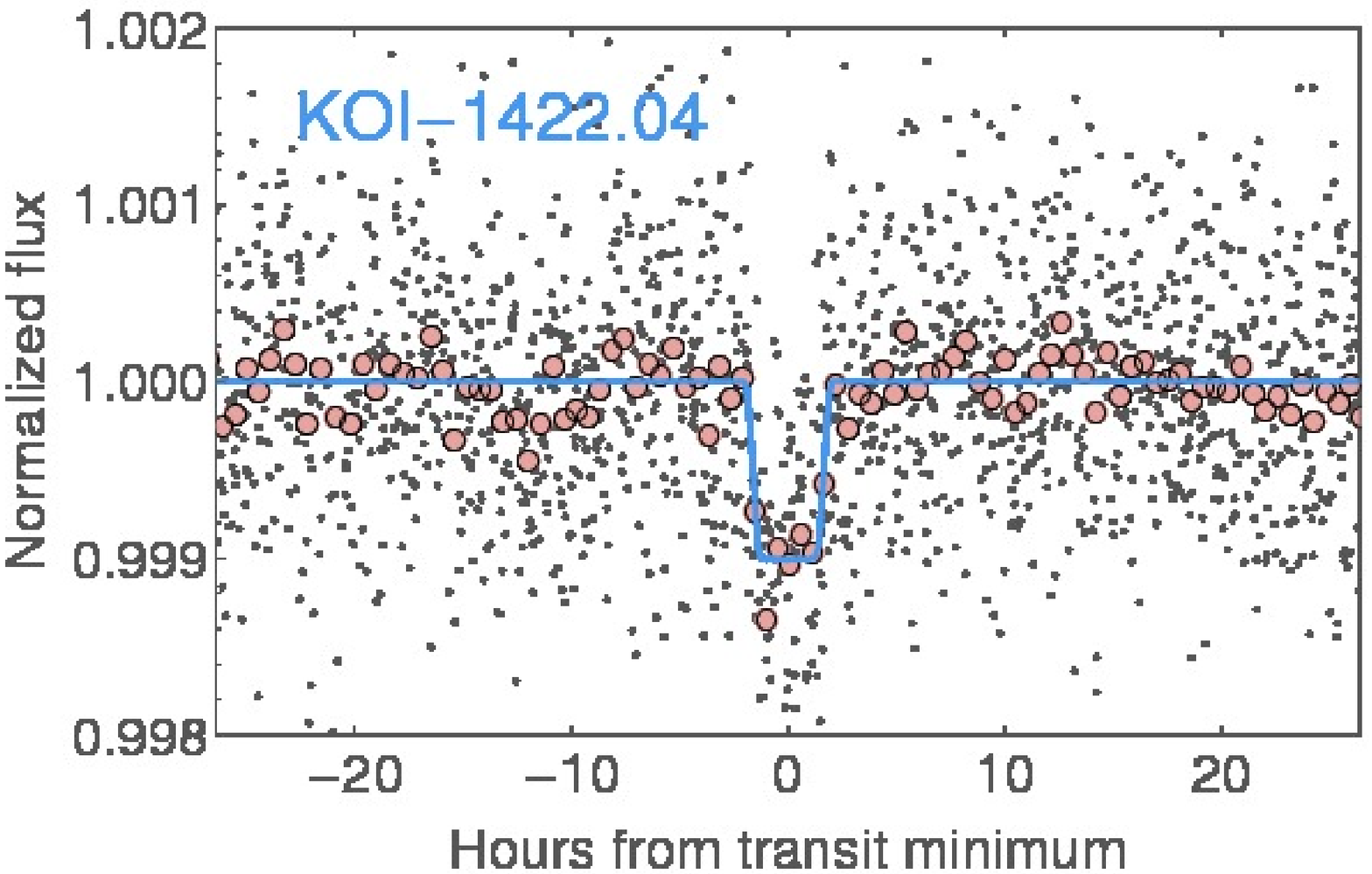} &
\includegraphics[width=5.5cm]{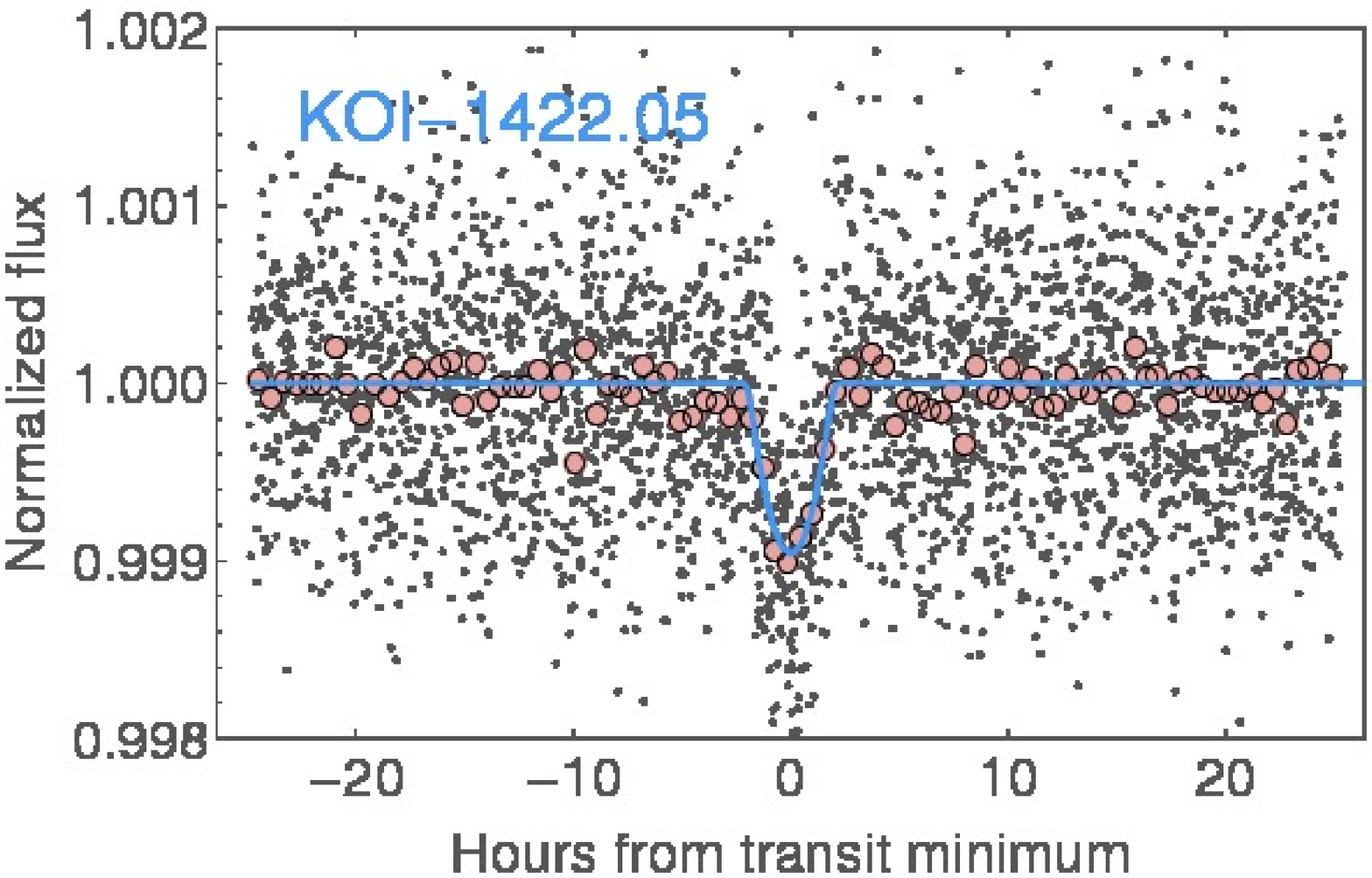}\\[+1ex]
\includegraphics[width=5.5cm]{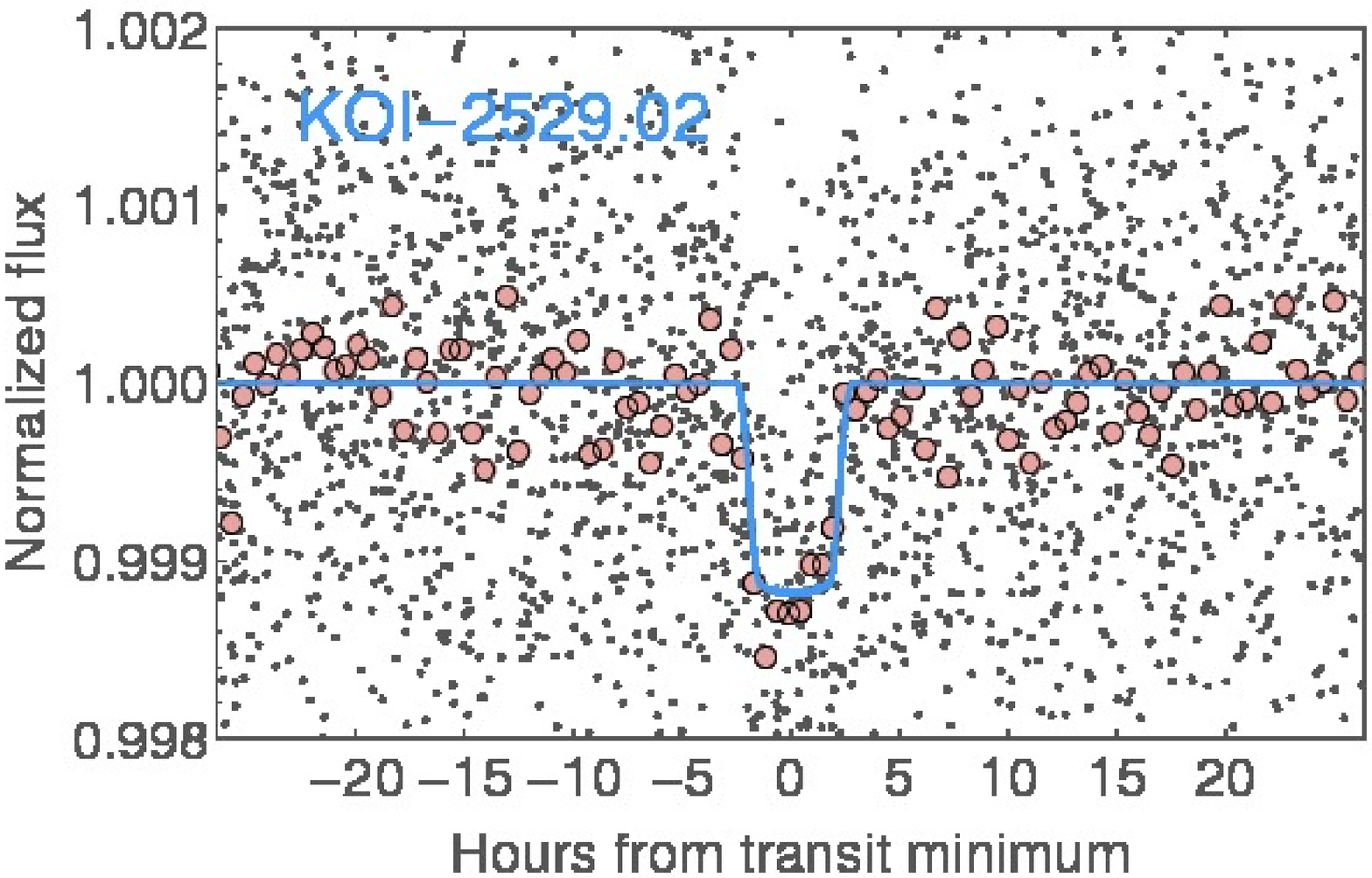} &
\includegraphics[width=5.5cm]{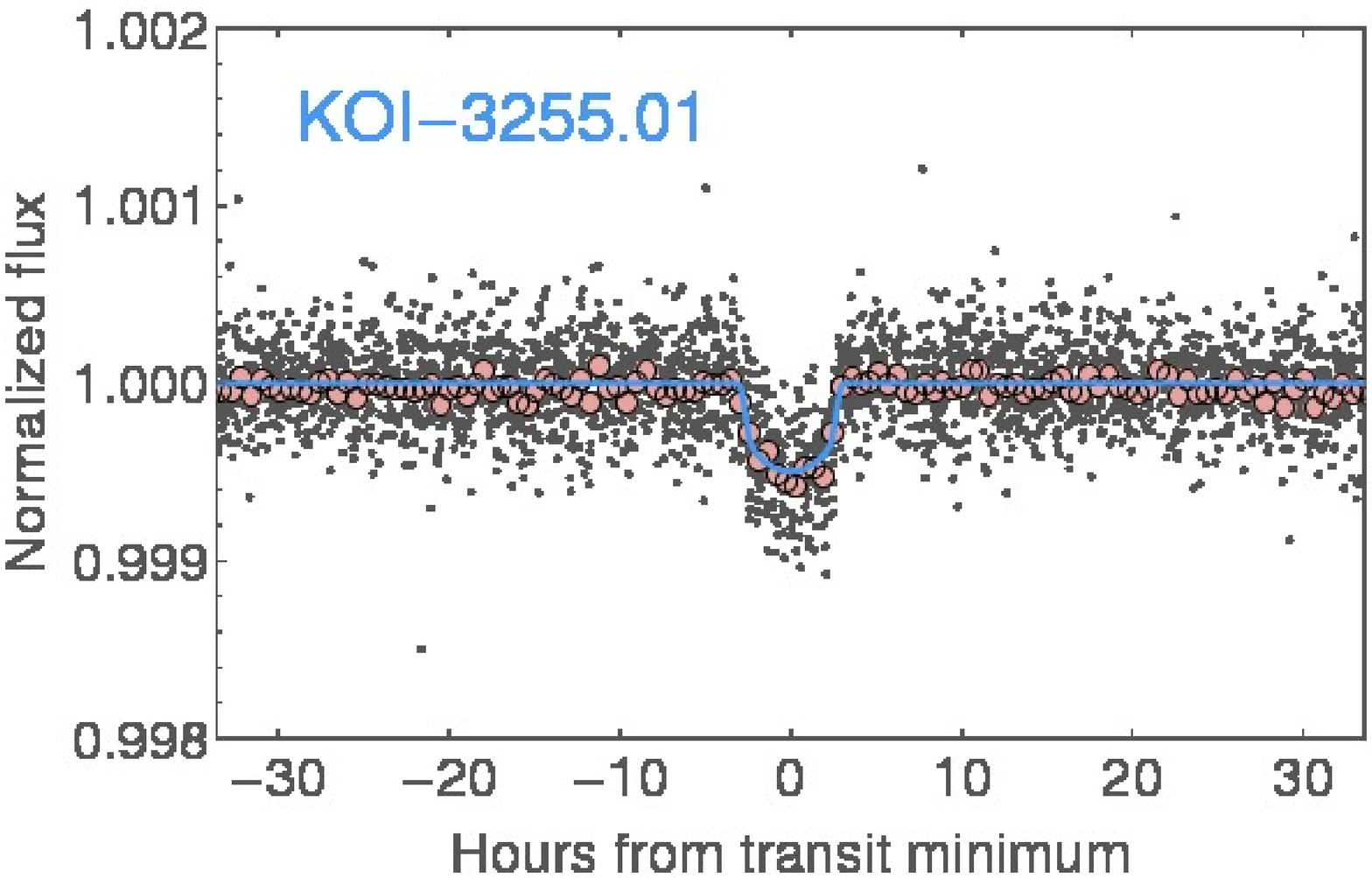} &
\includegraphics[width=5.5cm]{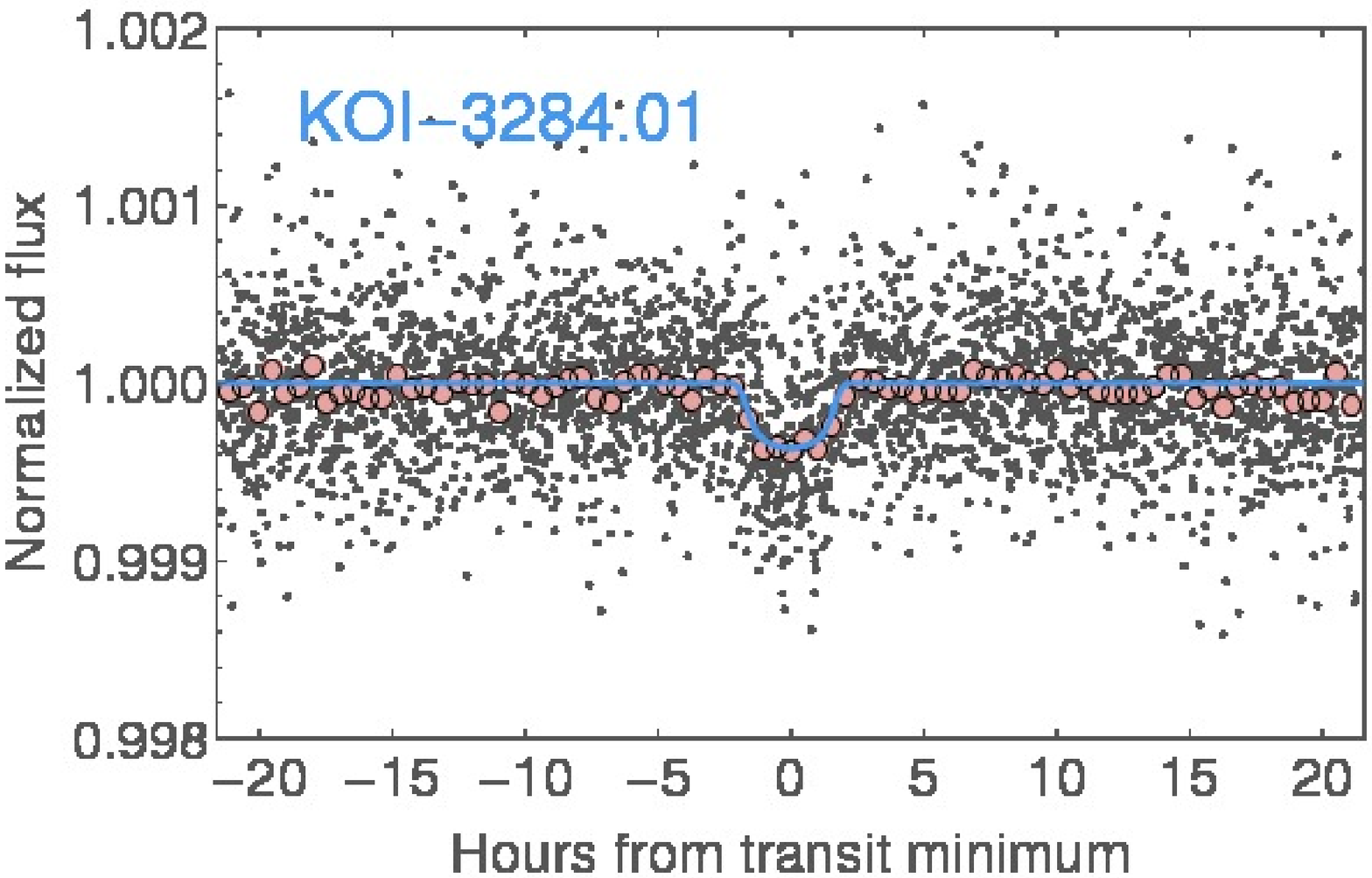}\\[+1ex]
\includegraphics[width=5.5cm]{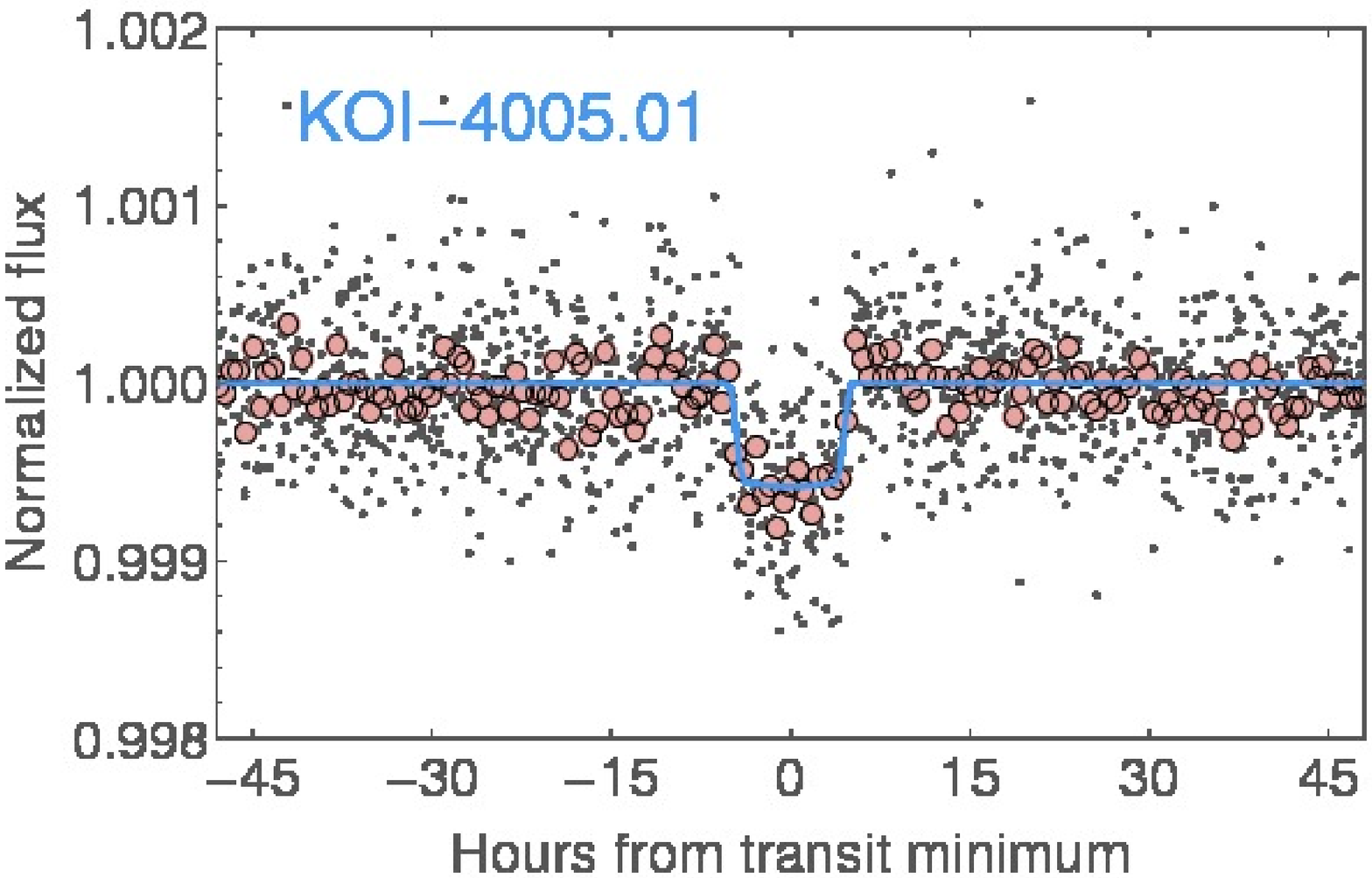} &
\includegraphics[width=5.5cm]{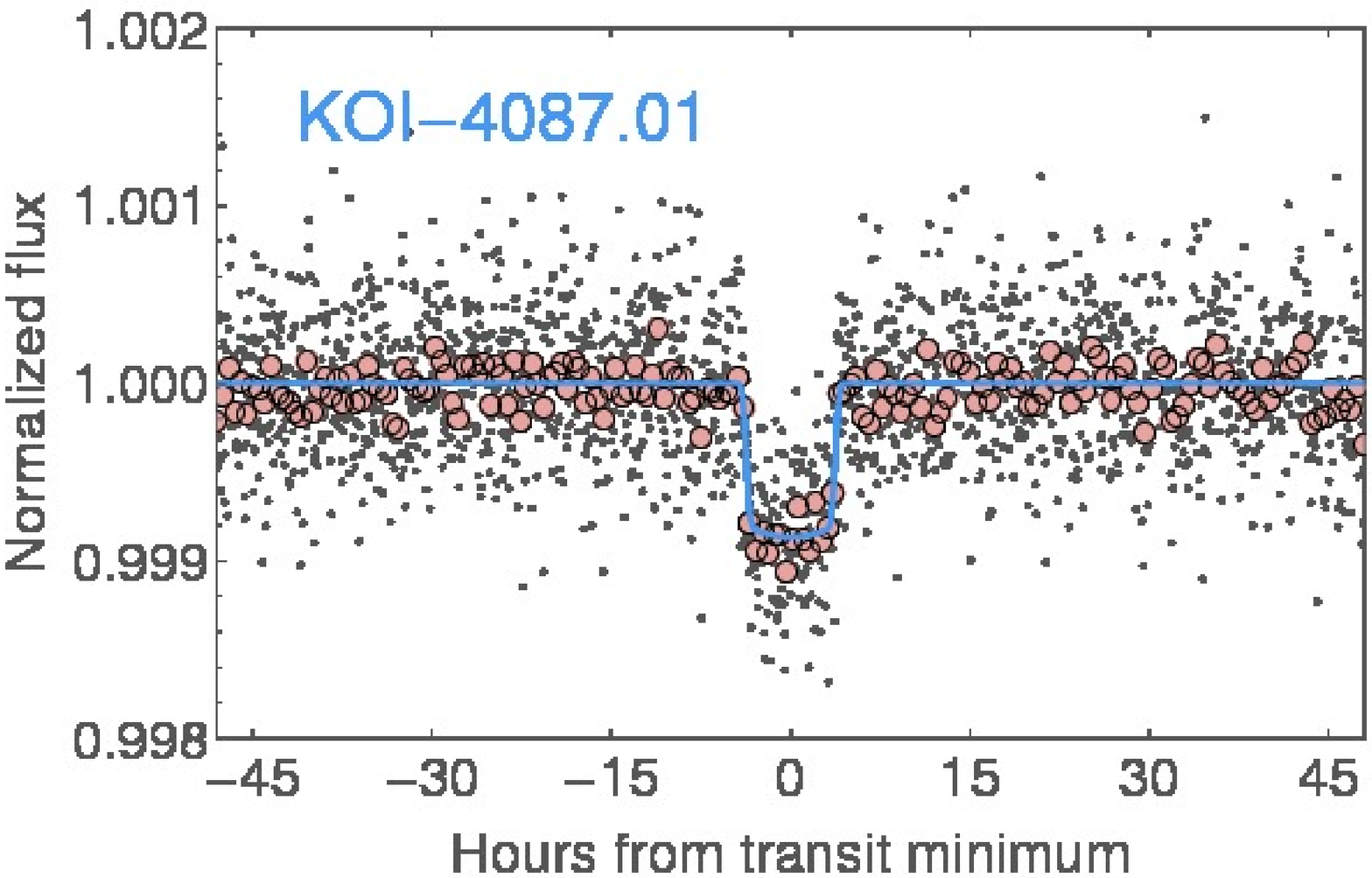} &
\includegraphics[width=5.5cm]{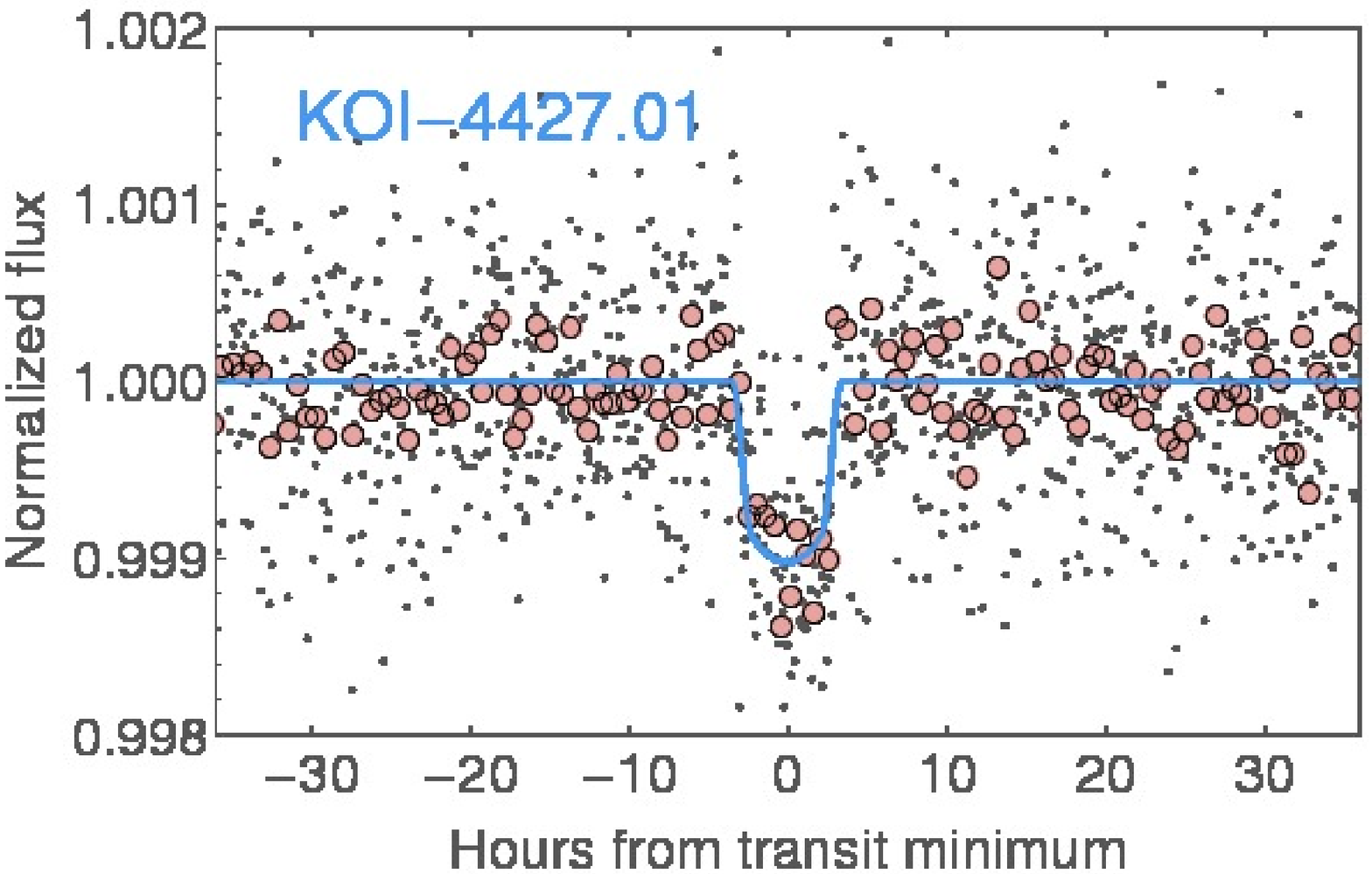}\\[+1ex]
\includegraphics[width=5.5cm]{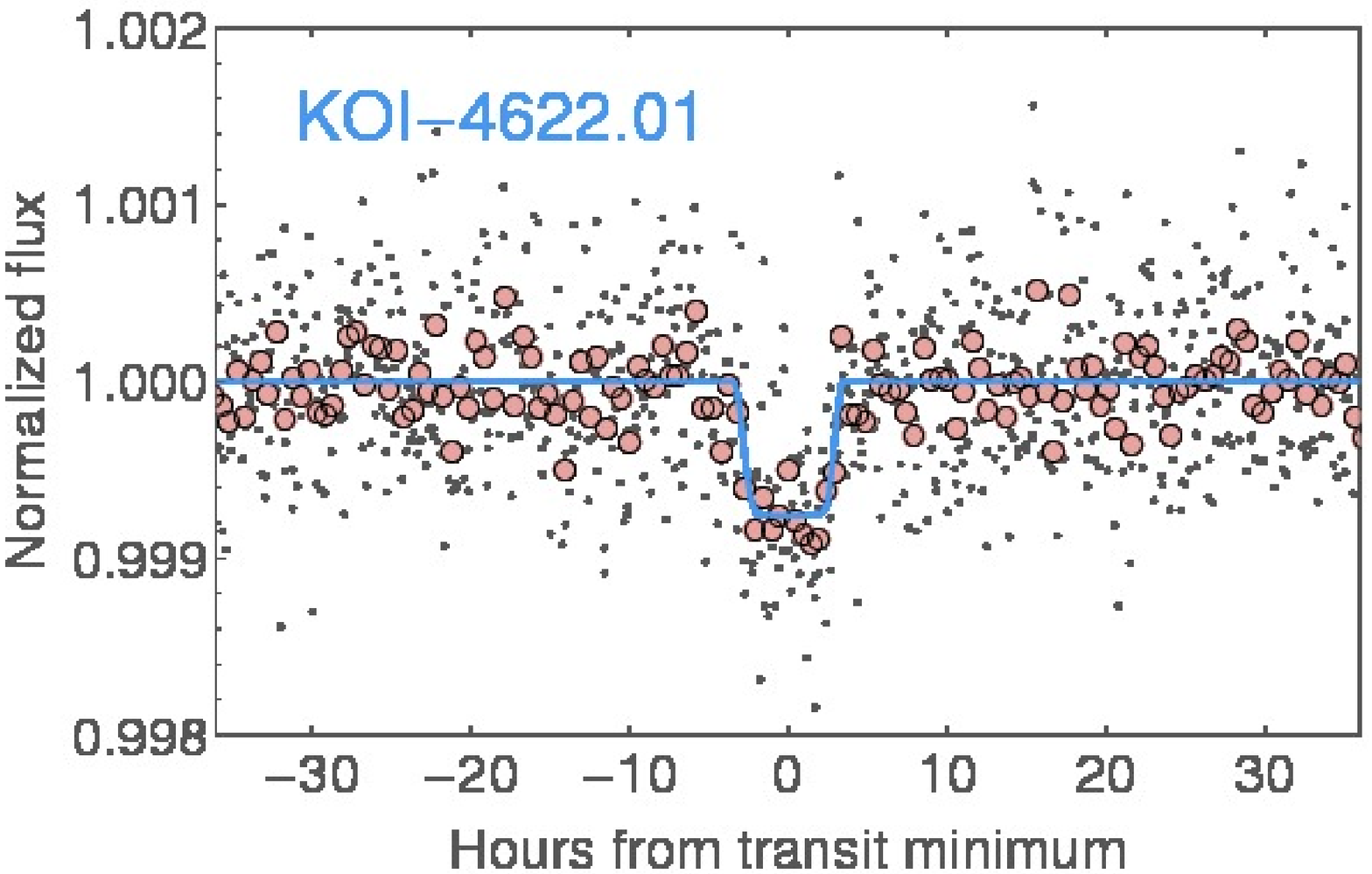} &
\includegraphics[width=5.5cm]{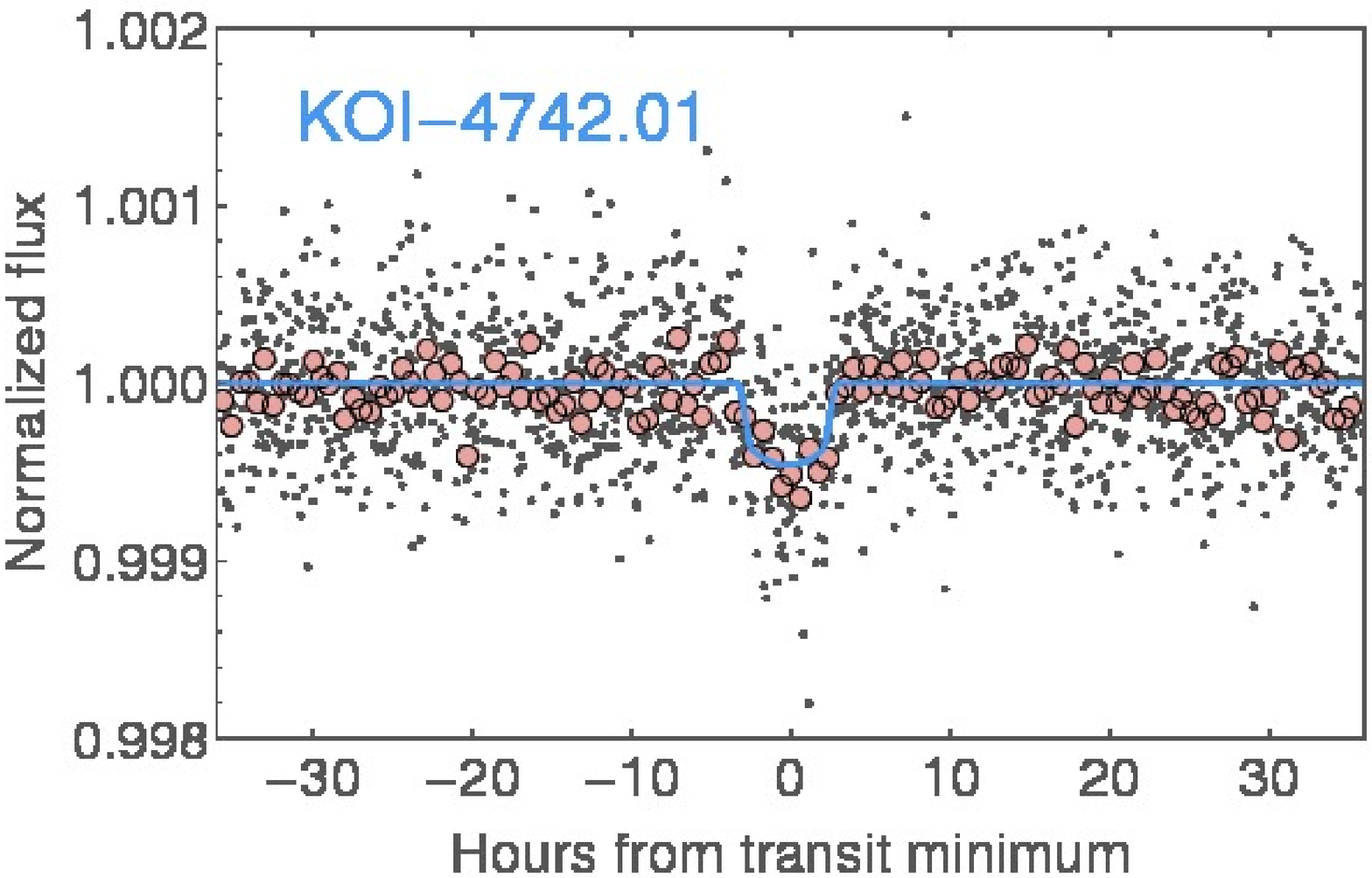} &
\includegraphics[width=5.5cm]{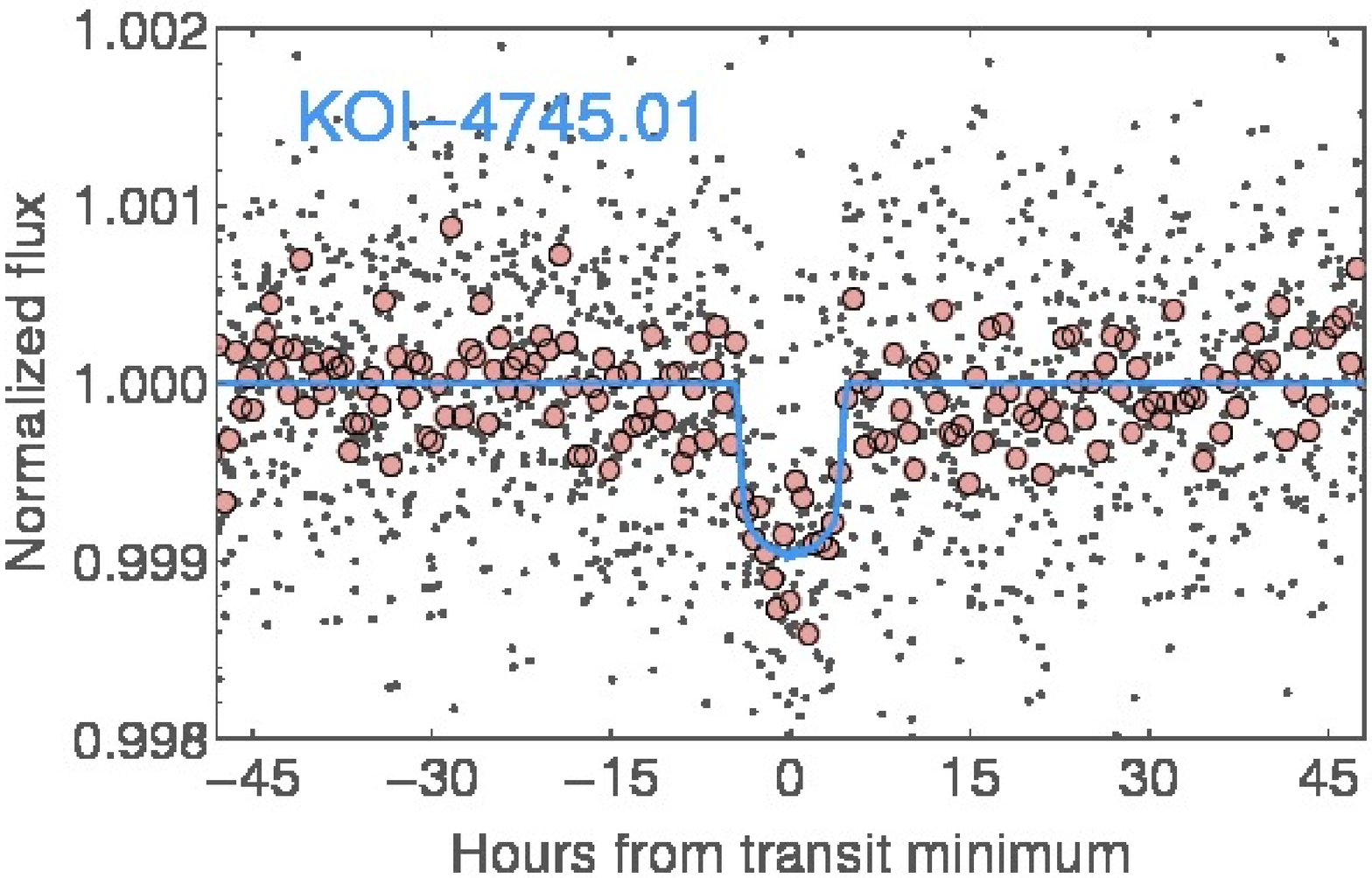}\\[+1ex]
\end{tabular}

\figcaption[]{Light curve fits and \kepler\ observations (small dots)
  for our twelve KOIs. Pink circles represent 10-point binned
  data.\label{fig:lcs}}

\end{figure*}

To regress our 8-parameter model to the observations, we employed the
multi-modal nested sampling algorithm {\sc MultiNest} described by
\citet{Feroz:08} and \cite{Feroz:09}. We used 4000 live points with
constant efficiency mode turned off, and set an enlargement factor of
0.1. The maximum \emph{a posteriori} model parameters and their
associated 68.3\% credible intervals are collected in
Tables~\ref{tab:finalparams1} and \ref{tab:finalparams2}. Also found
there are other derived parameters including the scaled semimajor axis
($a/R_{\star}$), the orbital inclination angle $i$, the transit
durations $T_{14}$ and $T_{23}$ (first-to-fourth and second-to-third
contacts), and the planetary radius and semimajor axis, $R_p$ and
$a_p$. Additional derived quantities of interest are discussed below.
The folded transit light curves are displayed graphically in
Figure~\ref{fig:lcs} along with the models.

The planetary radii for our targets are all under 2.4\,$R_{\earth}$
with the exception of KOI-2529.02, and the semimajor axes range from
about 0.15 to 0.64 AU. We find that in all cases our planetary radii
are larger than those currently held on the NASA Exoplanet
Archive. This may be partly due to the fact that our stellar radii
reported earlier also tend to be larger, and that in four cases we
have included dilution effects from close companions.

\vskip 20pt

[REST OF PAGE IS BLANK]

\clearpage

\begin{turnpage}

\begin{deluxetable}{c c c c c c c c c c c c} 
\tablecaption{
Final parameter estimates for the objects studied in this work.
\label{tab:finalparams1} 
} 
\tablehead{
\colhead{~~~~~~~~~~~~Parameter~~~~~~~~~~~~}
	& \colhead{KOI-0571.05}
	& \colhead{KOI-1422.04}
	& \colhead{KOI-1422.05}
	& \colhead{KOI-2529.02}
	& \colhead{KOI-3255.01}
	& \colhead{KOI-3284.01}
}
\startdata

\multicolumn{7}{c}{Fitted parameters} \\
\noalign{\vskip 2pt}
\hline
\noalign{\vskip 2pt}
$P$ [days]\dotfill   
	& $129.9441_{-0.0012}^{+0.0013}$		
	& $63.33600_{-0.00050}^{+0.00050}$		
	& $34.14204_{-0.00029}^{+0.00025}$		
	& $64.00205_{-0.00053}^{+0.00072}$		
	& $66.65062_{-0.00032}^{+0.00033}$		
	& $35.23319_{-0.00029}^{+0.00025}$		
	\\ [1.0ex]
$\tau$ [BJD$_{\mathrm{UTC}}-2,\!400,\!000$] \dotfill
	& $55789.4940_{-0.0041}^{+0.0036}$		
	& $55692.3051_{-0.0026}^{+0.0026}$		
	& $55686.0154_{-0.0035}^{+0.0030}$		
	& $55477.6307_{-0.0049}^{+0.0042}$		
	& $55670.6876_{-0.0023}^{+0.0023}$		
	& $55673.0307_{-0.0032}^{+0.0044}$		
	\\ [1.0ex]
$(R_p/R_{\star})$ \dotfill
	& $0.0205_{-0.0013}^{+0.0012}$			
	& $0.0362_{-0.0018}^{+0.0022}$ 			
	& $0.0297_{-0.0037}^{+0.0029}$ 			
	& $0.0354_{-0.0035}^{+0.0024}$			
	& $0.0279_{-0.0042}^{+0.0021}$			
	& $0.0200_{-0.0018}^{+0.0020}$			
	\\ [1.0ex]
$\log(\rho_{\star,\rm phot}$\,[g\,cm$^{-3}$])\tablenotemark{a} \dotfill
	& $0.80_{-0.17}^{+0.40}$			
	& $1.08_{-0.15}^{+0.29}$ 			
	& $0.77_{-0.19}^{+0.44}$ 			
	& $0.72_{-0.21}^{+0.36}$			
	& $0.50_{-0.19}^{+0.19}$			
	& $0.78_{-0.28}^{+0.38}$			
	\\ [1.0ex]
$b$ \dotfill
	& $0.13_{-0.13}^{+0.28}$			
	& $0.09_{-0.09}^{+0.26}$			
	& $0.18_{-0.18}^{+0.30}$			
	& $0.09_{-0.09}^{+0.29}$			
	& $0.56_{-0.31}^{+0.30}$			
	& $0.90_{-0.34}^{+0.08}$			
	\\ [1.0ex]
$q_1$ \dotfill 
	& $0.35_{-0.28}^{+0.31}$			
	& $0.03_{-0.03}^{+0.16}$			
	& $0.93_{-0.18}^{+0.07}$			
	& $0.21_{-0.21}^{+0.32}$			
	& $0.36_{-0.36}^{+0.26}$			
	& $0.85_{-0.30}^{+0.15}$			
	\\ [1.0ex]
$q_2$ \dotfill 
	& $0.08_{-0.08}^{+0.27}$			
	& $0.03_{-0.03}^{+0.22}$			
	& $0.97_{-0.15}^{+0.03}$			
	& $0.05_{-0.05}^{+0.30}$			
	& $0.39_{-0.30}^{+0.30}$			
	& $0.54_{-0.34}^{+0.32}$			
	\\ [1.0ex]
$\log \beta$ \dotfill
	& \nodata              					
	& $(-0.532\pm0.089)$\tablenotemark{b} 			
	& $(-0.532\pm0.089)$\tablenotemark{b} 			
	& $(-1.330\pm0.084)$\tablenotemark{b}			
	& $(-0.080\pm0.089)$\tablenotemark{b}	 		
	& $(-0.577\pm0.070)$\tablenotemark{b}			
\\
\noalign{\vskip 2pt}
\hline
\noalign{\vskip 2pt}
\multicolumn{7}{c}{Other transit parameters} \\
\noalign{\vskip 2pt}
\hline
\noalign{\vskip 2pt}    
$(a/R_{\star})$ \dotfill 
	& $178_{-21}^{+65}$				
	& $137_{-15}^{+34}$				
	& $71_{-10}^{+29}$				
	& $104_{-16}^{+34}$				
	& $91_{-12}^{+14}$				
	& $73_{-15}^{+26}$				
	\\ [1.0ex]
$i$\,[$\arcdeg$] \dotfill 
	& $89.96_{-0.10}^{+0.04}$			
	& $89.95_{-0.12}^{+0.05}$			
	& $89.89_{-0.26}^{+0.11}$			
	& $89.93_{-0.18}^{+0.07}$			
	& $89.90_{-0.23}^{+0.10}$			
	& $89.86_{-0.32}^{+0.14}$			
	\\ [1.0ex]
$u_1$ \dotfill
	& $0.70_{-0.38}^{+0.39}$			
	& $0.35_{-0.32}^{+0.31}$			
	& $1.52_{-0.34}^{+0.32}$			
	& $0.67_{-0.48}^{+0.44}$			
	& $0.92_{-0.31}^{+0.35}$			
	& $0.88_{-0.51}^{+0.47}$			
	\\ [1.0ex]
$u_2$ \dotfill 
	& $-0.16_{-0.30}^{+0.31}$			
	& $-0.01_{-0.20}^{+0.35}$			
	& $-0.70_{-0.27}^{+0.23}$			
	& $-0.14_{-0.31}^{+0.38}$			
	& $-0.31_{-0.33}^{+0.25}$			
	& $-0.42_{-0.56}^{+0.32}$			
	\\ [1.0ex]
$T_{14}$\,[hours] \dotfill 
	& $5.62_{-0.24}^{+0.34}$			
	& $3.60_{-0.18}^{+0.19}$			
	& $3.62_{-0.26}^{+0.26}$			
	& $4.67_{-0.34}^{+0.35}$			
	& $5.59_{-0.29}^{+0.21}$			
	& $3.48_{-0.41}^{+0.30}$			
	\\ [1.0ex]
$T_{23}$\,[hours] \dotfill 
	& $5.22_{-0.30}^{+0.25}$			
	& $3.26_{-0.19}^{+0.19}$			
	& $3.34_{-0.26}^{+0.31}$			
	& $4.16_{-0.43}^{+0.33}$			
	& $5.21_{-0.24}^{+0.27}$			
	& $3.24_{-0.39}^{+0.31}$			
	\\
\noalign{\vskip 2pt}
\hline
\noalign{\vskip 2pt}
\multicolumn{7}{c}{Physical parameters} \\
\noalign{\vskip 2pt}
\hline
\noalign{\vskip 2pt}
$R_p$\,[$R_{\oplus}$] \dotfill
	& $1.17_{-0.08}^{+0.08}$			
	& $1.75_{-0.19}^{+0.12}$			
	& $1.48_{-0.25}^{+0.16}$			
	& $2.73_{-0.24}^{+0.23}$			
	& $2.14_{-0.17}^{+0.22}$			
	& $1.12_{-0.17}^{+0.16}$			
	\\ [1.0ex]
$a_p$\,[AU] \dotfill
	& $0.432_{-0.053}^{+0.171}$			
	& $0.283_{-0.041}^{+0.046}$			
	& $0.149_{-0.024}^{+0.035}$			
	& $0.339_{-0.053}^{+0.134}$			
	& $0.288_{-0.040}^{+0.066}$			
	& $0.166_{-0.042}^{+0.051}$			
	\\ [1.0ex]
$\log(\rho_{\star}$\,[g\,cm$^{-3}$])\tablenotemark{c} \dotfill
	& $0.734_{-0.020}^{+0.022}$			
	& $0.915_{-0.064}^{+0.069}$			
	& $0.915_{-0.064}^{+0.069}$			
	& $0.482_{-0.036}^{+0.036}$			
	& $0.504_{-0.035}^{+0.035}$			
	& $0.717_{-0.156}^{+0.079}$			
\\ [1.0ex]
$e_{\mathrm{min}}$ \dotfill
	& $0.04_{-0.04}^{+0.07}$			
	& $0.12_{-0.09}^{+0.10}$			
	& $0.10_{-0.10}^{+0.14}$			
	& $0.19_{-0.11}^{+0.13}$			
	& $0.02_{-0.02}^{+0.08}$			
	& $0.03_{-0.03}^{+0.10}$			
	\\ [1.0ex]
$S_{\mathrm{eff}}$\,[$S_{\oplus}$] \dotfill 
	& $0.30_{-0.15}^{+0.10}$		
	& $0.39_{-0.16}^{+0.10}$		
	& $1.45_{-0.73}^{+0.50}$		
	& $1.69_{-0.79}^{+0.58}$		
	& $2.15_{-0.88}^{+0.74}$		
	& $1.40_{-0.77}^{+0.67}$		
	\\ [1.0ex]
$\mathcal{P}[\mathrm{HZ}]$\,[\%] \dotfill
	& $98.4$					
	& $99.7$					
	& $82.0$					
	& $72.3$					
	& $56.8$					
	& $71.8$					
        \\ [1.0ex]
$\mathcal{P}[\mathrm{rocky}]$\,[\%] \dotfill
	& $68.4$					
	& $30.6$					
	& $50.7$					
	&  $0.6$					
	& $11.7$					
	& $69.6$					
\enddata
\tablenotetext{a}{Mean stellar density derived from the light curve fit.}
\tablenotetext{b}{Gaussian prior.}
\tablenotetext{c}{Mean stellar density derived from our stellar evolution modeling.}
\tablecomments{All quantities correspond to the mode of the posterior distributions.}
\end{deluxetable}

\end{turnpage}

\clearpage

\begin{turnpage}

\begin{deluxetable}{c c c c c c c} 
\tablecaption{
Final parameter estimates for the objects studied in this work.
\label{tab:finalparams2} 
} 
\tablehead{
\colhead{~~~~~~~~~~~~Parameter~~~~~~~~~~~~}
        & \colhead{KOI-4005.01}
        & \colhead{KOI-4087.01}
        & \colhead{KOI-4427.01}
        & \colhead{KOI-4622.01}
        & \colhead{KOI-4742.01}
        & \colhead{KOI-4745.01}
}
\startdata
\multicolumn{7}{c}{Fitted parameters} \\
\noalign{\vskip 2pt}
\hline
\noalign{\vskip 2pt}
$P$ [days]\dotfill
        & $178.1396_{-0.0018}^{+0.0016}$                
	& $101.11141_{-0.00068}^{+0.00087}$		
	& $147.6606_{-0.0014}^{+0.0011}$		
	& $207.2482_{-0.0020}^{+0.0022}$		
	& $112.3053_{-0.0028}^{+0.0024}$		
	& $177.6693_{-0.0030}^{+0.0031}$		
	\\ [1.0ex]
$\tau$ [BJD$_{\mathrm{UTC}}-2,\!400,\!000$] \dotfill
        & $55399.3987_{-0.0054}^{+0.0039}$              
	& $55756.2986_{-0.0027}^{+0.0028}$		
	& $55815.2616_{-0.0042}^{+0.0047}$		
	& $55667.7764_{-0.0065}^{+0.0040}$		
	& $55849.5578_{-0.0056}^{+0.0067}$		
	& $55630.2460_{-0.0077}^{+0.0076}$		
	\\ [1.0ex]
$(R_p/R_{\star})$ \dotfill
        & $0.02392_{-0.00111}^{+0.00099}$               
	& $0.03038_{-0.00098}^{+0.00112}$		
	& $0.0339_{-0.0020}^{+0.0022}$			
	& $0.0280_{-0.0014}^{+0.0017}$			
	& $0.0211_{-0.0016}^{+0.0019}$			
	& $0.0304_{-0.0022}^{+0.0022}$			
	\\ [1.0ex]
$\log(\rho_{\star,\rm phot}$\,[g\,cm$^{-3}$])\tablenotemark{a} \dotfill
        & $0.23_{-0.15}^{+0.16}$                        
	& $0.25_{-0.13}^{+0.13}$			
	& $0.78_{-0.14}^{+0.36}$			
	& $0.89_{-0.16}^{+0.36}$			
	& $0.68_{-0.20}^{+0.56}$			
	& $0.32_{-0.20}^{+0.30}$			
	\\ [1.0ex]
$b$ \dotfill
        & $0.10_{-0.10}^{+0.28}$                        
	& $0.09_{-0.09}^{+0.27}$			
	& $0.08_{-0.08}^{+0.25}$			
	& $0.25_{-0.25}^{+0.30}$			
	& $0.22_{-0.22}^{+0.29}$			
	& $0.26_{-0.26}^{+0.32}$			
	\\ [1.0ex]
$q_1$ \dotfill 
        & $0.05_{-0.05}^{+0.16}$                        
	& $0.03_{-0.03}^{+0.10}$			
	& $0.16_{-0.16}^{+0.26}$			
	& $0.03_{-0.03}^{+0.15}$			
	& $0.92_{-0.31}^{+0.08}$			
	& $0.33_{-0.33}^{+0.32}$			
	\\ [1.0ex]
$q_2$ \dotfill 
        & $0.08_{-0.08}^{+0.24}$                        
	& $0.03_{-0.03}^{+0.20}$			
	& $0.08_{-0.08}^{+0.30}$			
	& $0.03_{-0.03}^{+0.21}$			
	& $0.90_{-0.30}^{+0.10}$			
	& $0.12_{-0.12}^{+0.29}$			
	\\ [1.0ex]
$\log \beta$ \dotfill
        & \nodata                                       
	& \nodata             					
	& $(-0.804\pm0.089)$\tablenotemark{b} 			
	& \nodata             					
	& \nodata                 				
	& \nodata              					
\\
\noalign{\vskip 2pt}
\hline
\noalign{\vskip 2pt}
\multicolumn{7}{c}{Other transit parameters} \\
\noalign{\vskip 2pt}
\hline
\noalign{\vskip 2pt}
$(a/R_{\star})$ \dotfill 
        & $142_{-15}^{+18}$                             
	& $99.0_{-9.4}^{+10.2}$				
	& $190_{-21}^{+62}$				
	& $260_{-31}^{+83}$				
	& $146_{-22}^{+80}$				
	& $151_{-22}^{+39}$				
	\\ [1.0ex]
$i$\,[$\arcdeg$] \dotfill 
        & $89.95_{-0.12}^{+0.05}$                       
	& $89.93_{-0.18}^{+0.07}$			
	& $89.97_{-0.08}^{+0.03}$			
	& $89.97_{-0.07}^{+0.03}$			
	& $89.94_{-0.12}^{+0.06}$			
	& $89.94_{-0.13}^{+0.06}$			
	\\ [1.0ex]
$u_1$ \dotfill
        & $0.39_{-0.34}^{+0.28}$                        
	& $0.34_{-0.23}^{+0.24}$			
	& $0.67_{-0.39}^{+0.41}$			
	& $0.28_{-0.28}^{+0.27}$			
	& $1.03_{-0.49}^{+0.50}$			
	& $0.65_{-0.48}^{+0.41}$			
	\\ [1.0ex]
$u_2$ \dotfill 
        & $-0.08_{-0.23}^{+0.25}$                       
	& $-0.05_{-0.19}^{+0.21}$			
	& $-0.23_{-0.34}^{+0.26}$			
	& $-0.02_{-0.20}^{+0.30}$			
	& $-0.43_{-0.43}^{+0.32}$			
	& $-0.15_{-0.31}^{+0.25}$			
	\\ [1.0ex]
$T_{14}$\,[hours] \dotfill 
        & $9.59_{-0.42}^{+0.30}$                        
	& $7.95_{-0.21}^{+0.25}$			
	& $5.99_{-0.29}^{+0.33}$			
	& $6.05_{-0.32}^{+0.30}$			
	& $5.62_{-0.52}^{+0.40}$			
	& $8.95_{-0.59}^{+0.54}$			
	\\ [1.0ex]
$T_{23}$\,[hours] \dotfill 
        & $9.04_{-0.35}^{+0.36}$                        
	& $7.27_{-0.33}^{+0.23}$			
	& $5.42_{-0.37}^{+0.28}$			
	& $5,61_{-0.32}^{+0.30}$			
	& $5.26_{-0.52}^{+0.40}$			
	& $8.28_{-0.56}^{+0.70}$			
	\\
\noalign{\vskip 2pt}
\hline
\noalign{\vskip 2pt}
\multicolumn{7}{c}{Physical parameters} \\
\noalign{\vskip 2pt}
\hline
\noalign{\vskip 2pt}
$R_p$\,[$R_{\oplus}$] \dotfill
        & $2.24_{-0.45}^{+0.16}$                        
	& $1.86_{-0.19}^{+0.24}$			
	& $1.84_{-0.24}^{+0.22}$			
	& $1.64_{-0.24}^{+0.22}$			
	& $1.34_{-0.18}^{+0.11}$			
	& $2.33_{-0.22}^{+0.19}$			
	\\ [1.0ex]
$a_p$\,[AU] \dotfill
        & $0.563_{-0.080}^{+0.165}$                     
	& $0.242_{-0.041}^{+0.066}$			
	& $0.419_{-0.081}^{+0.073}$			
	& $0.64_{-0.13}^{+0.32}$			
	& $0.409_{-0.060}^{+0.209}$			
	& $0.495_{-0.075}^{+0.186}$			
	\\ [1.0ex]
$\log(\rho_{\star}$\,[g\,cm$^{-3}$])\tablenotemark{c} \dotfill
        & $0.295_{-0.076}^{+0.084}$                     
	& $0.657_{-0.162}^{+0.084}$			
	& $0.751_{-0.095}^{+0.083}$			
	& $0.672_{-0.211}^{+0.091}$			
	& $0.605_{-0.035}^{+0.036}$			
	& $0.477_{-0.035}^{+0.043}$			
\\ [1.0ex]
$e_{\mathrm{min}}$ \dotfill
        & $0.03_{-0.03}^{+0.08}$                        
	& $0.34_{-0.19}^{+0.12}$			
	& $0.02_{-0.02}^{+0.07}$			
	& $0.10_{-0.10}^{+0.11}$			
	& $0.04_{-0.04}^{+0.08}$			
	& $0.11_{-0.11}^{+0.15}$			
	\\ [1.0ex]
$S_{\mathrm{eff}}$\,[$S_{\oplus}$] \dotfill 
        & $1.83_{-0.62}^{+0.51}$                
	& $1.20_{-0.65}^{+0.46}$			
	& $0.233_{-0.110}^{+0.069}$			
	& $0.21_{-0.11}^{+0.11}$			
	& $0.66_{-0.41}^{+0.23}$			
	& $0.86_{-0.37}^{+0.29}$			
	\\ [1.0ex]
$\mathcal{P}[\mathrm{HZ}]$\,[\%] \dotfill
        & $74.1$                                        
	& $89.0$					
	& $71.9$					
	& $48.6$					
	& $96.9$					
	& $89.9$					
	\\ [1.0ex]
$\mathcal{P}[\mathrm{rocky}]$\,[\%] \dotfill
        &  $6.5$                                        
	& $29.8$					
	& $27.3$					
	& $45.0$					
	& $60.7$					
	&  $4.9$					
\enddata
\tablenotetext{a}{Mean stellar density derived from the light curve fit.}
\tablenotetext{b}{Gaussian prior.}
\tablenotetext{c}{Mean stellar density derived from our stellar evolution modeling.}
\tablecomments{All quantities correspond to the mode of the posterior distributions.}
\end{deluxetable}









\end{turnpage}

\clearpage
\global\pdfpageattr\expandafter{\the\pdfpageattr/Rotate 90}

\setlength{\tabcolsep}{3pt}
\setlength{\tabcolsep}{6pt}   

\section{Constraints from asterodensity profiling}
\label{sec:ap}

A comparison between the mean stellar density inferred from our
light-curve fits above (photometric density $\rho_{\star,\rm phot}$)
and some independent measure of the density, such as that derived from
model isochrones, can provide very useful information about a
planetary system. This technique, referred to as Asterodensity
Profiling (AP), was discussed in detail by
\cite{Kipping:12}.\footnote{A similar concept was described by
  \cite{Tingley:11}, although the current implementation is
  considerably more advanced.} For a ``vanilla'' exoplanet (circular
Keplerian orbit, no blends, etc.)  these two density metrics should
agree. Deviations between them occur when one of the following
physical effects is not accounted for in the light curve analysis:
\emph{i)} orbital eccentricity (`photo-eccentric effect'); \emph{ii)}
contaminating light (`photo-blend effect'); \emph{iii)} unocculted
star spots (`photo-spot effect'); \emph{iv)} transit timing variations
(`photo-timing effect'); \emph{v)} transit duration variations
(`photo-duration effect'); \emph{vi)} substantial mass of the
transiting object (`photo-mass effect'); or \emph{vii)} the object
orbits a different star. These effects have been described in detail
by \citet{Kipping:14a}, who provided approximate expressions for each.

In what follows we apply AP to 1) test the hypothesis that all planets
in the multiple transiting planet systems KOI-0571 and KOI-1422 orbit
the same star \citep[Multi-body Asterodensity Profiling, or
  MAP;][]{Kipping:12}; 2) derive minimum eccentricities for the twelve
KOIs in our sample; and 3) evaluate the possibility that the validated
planet orbits the close companion instead of the target, in the four
cases where high-resolution imaging has identified such sources.

\subsection{Using AP to derive average stellar densities for the
multi-planet systems KOI-0571 and KOI-1422}
\label{sec:ap_multi}

Of the twelve candidates studied in this work, three reside in systems
with five transiting planets each: one orbits KOI-0571, and two are
associated with KOI-1422. These rare systems provide extra information
that we will exploit here. For each planetary candidate in each system
we performed an independent detrending and subsequent light curve fit
using the methods described earlier.

If all five planets in each of these two systems orbit the same star,
then we expect all of the planets to maintain low orbital
eccentricities in order for these compact systems to be dynamically
stable over Gyr timescales \citep[see also][]{Rowe:14}.  Therefore,
under the assumption that the planets orbit the same star, one would
expect negligible photo-eccentric effects, implying that the mean
stellar densities derived from the light curves should be consistent.
On the same assumption that all five planets orbit the same star, two
other AP effects listed above can also be neglected: the photo-blend
effect, and effects from the planets orbiting different stars. The
latter scenario is obviously not relevant under the posed assumption,
and the former can be negated since, to first order, contaminating
light that is unaccounted for disturbs the light-curve-derived stellar
density for all five planets to the same degree.

\begin{figure}
\epsscale{1.10}
\plotone{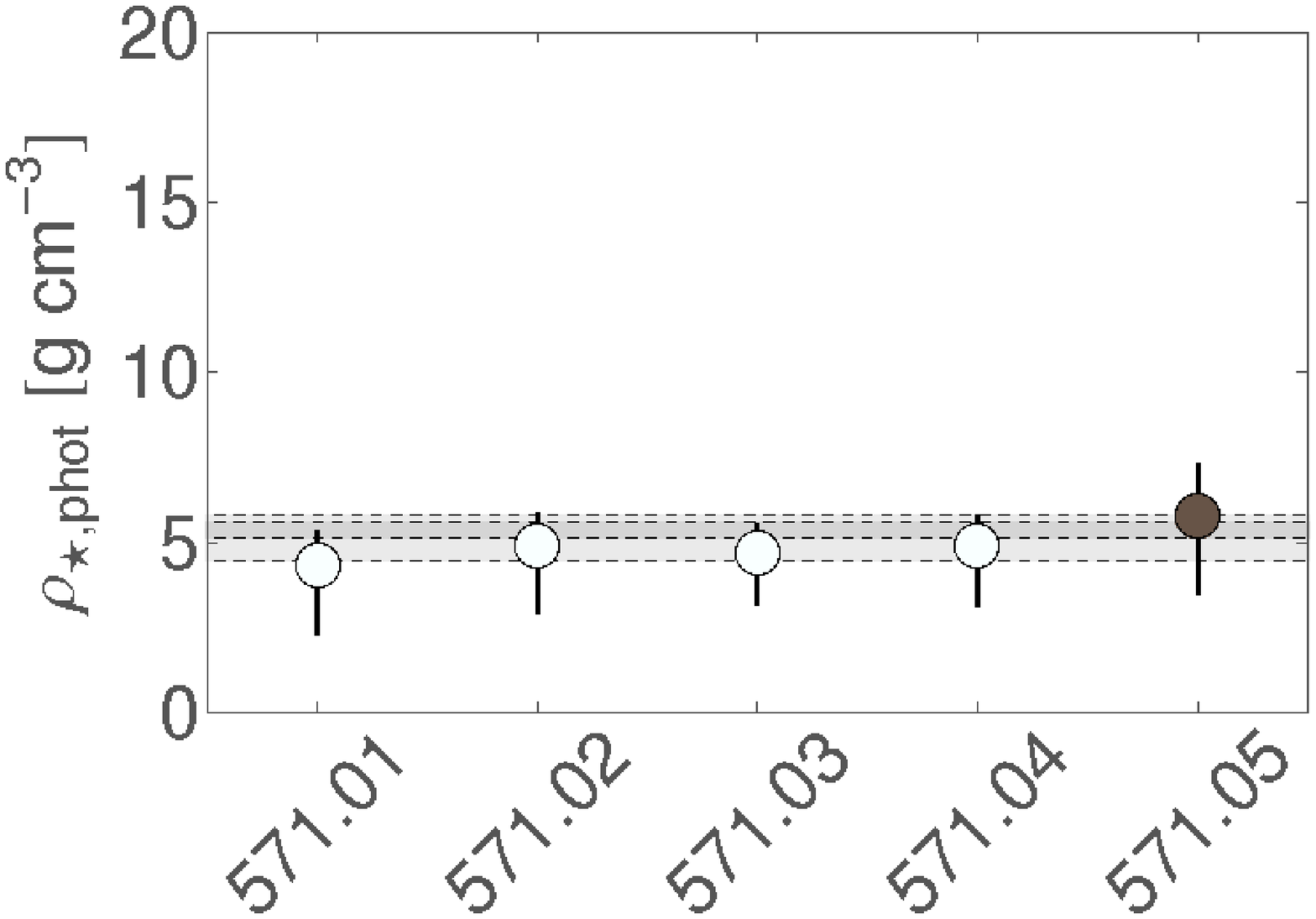}
\vskip 3pt
\epsscale{1.15}
\plotone{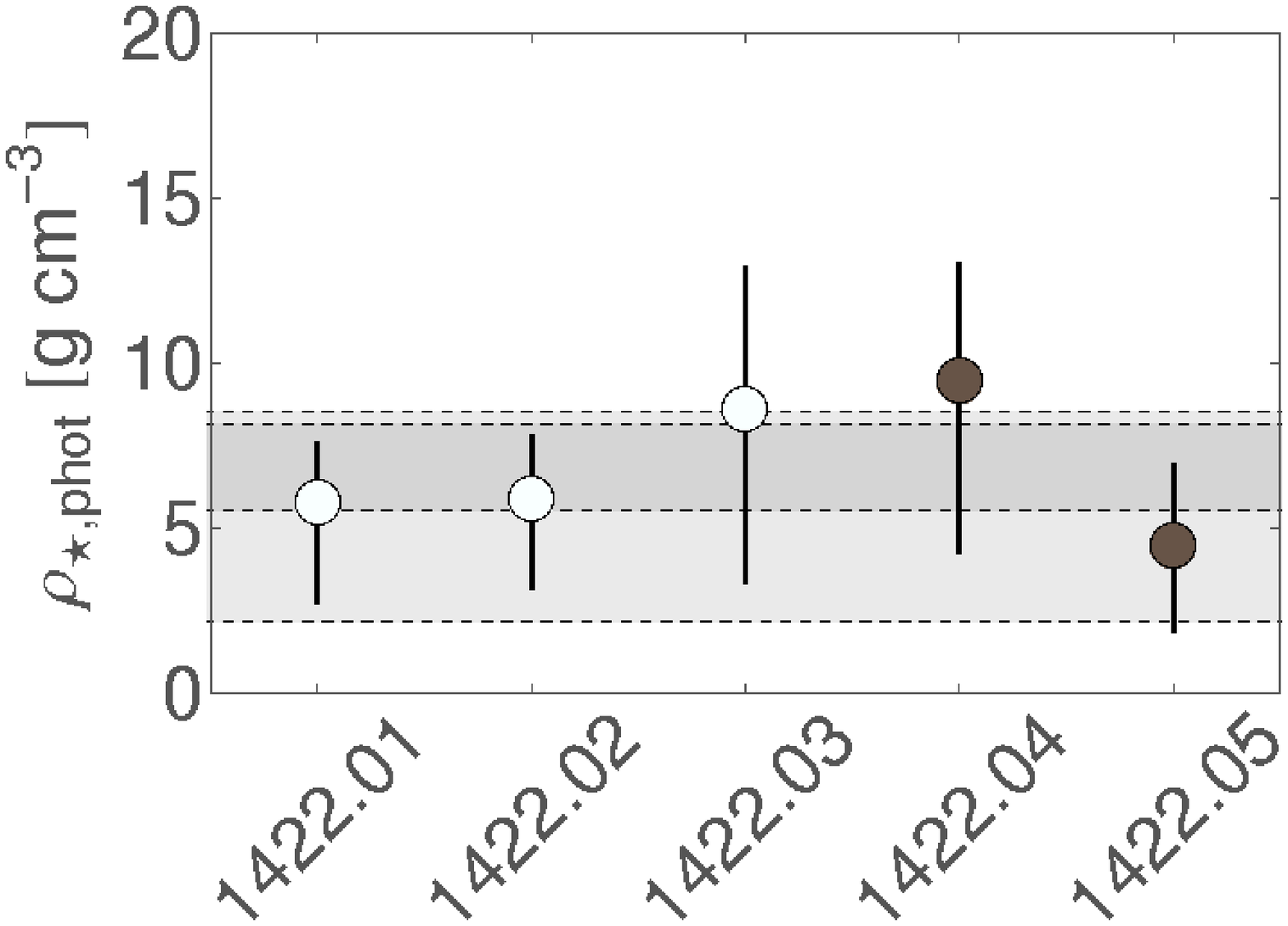}

\figcaption[]{Mean stellar densities from separate light curve fits to
  each of the planets in the five-planet systems KOI-0571 and 1422.
  The dashed lines and gray areas represent the 1$\sigma$ and
  2$\sigma$ confidence regions of the average density computed from
  the other companions in each system (open circles) excluding the
  ones we have validated here (filled circles). In both cases the
  agreement between this mean density and the separately determined
  one of the validated planets is excellent, supporting the notion
  that in each system all five planets orbit the same
  star.\label{fig:APcomps}}

\end{figure}

A comparison of the photometric stellar densities for the five planets
around KOI-0571 and KOI-1422 reveals excellent agreement in both cases,
as shown in Figure~\ref{fig:APcomps}. Furthermore, similar diagrams
(not shown) comparing the limb-darkening parameters $u_1$ and $u_2$
also show very good agreement among all planets in both systems.  We
conclude from this that each system is consistent with having five
planets orbiting the same star, given the available data.

Averaging over the four other planets in KOI-0571, and the three
others in KOI-1422, we may derive a mean photometric density for the
host stars that we denote $\rho_{\star,\rm MAP}$.  We obtained
$\rho_{\star,\rm MAP} = 5.41_{-0.24}^{+0.32}$\,g\,cm$^{-3}$ and
$7.2_{-1.2}^{+2.9}$\,g\,cm$^{-3}$, respectively, where the extra
dilution from the close companion to KOI-1422 has been properly
accounted for, as before. The 1$\sigma$ and 2$\sigma$ confidence
intervals of these values are marked in Figure~\ref{fig:APcomps} with
dashed lines. Both of these mean densities are consistent with the
corresponding photometric densities derived from the light curves of
KOI-0571.05, KOI-1422.04, and KOI-1422.05 individually, which are
$\rho_{\star,\rm{phot}} = 5.9_{-2.5}^{+1.4}$\,g\,cm$^{-3}$,
$11.4_{-4.0}^{+7.5}$\,g\,cm$^{-3}$, and
$5.3_{-2.4}^{+2.7}$\,g\,cm$^{-3}$, respectively.

The average stellar densities for these multi-planet systems (computed
from the remaining planets in each case, excluding the validated
planets) were used as luminosity indicators to strengthen the
determination of the host star parameters via our stellar evolution
modeling in Sect.~\ref{sec:stellarproperties} \citep{Sozzetti:07}.
This is particularly useful for these three KOIs as neither of the
host stars has a spectroscopic determination of $\log g$ available to
otherwise constrain the luminosity (see Table~\ref{tab:specmatch}).

\subsection{Using AP to measure minimum eccentricities}

A measure of the stellar density that is independent of that derived
from the individual light curves of the validated planets is available
for each of the twelve KOIs studied in this work (see
Sect.~\ref{sec:stellarproperties}).  For KOI-0571 and KOI-1422, this
comes from the MAP-based density ($\rho_{\star,\rm MAP}$) refined by
our isochrone analysis by making use of additional constraints on age
from gyrochronology, along with the spectroscopic temperature and
metallicity estimates. For KOI-3284, 4087, 4427, and 4622 our stellar
evolution modeling used only the age along with $T_{\rm eff}$ and
[Fe/H], and for the remaining targets that are earlier in spectral
type we used $\log g$ as the luminosity indicator, which for these
stars we were able to determine spectroscopically using {\tt
SpecMatch}.

These independent stellar densities, denoted here simply by
$\rho_{\star}$, may be compared with the light-curve-derived stellar
density of each validated planet, $\rho_{\star,\rm phot}$. In these
cases we expect the photo-eccentric effect to be the dominant AP
effect since contamination has been accounted for in the fits
(photo-blend effect), the stars show relatively low activity
(photo-spot effect), there are no known timing effects for these
objects (photo-timing and photo-duration effects), and their small
sizes suggest they should be of low mass (photo-mass effect). We
therefore attribute any differences between $\rho_{\star,\rm{phot}}$
and $\rho_{\star}$ as being due to orbital eccentricity. The minimum
eccentricity for each KOI is directly given by this comparison via the
expression presented by \citet{Kipping:14a},
\begin{align*}
e_{\mathrm{min}} = \frac{ | 1 - (\rho_{\star,\rm phot}/\rho_{\star})^{2/3} |}
{ 1 + (\rho_{\star,\rm phot}/\rho_{\star})^{2/3} }~.
\label{eqn:emin}
\end{align*}
The results of this calculation, shown in
Tables~\ref{tab:finalparams1} and \ref{tab:finalparams2}, reveal that
most of the KOIs in the sample are consistent with orbiting their parent
stars on low-eccentricity orbits.

\subsection{Using AP to test blend scenarios}
\label{sec:APblends}

For the four KOIs found to have close companions from our
high-resolution imaging (KOI-1422, 2529, 3255, and 3284) the same
techniques from the preceding section enable us to revisit the
possibility that the planets orbit the companions rather than the
primary stars. As done earlier, we made the reasonable assumption that
these close neighbors are bound to the targets. This allows us to
reuse the target isochrones to infer approximate stellar properties
for the companions, including their mean densities, on the basis of
their brightness differences compared to the target.  We re-fitted the
transits adopting revised blend factors ($\beta$) appropriate for the
different host stars to derive photometric densities for the
companions, $\rho_{\star,\rm phot}$. We then used these densities to
estimate a minimum eccentricity for the scenarios in which the planets
orbit the companions rather than the brighter stars.

\begin{deluxetable}{l c c }
\tablecaption{AP-derived minimum eccentricities and other properties for KOIs with close
  companions.\label{tab:APblends}}
\tablehead{
\colhead{~~~~~~~~~Parameter~~~~~~~~~} &
\colhead{Primary} &
\colhead{Companion} 
}
\startdata

\multicolumn{3}{c}{KOI-1422.04} \\
\noalign{\vskip 1pt}
\hline
\noalign{\vskip 2pt}
$K\!p$ [mag]\dotfill & $16.15\pm0.03$ & $17.72\pm0.15$ \\[0.5ex]     
$\log \beta$\dotfill & $(-0.628\pm0.061)$ & $(0.628\pm0.061)$ \\[0.5ex]     
$\log (\rho_{\star}$\,[g\,cm$^{-3}$])\dotfill & $0.915_{-0.064}^{+0.069}$ & $1.342_{-0.036}^{+0.037}$ \\[0.5ex]     
$\log (\rho_{\star,\rm phot}$\,[g\,cm$^{-3}$])\dotfill & $1.08_{-0.15}^{+0.29}$ & $1.08_{-0.13}^{+0.27}$ \\[0.5ex]     
$\log(\rho_{\star,\rm phot}/\rho_{\star})$\dotfill & $0.16_{-0.17}^{+0.37}$ & $-0.27_{-0.14}^{+0.34}$ \\[0.5ex]     
$e_{\mathrm{min}}$\dotfill & $0.12_{-0.09}^{+0.10}$ & $0.20_{-0.20}^{+0.10}$ \\[0.5ex]     
$R_p$\,[$R_{\oplus}$]\dotfill & $1.76_{-0.14}^{+0.16}$ & $2.01_{-0.18}^{+0.18}$ \\[0.5ex]     
$S_{\mathrm{eff}}$\,[$S_{\oplus}$]\dotfill & $0.38_{-0.16}^{+0.11}$ & $0.28_{-0.10}^{+0.06}$ \\     
\noalign{\vskip 2pt}
\hline
\noalign{\vskip 2pt}

\multicolumn{3}{c}{KOI-1422.05} \\
\noalign{\vskip 1pt}
\hline
\noalign{\vskip 2pt}
$K\!p$ [mag]\dotfill & $16.15\pm0.03$ & $17.72\pm0.15$ \\[0.5ex]     
$\log \beta$\dotfill & $(-0.628\pm0.061)$ & $(0.628\pm0.061)$ \\[0.5ex]     
$\log (\rho_{\star}$\,[g\,cm$^{-3}$])\dotfill & $0.915_{-0.064}^{+0.069}$ & $1.342_{-0.036}^{+0.037}$ \\[0.5ex]     
$\log (\rho_{\star,\rm phot}$\,[g\,cm$^{-3}$])\dotfill & $0.77_{-0.19}^{+0.44}$ & $0.77_{-0.28}^{+0.43}$ \\[0.5ex]     
$\log(\rho_{\star,\rm phot}/\rho_{\star})$\dotfill & $-0.14_{-0.21}^{+0.58}$ & $-0.56_{-0.27}^{+0.42}$ \\[0.5ex]     
$e_{\mathrm{min}}$\dotfill & $0.10_{-0.10}^{+0.14}$ & $0.79_{-0.22}^{+0.05}$ \\[0.5ex]     
$R_p$\,[$R_{\oplus}$]\dotfill & $1.42_{-0.18}^{+0.19}$ & $1.68_{-0.53}^{+0.27}$ \\[0.5ex]     
$S_{\mathrm{eff}}$\,[$S_{\oplus}$]\dotfill & $1.35_{-0.66}^{+0.49}$ & $1.01_{-0.47}^{+0.50}$ \\     
\noalign{\vskip 2pt}
\hline
\noalign{\vskip 2pt}

\multicolumn{3}{c}{KOI-2529.02} \\
\noalign{\vskip 1pt}
\hline
\noalign{\vskip 2pt}
$K\!p$ [mag]\dotfill & $15.86\pm0.10$ & $22.00\pm0.20$  \\[0.5ex]     
$\log \beta$\dotfill & $(-1.330\pm0.084)$ & $(2.474\pm0.089)$ \\[0.5ex]    
$\log (\rho_{\star}$\,[g\,cm$^{-3}$])\dotfill & $0.482_{-0.036}^{+0.036}$ & $1.866_{-0.039}^{+0.046}$ \\[0.5ex]     
$\log (\rho_{\star,\rm phot}$\,[g\,cm$^{-3}$])\dotfill & $0.72_{-0.21}^{+0.36}$ & $0.88_{-0.13}^{+0.14}$ \\[0.5ex]     
$\log(\rho_{\star,\rm phot}/\rho_{\star})$\dotfill & $0.24_{-0.21}^{+0.40}$ & $-1.01_{-0.15}^{+0.13}$ \\[0.5ex]     
$e_{\mathrm{min}}$\dotfill & $0.19_{-0.11}^{+0.13}$ & $0.65_{-0.06}^{+0.06}$ \\[0.5ex]     
$R_p$\,[$R_{\oplus}$]\dotfill & $2.73_{-0.24}^{+0.23}$ & $8.2_{-2.8}^{+2.1}$ \\[0.5ex]     
$S_{\mathrm{eff}}$\,[$S_{\oplus}$]\dotfill & $1.69_{-0.79}^{+0.58}$ & $0.30_{-0.08}^{+0.06}$ \\     
\noalign{\vskip 2pt}
\hline
\noalign{\vskip 2pt}

\multicolumn{3}{c}{KOI-3255.01} \\
\noalign{\vskip 1pt}
\hline
\noalign{\vskip 2pt}
$K\!p$ [mag]\dotfill & $15.01\pm0.10$ & $15.21\pm0.20$ \\[0.5ex]     
$\log \beta$\dotfill & $(-0.080\pm0.089)$ & $(0.080\pm0.089)$ \\[0.5ex]     
$\log (\rho_{\star}$\,[g\,cm$^{-3}$])\dotfill & $0.504_{-0.035}^{+0.035}$ & $0.524_{-0.010}^{+0.009}$ \\[0.5ex]     
$\log (\rho_{\star,\rm phot}$\,[g\,cm$^{-3}$])\dotfill & $0.50_{-0.19}^{+0.19}$ & $0.51_{-0.18}^{+0.16}$ \\[0.5ex]     
$\log(\rho_{\star,\rm phot}/\rho_{\star})$\dotfill & $0.01_{-0.18}^{+0.22}$ & $0.00_{-0.16}^{+0.17}$ \\[0.5ex]     
$e_{\mathrm{min}}$\dotfill & $0.02_{-0.02}^{+0.08}$ & $0.02_{-0.02}^{+0.08}$ \\[0.5ex]     
$R_p$\,[$R_{\oplus}$]\dotfill & $2.14_{-0.17}^{+0.22}$ & $2.25_{-0.20}^{+0.20}$ \\[0.5ex]     
$S_{\mathrm{eff}}$\,[$S_{\oplus}$]\dotfill & $2.15_{-0.88}^{+0.74}$ & $1.91_{-0.49}^{+0.57}$ \\     
\noalign{\vskip 2pt}
\hline
\noalign{\vskip 2pt}

\multicolumn{3}{c}{KOI-3284.01} \\
\noalign{\vskip 1pt}
\hline
\noalign{\vskip 2pt}
$K\!p$ [mag]\dotfill & $14.58\pm0.10$ & $17.01\pm0.20$ \\[0.5ex]     
$\log \beta$\dotfill & $(-0.577\pm0.070)$ & $(1.036\pm0.088)$ \\[0.5ex]     
$\log (\rho_{\star}$\,[g\,cm$^{-3}$])\dotfill & $0.717_{-0.156}^{+0.079}$ & $1.321_{-0.011}^{+0.020}$ \\[0.5ex]     
$\log (\rho_{\star,\rm phot}$\,[g\,cm$^{-3}$])\dotfill & $0.78_{-0.28}^{+0.38}$ & $0.77_{-0.28}^{+0.43}$ \\[0.5ex]     
$\log(\rho_{\star,\rm phot}/\rho_{\star})$\dotfill & $-0.03_{-0.35}^{+0.31}$ & $-0.54_{-0.27}^{+0.42}$ \\[0.5ex]     
$e_{\mathrm{min}}$\dotfill & $0.03_{-0.03}^{+0.10}$ & $0.40_{-0.31}^{+0.16}$ \\[0.5ex]     
$R_p$\,[$R_{\oplus}$]\dotfill & $1.12_{-0.17}^{+0.16}$ & $1.77_{-0.21}^{+0.23}$ \\[0.5ex]     
$S_{\mathrm{eff}}$\,[$S_{\oplus}$]\dotfill & $1.40_{-0.77}^{+0.67}$ & $0.86_{-0.39}^{+0.39}$ 
\enddata

\tablecomments{The $\log \beta$ values indicated in parentheses were
  used as Gaussian priors. For KOI-2529.02 and KOI-3284.01 the $\beta$
  factors for the companions are not the reciprocal of the values for
  the primaries because of the presence of additional companions in
  the aperture (see Table~\ref{tab:ao}). Those additional stars are
  ruled out as the source of the transit signals by our centroid
  analysis of Sect.~\ref{sec:centroids}, but still cause dilution. The
  symbol $\rho_{\star}$ represents the mean stellar density based on
  our isochrone analysis (and other constraints), while
  $\rho_{\star,\rm phot}$ is the photometric density from the
  individual light curve fit for the planet.}

\end{deluxetable}

We report the results of this calculation in Table~\ref{tab:APblends},
where we list also the planetary radii that we infer if the planets
transited the companions, as well as their effective insolation level
(see next section).  Although the eccentricity distribution of small
planets around M dwarf stars is presently unknown \citep{Kipping:14b},
in general we expect circular orbits to be more likely than eccentric
orbits, based on evidence from other sub-populations of planets
\citep[see, e.g.,][]{Kipping:13a}. We may thus use $e_{\rm min}$ as a
rough criterion to decide which location for the planet is favored.
Additionally, the sign of the quantity $\log(\rho_{\star,\rm
  phot}/\rho_{\star})$ conveys useful information.  From geometric
arguments, a planet in an eccentric orbit has a higher probability of
transiting near periapsis than apoapsis \citep[e.g.,][]{Barnes:07}.
For the more probable case of $0 < \omega < \pi$ (``near'' periapsis
transits), one expects $\log(\rho_{\star,\rm phot}/\rho_{\star}) >
0$. In contrast, for the less probable scenario of $\pi < \omega <
2\pi$ (``near'' apoapsis transits), one expects $\log(\rho_{\star,\rm
  phot}/\rho_{\star}) < 0$ \citep[see][]{Kipping:14a}.  Therefore, in
general the photo-eccentric effect is more likely to (but will not
necessarily) produce a positive sign for this logarithmic
ratio. Additionally, a diluted transit in a circular orbit will cause
$\log(\rho_{\star,\rm phot}/ \rho_{\star})$ to be negative via the
photo-blend effect \citep{Kipping:14a}, but can never yield a positive
sign. For these reasons, finding $\log(\rho_{\star,\rm
  phot}/\rho_{\star})>0$ is more consistent with a bona-fide planet
than a negative sign.

As seen in Table~\ref{tab:APblends}, despite the large uncertainties
the minimum eccentricities are considerably higher if KOI-1422.05,
KOI-2529.02 and KOI-3284.01 orbit the secondary star, which seems
\emph{a priori} less likely, although quantifying an accurate odds
ratio is not possible without knowing the eccentricity distribution of
terrestrial-sized planets around M dwarfs. For the first two cases
this assessment agrees with our conclusion from the \blender\ study
(Table~\ref{tab:frequencies}), which also favored the planet
transiting the primary.  For KOI-2529.02 and KOI-3284.01 the sign of
the log-density ratio $\log(\rho_{\star,\rm phot}/\rho_{\star})$ is
also negative if the planet is around the companion, requiring either
geometrically disfavored near-apoapsis transiting planets, or simply
the photo-blend effect. For KOI-1422.04 the minimum eccentricities
between the two scenarios are comparable, but the sign of the
log-density ratio favors the primary star as the host. Finally, the
test for KOI-3255.01 is ambiguous according to both metrics, in
agreement with the indications from \blender\ that were also
inconclusive for this planet.

\section{Habitability}
\label{sec:habitability}

While a broadband transit light curve alone does not reveal the
``habitability'' of a given planet, it does allow us to determine how
close to the ``habitable zone'' a particular planet is. Therefore, an
important distinction to make is that we do not claim any of the
planets in our sample are truly habitable, but merely that the
insolation they receive from the host star is suitable for the
presence of liquid water on the surface across the range of plausible
atmospheric conditions.

\citet{Kopparapu:13} have presented calculations of the habitable zone
for a wide variety of atmospheric conditions, and considered also
empirical boundaries for the hot inner edge modeled by a
recent-Venus-like world, and a cold outer edge modeled by an
early-Mars-like world. More recently \cite{Zsom:13} took a broader
view of the conditions that might provide habitable surface conditions
closer to the host star, and postulated an inner boundary for the HZ
that can be as close as 0.38\,AU for a solar-type star if the relative
humidity of the planetary atmosphere is low and the albedo high. Given
the complexity of the problem and our still limited state of
knowledge, in this work we have chosen to adopt the broadest possible
limits that can plausibly lead to suitable conditions for life, in
order to avoid prematurely dismissing as uninteresting planets that
may yet be habitable.  Specifically, we adopted the empirical cold
outer boundary for the HZ of \citet{Kopparapu:13} modeled by an
early-Mars-like world, and the inner edge for dry desert worlds from
\cite{Zsom:13}.  These boundaries depend upon two key input
parameters: $S_{\rm eff}$, the effective insolation received by the
planet, and the spectral type of the star, for which the usual proxy
is $T_{\rm eff}$, the star's effective temperature. We define $S_{\rm
  eff}$ in terms of the Earth's insolation, $S_{\earth}$, as
\begin{align*}
{S_{\rm eff}\over S_{\earth}} = {L_{\star}/L_{\sun}\over (a_p/{\rm AU})^2 \sqrt{1-e^2}}~,
\end{align*}
where $L_{\star}$ is the stellar luminosity and $a_p$ the planetary
semimajor axis in astronomical units.  For each planet in our sample
we computed a joint posterior for $S_{\rm eff}$ and $T_{\rm eff}$
based on the stellar properties inferred in
Sect.~\ref{sec:stellarproperties}, assuming the planet's orbit is
circular. At each realization in this posterior we evaluated whether
the sample falls inside or outside the adopted HZ boundaries.  Counting
the total number of cases inside allows us to assign a statistical
confidence that the planet in question is in the HZ, denoted
$\mathcal{P}[{\rm HZ}]$. We provide those results in
Tables~\ref{tab:finalparams1} and \ref{tab:finalparams2}. This
statistical approach to the habitable zone was first described by
\citet{Kippingetal:13} for the planet Kepler-22\,b.

\begin{figure}
\epsscale{1.20}
\plotone{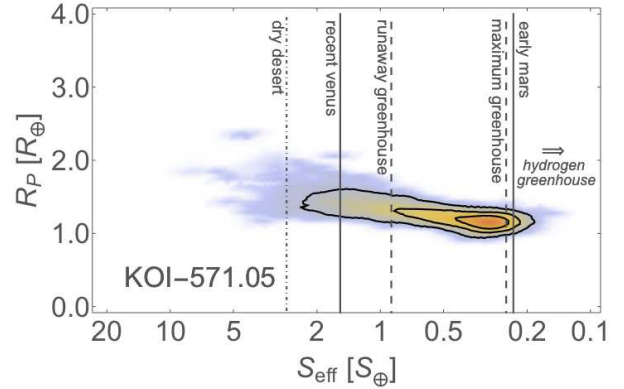}

\figcaption[]{Joint posterior distribution of the planetary radius
  ($y$ axis) and effective insolation ($x$ axis) for KOI-0571.05
  (Kepler-186\,f).  Labeled vertical lines mark the various boundaries
  of the habitable zone as defined by \cite{Kopparapu:13} and
  \cite{Zsom:13}. Here we adopt the optimistic `dry desert' and 'early
  Mars' inner and outer edges, respectively. These lines are drawn
  assuming an effective stellar temperature equal to the mode of the
  associated posteriors. Contours enclose 68.3\%, 95.4\% and 99.7\% of
  the Monte Carlo samples (corresponding to the 1$\sigma$, 2$\sigma$,
  and 3$\sigma$ levels).\label{fig:hab571}}

\end{figure}

\begin{figure*}
\centering
\begin{tabular}{ccc}
\includegraphics[width=5.9cm]{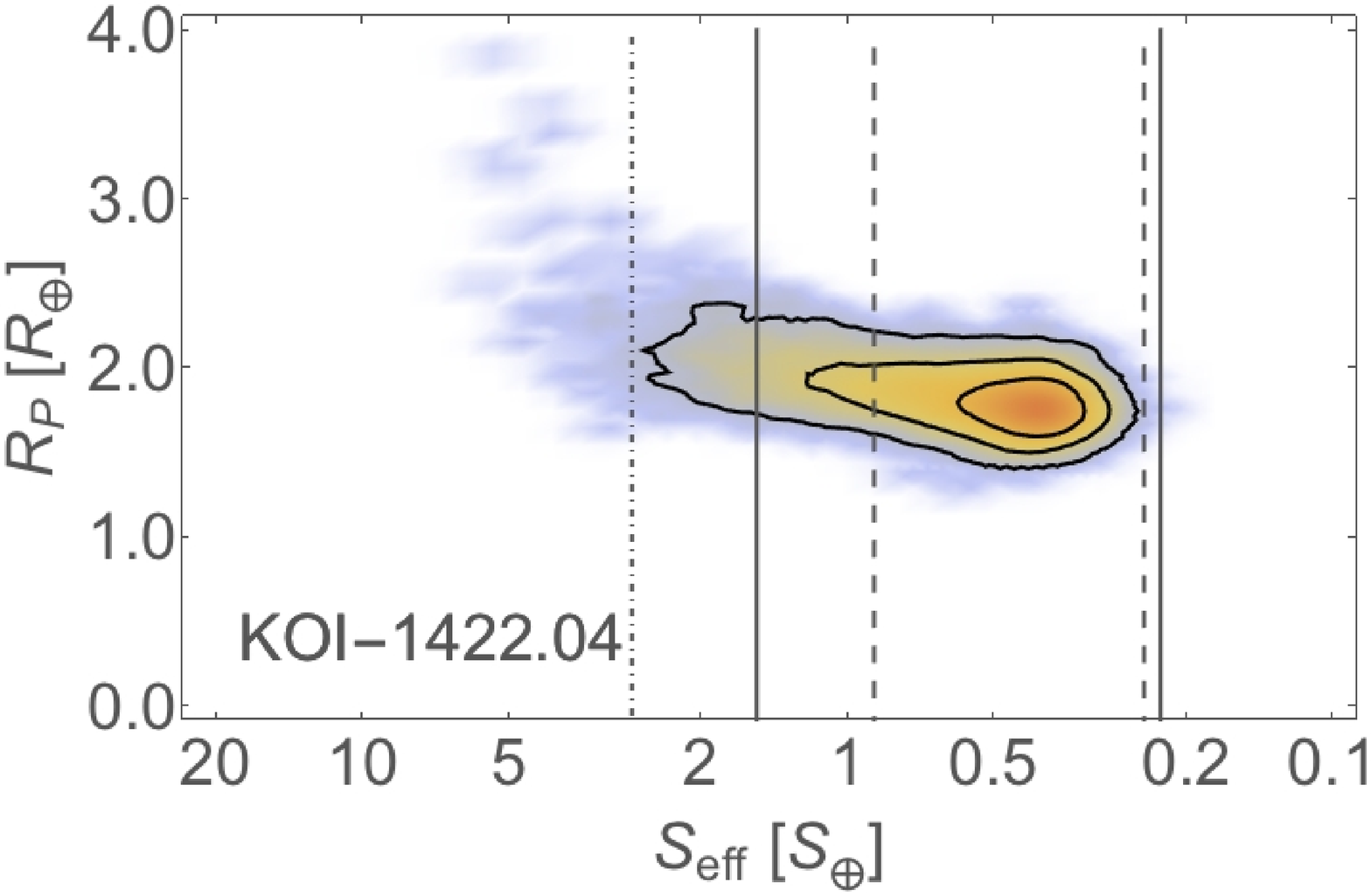} &
\includegraphics[width=5.9cm]{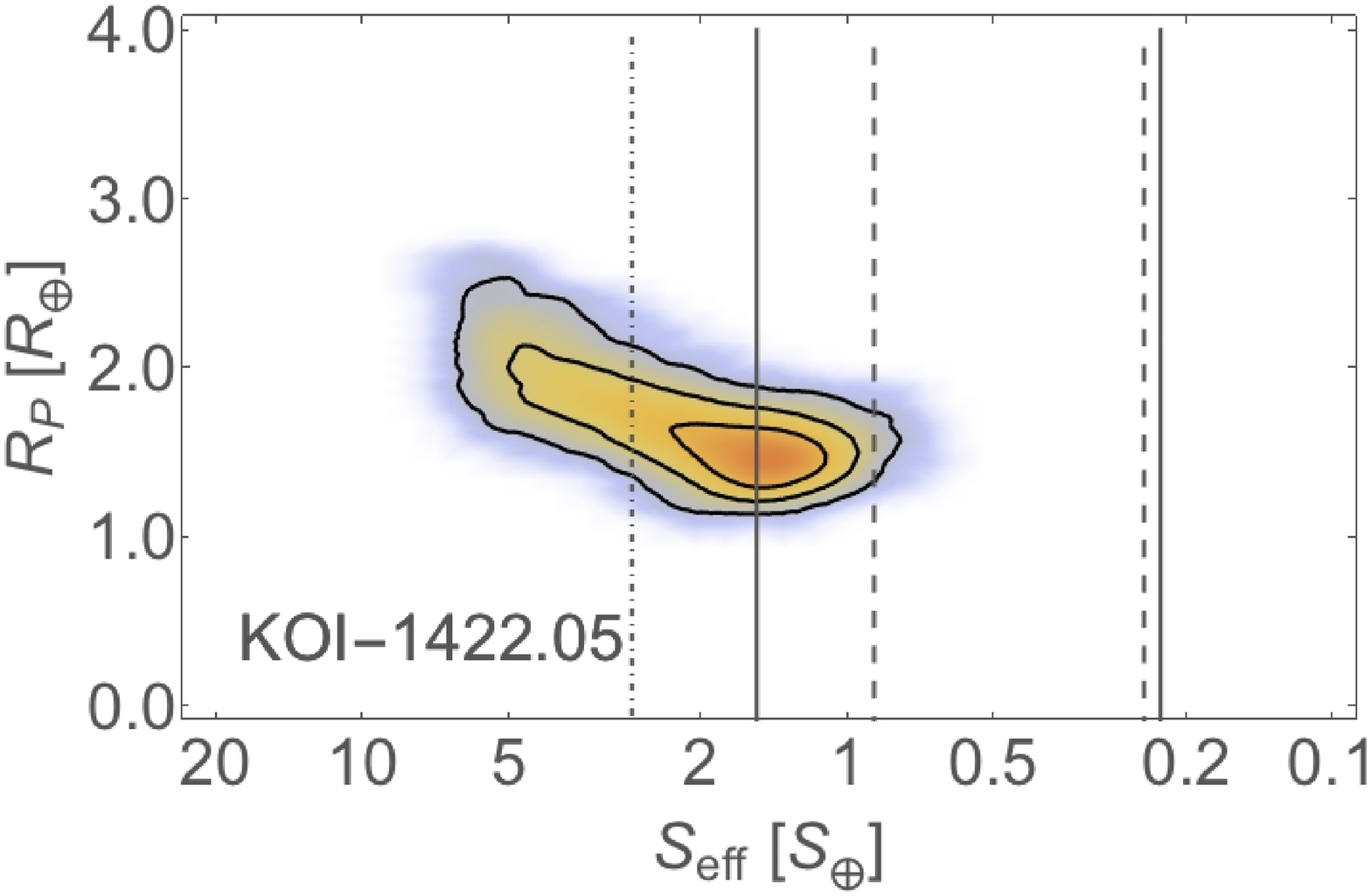} &
\includegraphics[width=5.9cm]{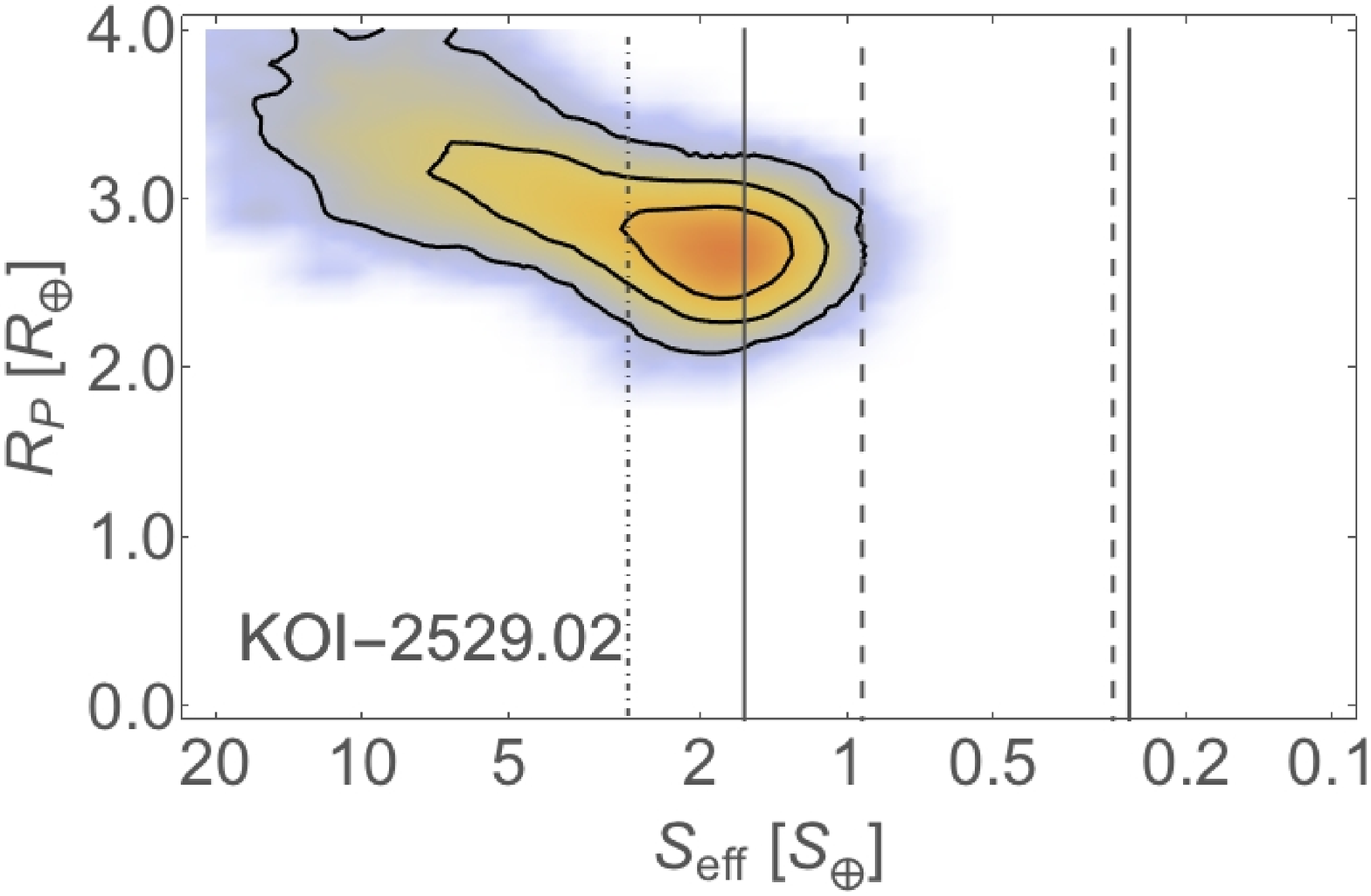}\\[+1ex]
\includegraphics[width=5.9cm]{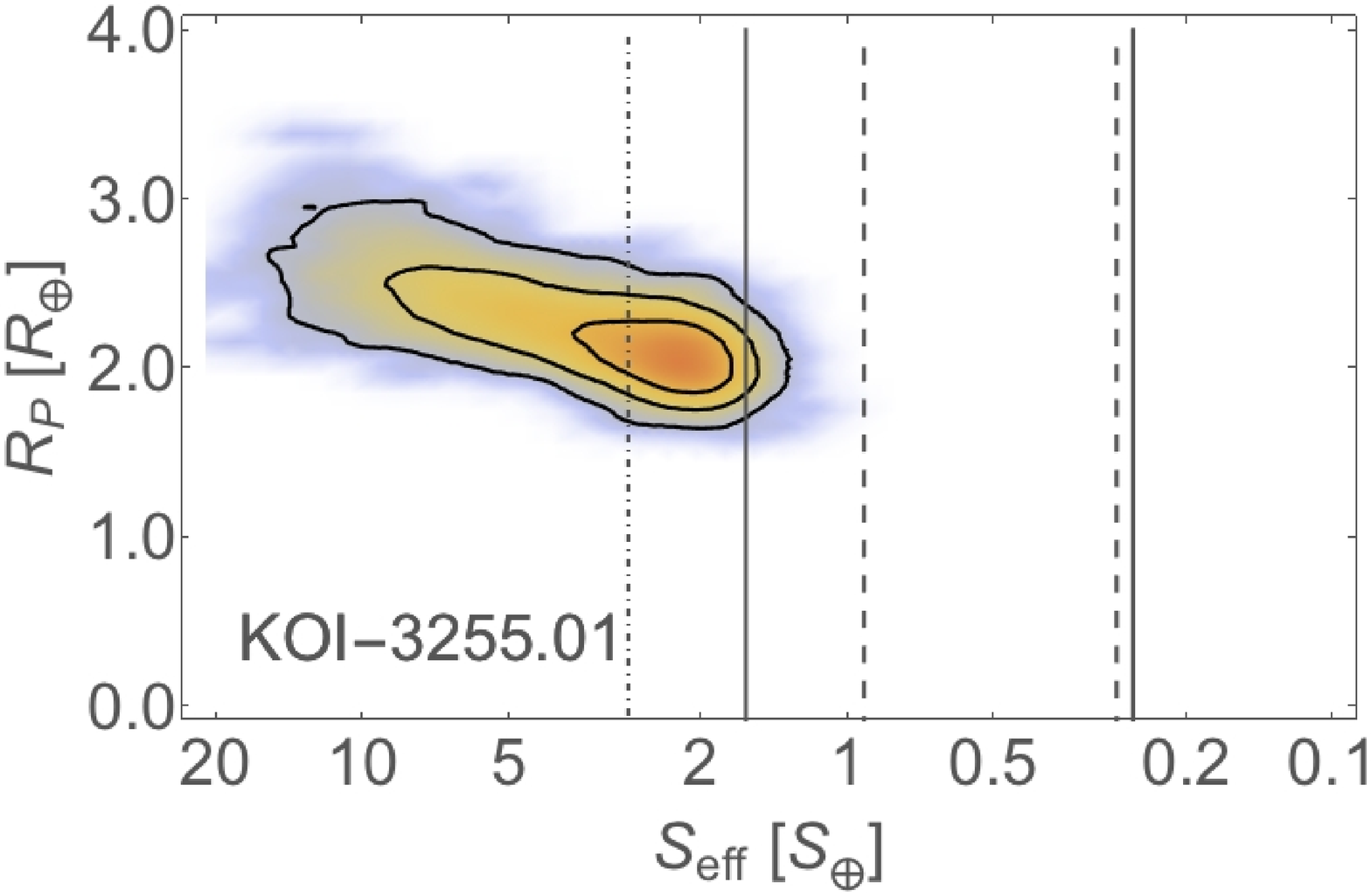} &
\includegraphics[width=5.9cm]{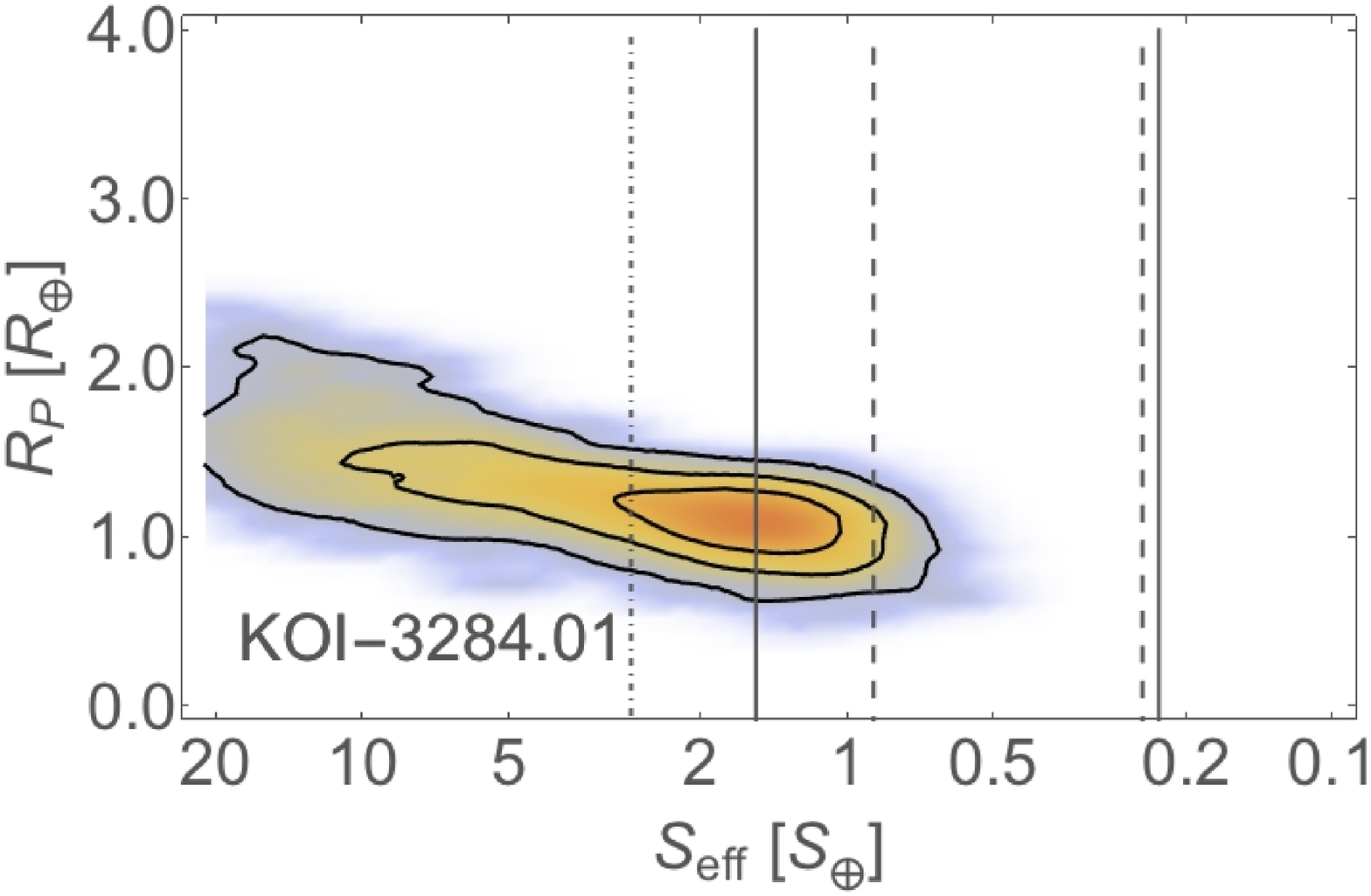} &
\includegraphics[width=5.9cm]{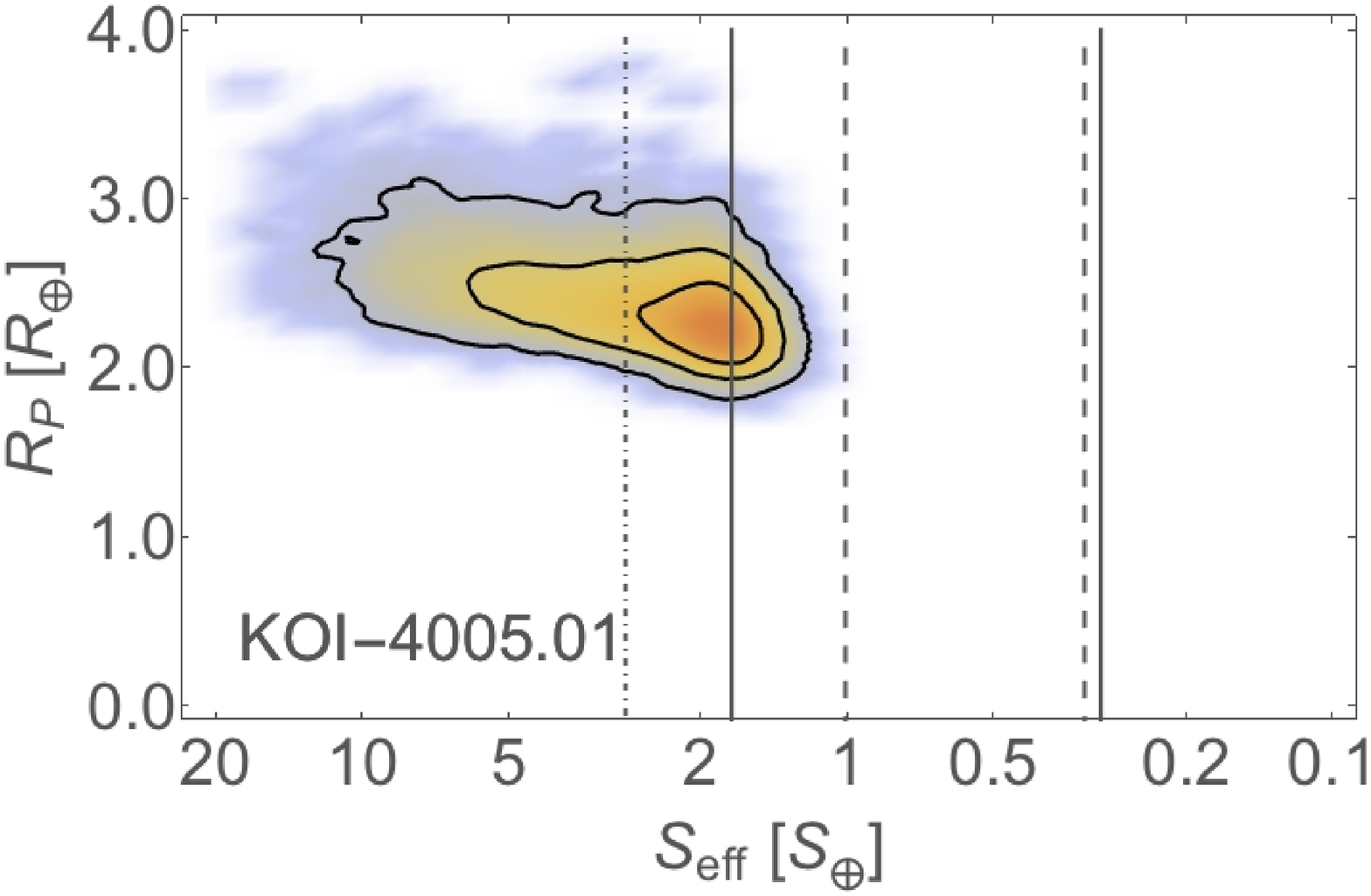}\\[+1ex]
\includegraphics[width=5.9cm]{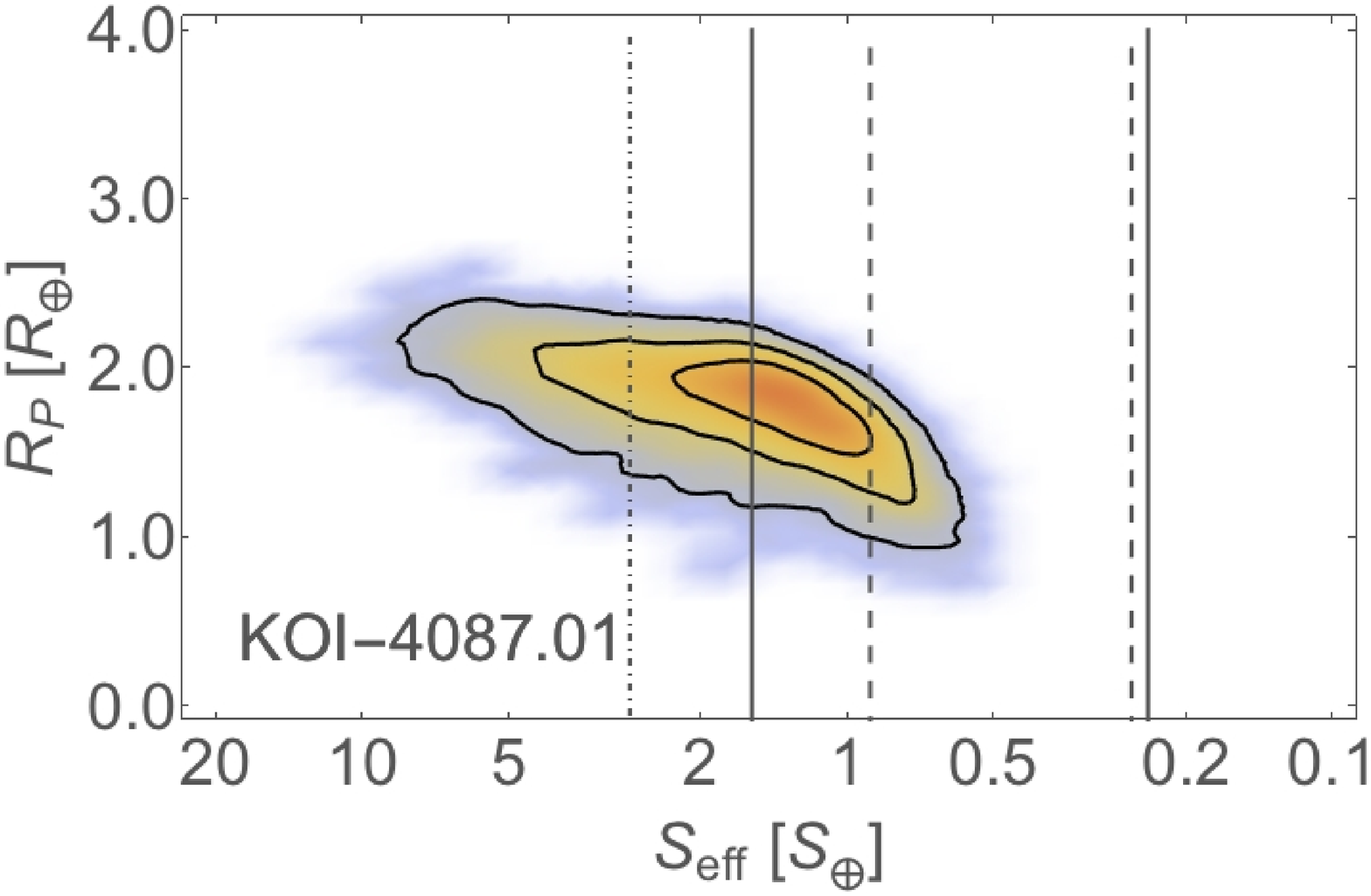} &
\includegraphics[width=5.9cm]{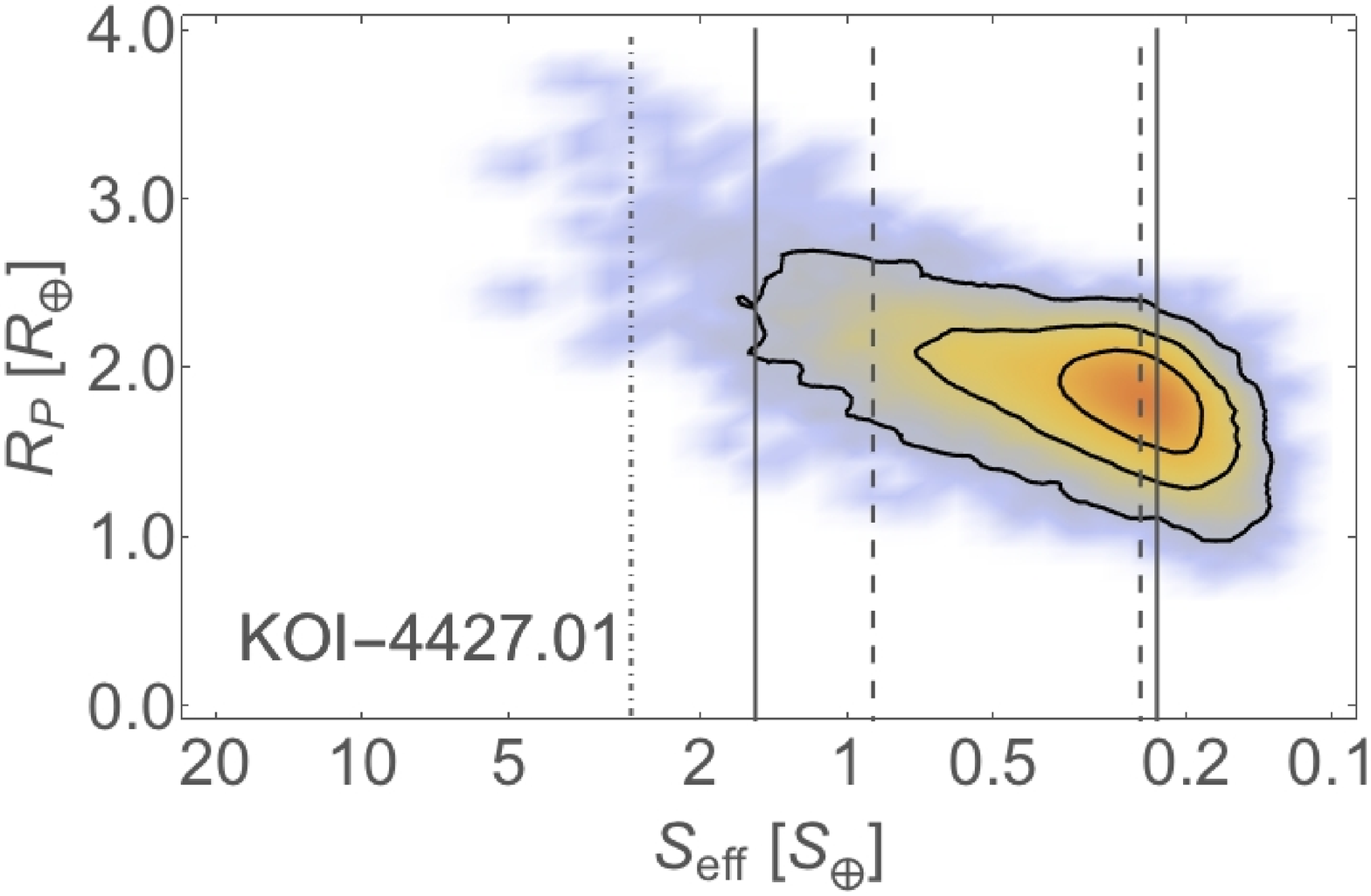} &
\includegraphics[width=5.9cm]{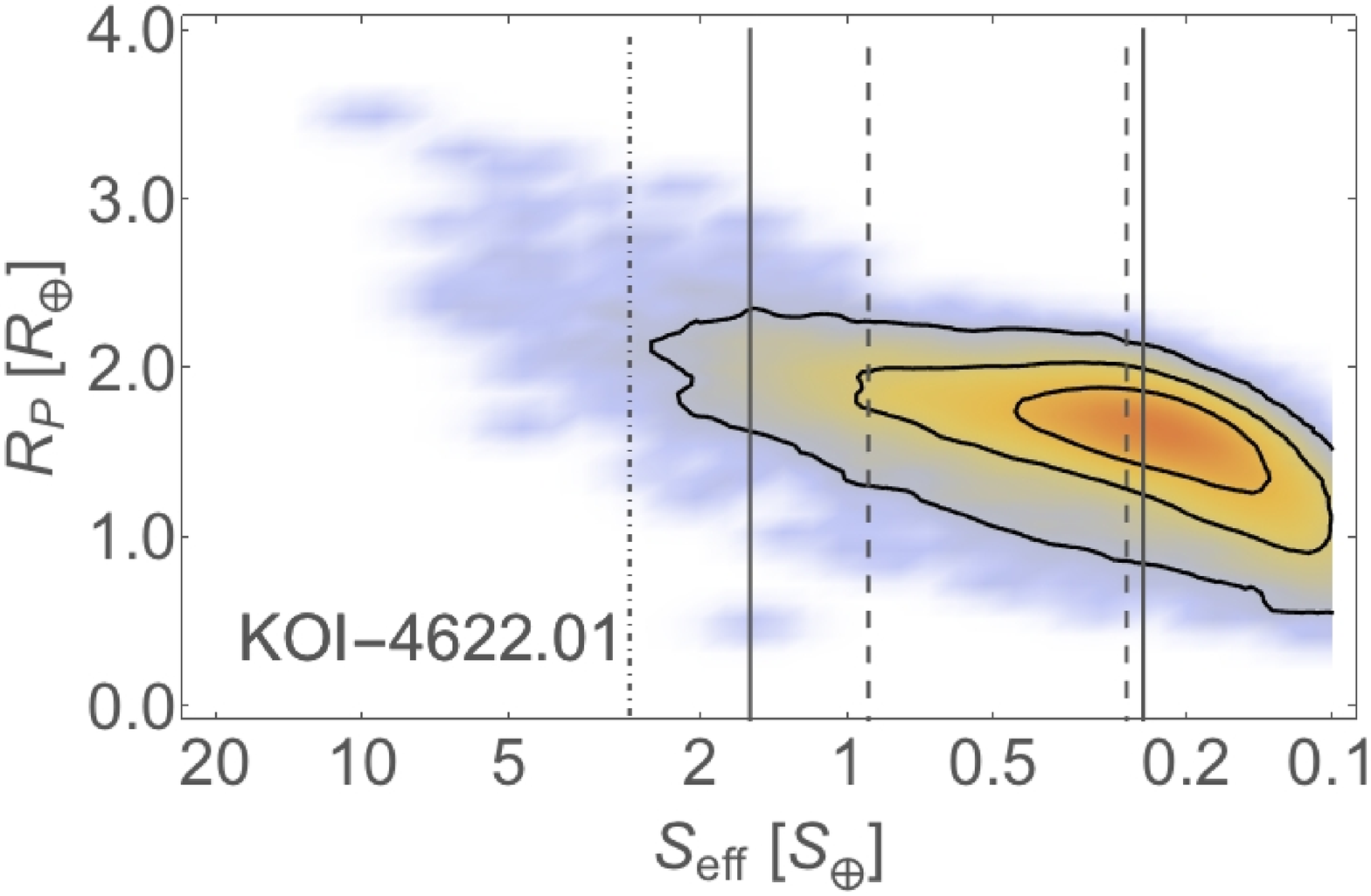}\\[+1ex]
\includegraphics[width=5.9cm]{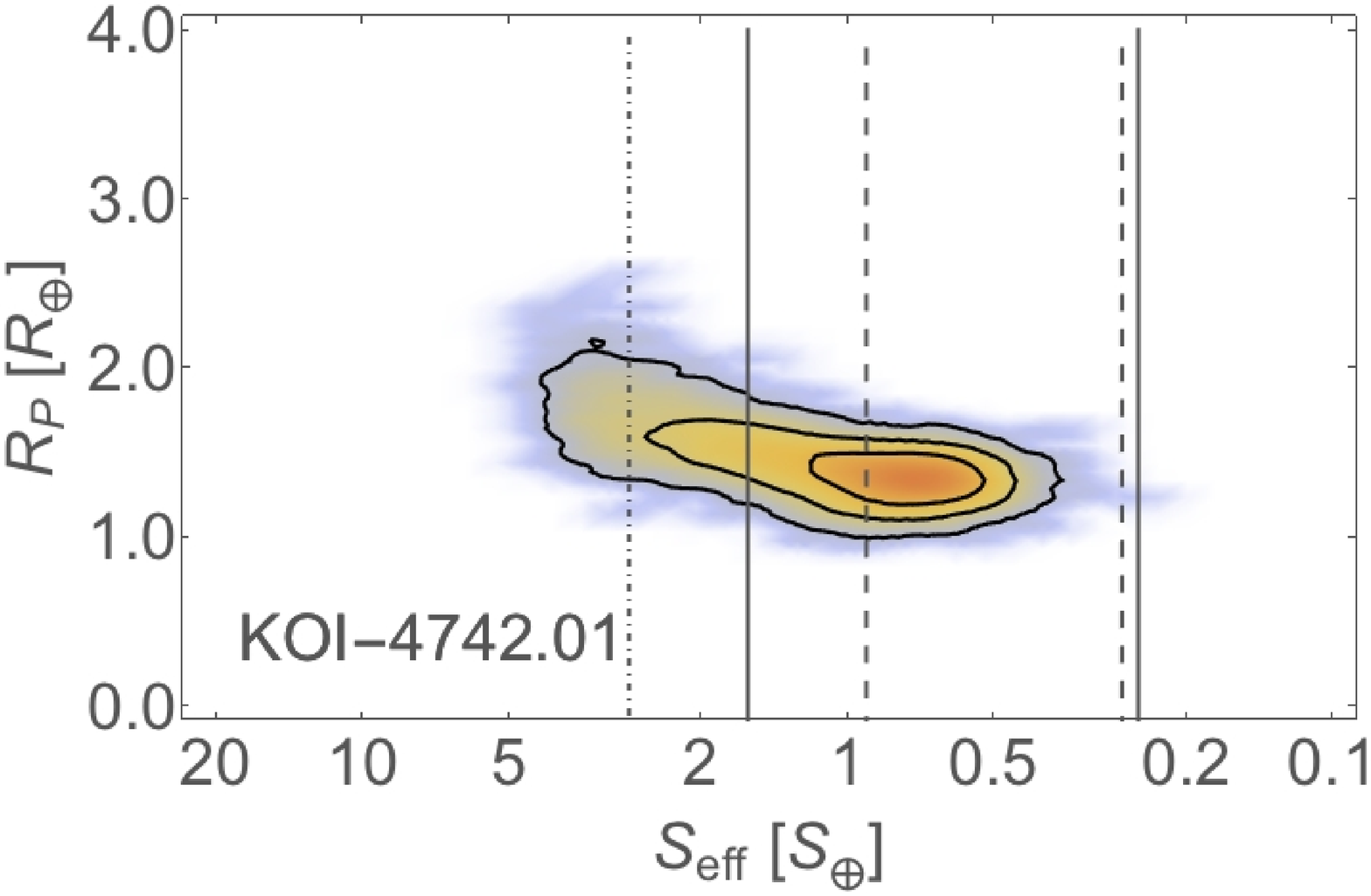} &
\includegraphics[width=5.9cm]{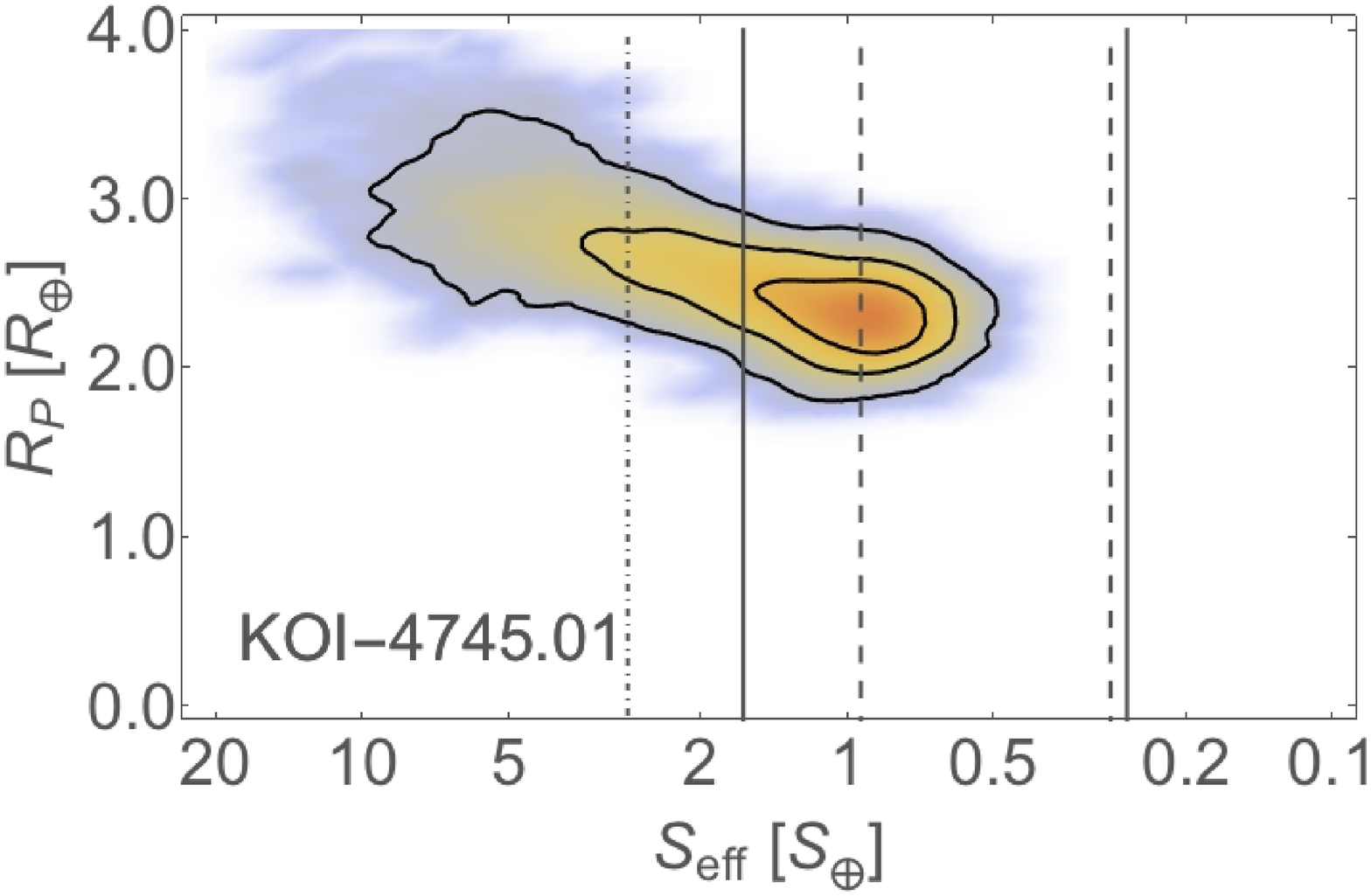} &
\end{tabular}

\figcaption[]{Similar to Fig.~\ref{fig:hab571} for the other planets
  in this study.\label{fig:seff}}

\end{figure*}

Additionally, we consider the radius of the planet to be an important
parameter in assessing its potential for habitability, as this can
determine whether the planet is likely to have a solid surface that
can support liquid water. In the past it was often conventionally
assumed that there likely existed some boundary in radius between
rocky planets and mini-Neptunes \citep[e.g.,][]{Weiss:13}, perhaps
around 1.75\,$R_{\earth}$ \citep{Lopez:13}. However, recent
counter-examples to this paradigm make such a simplistic division
questionable. For example, KOI-314\,c is a mini-Neptune despite being
1.6\,$R_{\earth}$ \citep{Kippingetal:14a}, and Kepler-10\,c has been
claimed to be rocky despite being 2.35\,$R_{\earth}$
(\citealt{Dumusque:14}; for a differing view see \citealt{Rogers:14}
and \citealt{Wolfgang:14}).  Nevertheless, it is reasonable to assume
that smaller planets are more likely to be rocky, and thus we present
joint posterior distributions in radius and insolation. An example of
this diagram for KOI-0571.05 is given in Figure~\ref{fig:hab571}, in
which we indicate various boundaries for the HZ as defined by
\citet{Kopparapu:13} and \cite{Zsom:13}. Similar graphs for the
remaining targets in our sample are shown in Figure~\ref{fig:seff}.

\setlength{\tabcolsep}{-3pt}
\setlength{\tabcolsep}{6pt}   

From the standpoint of habitability it is of interest also to quantify
the likelihood that the planets are rocky, given their measured sizes
and corresponding uncertainties. To do this we have considered an
empirical density model giving a statistical estimate of the mass of a
planet as a function of its radius, constructed based on the available
sample of transiting planets with measured densities (i.e., with
masses measured either via the Doppler technique, or from TTVs). This
model is expected to be biased towards larger and more massive planets
that are easier to detect, but it is possible to remove the bias by
convolving the density model with estimates of the rate of occurrence
of planets as a function of their size \citep{Fressin:13}, which we
have done in a Monte Carlo fashion \citep{Fressin:15}. In this
way we may infer the true distribution of planets in the mass-radius
diagram, which can then be used to establish the fraction of rocky or
gaseous planets at a given radius or mass.  For the purposes of this
work we have considered any planet denser than a composition of 100\%
MgSiO$_3$ perovskite to be rocky \citep{Seager:07}. Using the above
model and our posterior distributions for the radius of the planets in
our sample, we have computed the probability that they are rocky,
$\mathcal{P}[{\rm rocky}]$, by assigning a mass to each link in the
chain and counting the number of samples denser than the threshold,
analogous to what was done previously for the HZ. The
$\mathcal{P}[{\rm rocky}]$ values are listed for each object in
Tables~\ref{tab:finalparams1} and \ref{tab:finalparams2}. Estimates of
$\mathcal{P}[{\rm rocky}]$ using an alternate density model proposed
recently by \cite{Rogers:14} lead to similar assessments, at least
qualitatively.

The results of our calculations indicate that all of the planets in
our sample have a $\sim$50\% or greater chance of being in the
temperate region of their parent stars.  KOI-0571.05, 1422.04, and
4742.01 are the best candidates for residing within the HZ boundaries,
with $\mathcal{P}[{\rm HZ}]$ equal to 98.4\%, 99.7\%, and 96.9\%,
respectively. The first two receive roughly one third of the flux of
the Earth, while the latter receives two thirds. Based on their
measured sizes, KOI-0571.05 and KOI-4742.01 have the best chance of
being rocky; KOI-1422.04 is larger and has a somewhat smaller
chance. KOI-3284.01 is the object in our sample with the highest
probability of being rocky in nature (69.6\%); its size is consistent
with being the same as that of the Earth: $R_p =
1.12_{-0.17}^{+0.16}$\,$R_{\earth}$. Other planets with non-negligible
chances of being rocky are KOI-1422.05, 4087.01, 4622.01, and
4742.01.\footnote{KOI-4427.01 would otherwise join this group, but it
  is not formally validated as a planet to the same level as the
  others. Its insolation is only about a quarter of that of the
  Earth.} On the other hand, the larger planets KOI-2529.02, 4005.01,
and 4745.01 have $\mathcal{P}[{\rm rocky}] \sim 5$\% or smaller, and
are most likely mini-Neptunes. KOI-3255.01 may still be rocky despite
its radius of 2.14\,$R_{\earth}$, with $\mathcal{P}[{\rm rocky}] =
11.7$\%.

In terms of their insolation, four of our validated planets with a
sizable chance of being rocky have $S_{\earth}$ values within 50\% of
that of the Earth: KOI-1422.05, 3284.01, 4087.01, and 4742.01. Of
these worlds, KOI-3284.01 and KOI-4742.01 appear to be the most
``Earth-like'' when considering both their size and amount of incident
flux jointly, surpassing KOI-0571.05, which is also rocky but receives
only one third of the incident flux of our planet.

\section{Discussion and concluding remarks}
\label{sec:discussion}

Eleven of the twelve candidates in our sample have been validated here
as planets to a very high degree of confidence exceeding 99.73\%
(3$\sigma$). The high bar we have set in this work is commensurate
with the importance we assign to these planets, which are all small
and potentially in the HZ of their host stars. Few such objects
have been verified so far, and our work now essentially doubles the number of cases listed in Sect.~\ref{sec:introduction}.

Three of our targets, KOI-0571.05, KOI-1422.04, and KOI-1422.05 are in
systems of five planets each, and have been previously validated by
others using very different methodologies and assumptions. Our
\blender\ validations confirm those assessments using conservative and
realistic hypotheses, generally with a higher degree of
confidence. KOI-1422.04 and KOI-1422.05 were included in the studies
of \cite{Lissauer:14} and \cite{Rowe:14}, who considered them to be
validated based on the general conclusion that the vast majority ($>
99$\%) of \kepler\ candidates in multiple systems are likely to be
bona-fide planets, as they showed statistically.\footnote{We note that
  another of our targets, KOI-2529.02, is a candidate in a two-planet
  system that was also included in the statistical studies above, but
  it was not considered to be validated there due to unresolved issues
  with the follow-up observations.}  In the present work we are able
to assign a specific confidence level (99.74\% and 99.89\%) to these
two planets using \blender. KOI-0571.05 was validated separately by
\cite{Quintana:14} through statistical arguments somewhat similar to
those in \blender\ in some respects, but not in detail.  For example,
\cite{Quintana:14} included short period planets in their estimate of
the planet prior, whereas we chose to count only planets with similar
period as the candidate, because the rates of occurrence of transiting
planets are a strong function of orbital period. For a planet with a
long period of $\sim$130 days such as KOI-0571.05, including short
periods overestimates PL by about a factor of 50 according to our
calculations. On the other hand, including short periods also for the
false positives (both for planets and EBs), as \cite{Quintana:14} have
done, will tend to make the blend frequencies higher. This partly
offsets the bias from the larger planet prior, but still results in a
final odds ratio a factor of 2 greater than if short periods are not
included.  An additional difference is that in order to reach a
comfortable validation level \cite{Quintana:14} applied a
``multiplicity boost'' to their planet prior (increasing it by a
factor of 30), invoking the fact that KOI-0571.05 is in a five-planet
system, that the other four planets were considered validated by
\cite{Lissauer:14} and \cite{Rowe:14}, and that false positives are
significantly less common in multis. While this is a perfectly
legitimate argument, the odds ratio achieved with \blender\ is high
enough (for all our multis) that this boost was not required here. Had
we used it, our confidence level for KOI-0571.05 would be 30 times
higher.

One of our candidates, KOI-4427.01, does not quite reach our threshold
for validation, although in other contexts a 99.16\% confidence level
such as we achieve might be considered high enough to declare it a
bona-fide planet. Interestingly, this is the same candidate for which
our centroid motion analysis indicated less confidence than in other
cases that the transit signal comes from the target itself as opposed
to a nearby background location, consistent with indications from
\blender. We note also that the host star does not show a clear
rotational signature, which causes its properties to be less well
determined than others for which we were able to use an age estimate
from gyrochronology to strengthen the model fits. For these reasons we
do not include this object among our formally validated planets,
although it is entirely possible that additional follow-up
observations may enable a higher level of validation in the future. It
remains a very interesting object, as it is likely to be within the HZ
and also has a fair chance of being rocky.

We find that the mean densities derived photometrically from the
individual planets in each of the five-planet systems KOI-0571 and
KOI-1422 show very good agreement, supporting the notion (for the
first time in KOI-1422) that in each case the five planets orbit the
same star. This is further supported by a similar consistency found
among the limb-darkening coefficients ($u_1$ and $u_2$), which were
fitted for separately for each of the planets. For the four targets in
our sample with close companions the location of the planets is more
uncertain, however.  In one case (KOI-2529.02) both \blender\ and AP
strongly favor the planet transiting the primary star. For
KOI-1422.04, KOI-1422.05, and KOI-3284.01 the evidence is in the same
direction but not as strong, and for KOI-3255.01 both \blender\ and AP
are inconclusive as to which star has the planet.  In these four cases
even if the planet orbits the companion the inferred radii are still
relatively small (1.68--2.25\,$R_{\earth}$), and at least for
KOI-1422.05 and KOI-3284.01 there is still a good chance they are
rocky.

Except for KOI-2529.02, the inferred radii of all our validated
planets are nominally smaller than 2.4\,$R_{\earth}$. In some cases
they are rather different than those reported previously, either
because of differences in the stellar radii, or because some of the
previous studies have not accounted for dilution by nearby
companions. The stellar properties in this work are all based on
spectroscopic estimates of the temperature and metallicity, as well as
$\log g$ for the hotter stars and other constraints on luminosity for
the cooler ones, superseding properties reported previously that have
relied only on photometry.

This study has significantly increased the number of small validated
transiting planets ($< 2.5$\,$R_{\earth}$) that are potentially in the
HZ of their host star.  Excluding those in our sample that have been
announced previously, as well as KOI-4427.01, we now add to the list
KOI-1422.05(*), KOI-2529.02, KOI-3255.01(*), KOI-3284.01(*),
KOI-4005.01, KOI-4087.01(*), KOI-4622(*), KOI-4742.01(*), and
KOI-4745.01. The ones followed by an asterisk are small enough to be
rocky, with a somewhat reduced chance of that being the case for
KOI-3255.01.  We note that rather than relying on an estimate of the
equilibrium temperature for our assessments regarding the HZ, here we
have used the effective insolation, which, as pointed out by
\cite{Kopparapu:13}, dispenses with the uncertainty in the albedo. Our
HZ statements are quantified by providing a probability that the
planet lies within the adopted inner and outer edges of the region,
which properly accounts for the uncertainties in all measured stellar
and planetary properties.  Similarly, we have provided a quantitative
indication of the probability that each of these planets is rocky,
also accounting for uncertainties.

A number of the previously announced small HZ planets (see
Sect.~\ref{sec:introduction}) have radii in excess of 2\,$R_{\earth}$,
and according to our prescription for $\mathcal{P}[{\rm rocky}]$ they
are significantly less likely to be rocky than perhaps deemed to be
the case in the original publications (particularly Kepler-22\,b and
Kepler-296\,f, with radii of 2.38\,$R_{\earth}$ and
2.31\,$R_{\earth}$, respectively). Thus, with the new examples added
here, the list of rocky planets in the HZ is enlarged by a factor of
two.  KOI-3284.01 and KOI-4742.01 are now the validated, rocky, HZ
planets that appear most similar to the Earth when considering both
their size ($R_p = 1.12_{-0.17}^{+0.16}$\,$R_{\earth}$ and $R_p =
1.34_{-0.18}^{+0.11}$\,$R_{\earth}$) and insolation ($S_{\rm eff} =
1.40_{-0.77}^{+0.67}$\,$S_{\earth}$ and $S_{\rm eff} =
0.66_{-0.41}^{+0.23}$\,$S_{\earth}$) jointly.

\acknowledgements

We thank the referee for helpful comments on the original manuscript.
This paper includes data collected by the \kepler\ spacecraft. Funding
for the \kepler\ Mission is provided by NASA's Science Mission
Directorate. The research has also made use of the Michael Dodds
Computing Facility, of NASA's Astrophysics Data System (ADS), and of
data products from the Mikulski Archive for Space Telescopes (MAST).
Some of the data presented herein were obtained at the W.\ M.\ Keck
Observatory, which is operated as a scientific partnership among the
California Institute of Technology, the University of California, and
NASA. We extend special thanks to those of Hawaiian ancestry on whose
sacred mountain of Mauna Kea we are privileged to be guests.  GT
acknowledges partial support for this work from NASA grant NNX14AB83G
(\kepler\ Participating Scientist Program). DMK is supported by the
Harvard College Observatory Menzel Fellowship.

\end{document}